\documentclass[11pt,oneside]{article}

\usepackage[top=1in, bottom=1in, left=1in, right=1in]{geometry}
\usepackage{pdflscape}

\usepackage[T1]{fontenc}
\usepackage[utf8]{inputenc}
\usepackage{newtxtext}
\usepackage{booktabs}
\usepackage[patch={item,toc,eqnum,verbatim}]{microtype}
\usepackage{amsmath,amsthm,amssymb,mathtools}
\usepackage{bbm}
\usepackage{dsfont}
\usepackage[mathscr]{euscript}
\usepackage{subcaption}
\usepackage{natbib}
\usepackage{verbatim}
\usepackage{fancyhdr}
\usepackage{titlesec}
\usepackage{ushort}

\usepackage{graphicx}

\usepackage{rotating}
\usepackage{float}
\usepackage[percent]{overpic}

\usepackage{tikz}
\usetikzlibrary{chains,shapes.multipart,shapes,calc,fit}
\usetikzlibrary{automata,positioning}

\usepackage{array}
\usepackage{booktabs}
\usepackage{longtable}
\usepackage{tabularx}
\usepackage{multirow}
\usepackage[para,online,flushleft]{threeparttable}

\usepackage{enumerate}
\usepackage[inline]{enumitem}

\usepackage[font=small]{caption}

\usepackage{color}

\usepackage{natbib}
\setcitestyle{round}
\usepackage[hidelinks]{hyperref}
\hypersetup{
    colorlinks=true,
    linkcolor=blue,        
    citecolor=blue,        
    urlcolor=blue,         
    filecolor=blue
}
\usepackage{cleveref}
\usepackage{refcount}

\usepackage{algorithm}
\usepackage{algpseudocode}
\newfloat{subroutine}{tbp}{losub}
\floatname{subroutine}{Subroutine}

\usepackage{appendix}
\usepackage{apptools}
\AtAppendix{}

\usepackage[pagewise]{lineno}

\usepackage{setspace}
\usepackage{changepage}
\usepackage[bottom]{footmisc}
\usepackage[all]{xy}
\usepackage{authblk}

\newtheoremstyle{ModifiedStyle}
	{\topsep}{3pt}{}{}{\bfseries}{.}{.5em}{}
\theoremstyle{ModifiedStyle}

\newtheorem{remark}{Remark}

\newtheorem{assumption}{Assumption}

\newtheorem{to-do}{To-Do}

\newcommand{\E}{\mathbb{E}}
\newcommand{\calI}{\mathcal{I}}
\newcommand{\calK}{\mathcal{K}}
\newcommand{\calJ}{\mathcal{J}}
\newcommand{\calB}{\mathcal{J}_{\mathrm b}} 
\newcommand{\calN}{\mathcal{J}_{\mathrm n}} 
\newcommand{\R}{\mathbb{R}}
\newcommand{\bv}{\bar v}

\newcommand{\BSmall}[1]{\mkern-1.7mu\raisebox{-1.2pt}{\scalebox{0.9}{$\scriptscriptstyle B$}}}
\newcommand{\NSmall}[1]{\mkern-1.7mu\raisebox{-1.2pt}{\scalebox{0.9}{$\scriptscriptstyle N$}}}

\makeatletter
\newcommand{\Biggg}{\bBigg@{2.5}}
\newcommand{\vast}{\bBigg@{3}}
\newcommand{\Vast}{\bBigg@{3.5}}
\newcommand{\massive}{\bBigg@{4.5}}
\newcommand{\Massive}{\bBigg@{6}}
\makeatother

\newcommand{\nquad}{\kern-1em}
\newcommand{\thickhline}{%
	\noalign{\ifnum 0=`}\fi \hrule height 1pt
	\futurelet \reserved@a \@xhline}
\newcolumntype{"}{@{\hskip\tabcolsep\vrule width 1pt\hskip\tabcolsep}}
\newcommand{\PreserveBackslash}[1]{\let\temp=\\#1\let\\=\temp}
\newcolumntype{C}[1]{>{\PreserveBackslash\centering}p{#1}}
\newcolumntype{R}[1]{>{\PreserveBackslash\raggedleft}p{#1}}
\newcolumntype{L}[1]{>{\PreserveBackslash\raggedright}p{#1}}

\DeclareMathAlphabet{\mathpzc}{OT1}{pzc}{m}{it}

\makeatletter

\newcommand{\Rom}[1]{\expandafter\@slowromancap\romannumeral #1@}
\makeatother
\newcommand{\customappendixref}[1]{\hyperref[#1]{\Alph{section}}}

\makeatletter
\let\LN@align\align
\let\LN@endalign\endalign
\renewcommand{\align}{\linenomath\LN@align}
\renewcommand{\endalign}{\LN@endalign\endlinenomath}
\let\LN@gather\gather
\let\LN@endgather\endgather
\renewcommand{\gather}{\linenomath\LN@gather}
\renewcommand{\endgather}{\LN@endgather\endlinenomath}
\makeatother

\title{\vspace{-20mm}\textbf{Dynamic Skill-Based Routing for Call Centers with Transfers: A Value-Based Framework}}
\author{Barış Ata\thanks{\texttt{baris.ata@chicagobooth.edu}} \ and Ebru Kaşıkaralar\thanks{\texttt{ebrukasikaralar@gmail.com}}}
\date{\today}
\begin{document}
\maketitle
\vspace{-10mm}
\begin{abstract}
We consider a multi-class, multi-pool queueing model of a telephone call center in which a service completion does not always end the call. The caller may be transferred to another agent for a second service, so the assignment of an agent to a call affects not only the caller's waiting time but also whether the call is resolved and how much further work it creates. We model these transfers as Markovian routing from original to rework classes, and the system manager dynamically assigns agent pools to classes so as to maximize the value generated by service, net of transfer penalties and waiting and abandonment costs, over a finite horizon.  Focusing on the Halfin–Whitt heavy-traffic regime, we derive an approximating Brownian control problem. Its Hamilton–Jacobi–Bellman equation is a semilinear partial differential equation whose dimension equals the number of classes. Building on the deep splitting method of \citet{beck2021deep}, we develop a simulation-based computational method that approximates the value function and its gradient by neural networks, and we use it to construct an implementable routing policy for the original (prelimit) system. We calibrate the model to publicly available data from a large U.S. bank call center. In all test problems we consider, the proposed policy outperforms the best available benchmark policy we could find.
\end{abstract}

\section{Introduction}\label{sec:intro}
In modern call centers, callers are grouped into distinct classes according to their service needs, while agents are organized into server pools according to their skills. Each server pool can serve a subset of caller classes, and skill-based routing refers to the real-time allocation of these pools to classes. In practice, however, a caller’s needs may not be resolved by a single agent. A call may consist of multiple sub-calls, with the caller interacting with several agents before service is complete. This paper studies the dynamic control of such systems in the Halfin–Whitt heavy-traffic regime. Two novel features are central to our model. First, a call may require multiple agent services. Whereas most queueing models of call centers consider single-pass systems, in which a service completion results in departure from the system, our model allows transfers. A service completion may generate additional work for another agent rather than a departure. Second, rather than focusing solely on minimizing congestion costs, we consider a system manager who seeks to maximize the value generated by service.

The first feature reflects the fact that a caller’s true service need may not be known when the call is initially routed. After interacting with an interactive voice response (IVR) system, callers who require further assistance are classified by an automatic call distributor (ACD) and routed to a class queue. This initial classification may be incorrect, sending a call to the wrong class. Alternatively, an agent may discover during the interaction that the caller requires an additional service. In either case, the caller is transferred and receives a second, rework service. We model these handoffs as Markovian routing from original to rework classes.

In our U.S. Bank data, 97.36\% of agent-handled contacts are resolved with a single service, while 2.59\% require two services. Accordingly, we organize the network into original classes, which receive external arrivals, and rework classes, which receive only calls routed after completion of the first service. We focus on the empirically relevant case in which a caller requires at most one rework, so that each contact entails at most two agent services. The system manager dynamically determines which server pool serves which class as new and rework calls accumulate. Although only 2.6\% of contacts in our data involve transfers, transfer rates can be substantially higher in other settings. For a single call type at a comparable U.S. bank call center, \citet{hathaway2023gatekeeper} report agent-level transfer rates ranging from 3.8\% to 36.9\%. Our framework can be modified to accommodate further transfers.

The second feature concerns our objective. Call centers have traditionally been managed as cost centers, with the system manager minimizing operating costs subject to service-level constraints, such as a target requiring that 80\% of calls be answered within 20 seconds \citep{gans2003telephone}. Such formulations capture delays but do not fully account for the quality or value of the service customers ultimately receive. Service quality depends not only on how long a customer waits but also on whether the interaction successfully addresses the customer’s needs. Moreover, because call centers often serve as both the public face of a firm and a source of incremental revenue, the value generated through service has become an increasingly important managerial consideration \citep{aksin2007modern}.

We therefore adopt a value-based objective that assigns a value to the outcome of each service completion. A completion that resolves the caller’s need generates a resolution value, whereas a completion that leads to a transfer incurs a rework penalty. Over a finite horizon, the manager seeks to maximize expected net value, defined as the value generated by service completions, less rework penalties and the costs associated with waiting and abandonment.

To analyze the system, we formulate an approximating Brownian control problem under the Halfin–Whitt heavy-traffic regime \citep{halfin1981heavy}. The resulting drift-rate control problem is characterized by a Hamilton–Jacobi–Bellman (HJB) equation that takes the form of a semilinear partial differential equation (PDE); see \citet{fleming2006controlled}. Conventional numerical methods for solving PDEs, such as finite-element methods, suffer from the curse of dimensionality and become computationally intractable as the state dimension grows. This limitation has restricted many applications of Brownian-control approximations to relatively small systems. To address it, we build on the deep learning literature initiated by \citet{han2018solving}. We describe the computational method in Section~\ref{sec:computational}.

This paper makes two contributions. First, we introduce and analyze a many-server stochastic processing network model that integrates customer transfers with value-based scheduling. We represent transfers through Markovian routing, formulate a total-value objective with outcome-dependent rewards and penalties, and derive the corresponding Brownian control problem. Second, we develop a computational method for solving the resulting high-dimensional control problem. Building on the deep-learning approach for semilinear PDEs developed by \citet{beck2021deep}, we approximate the value function and its gradient and use them to construct an implementable scheduling policy for a queueing network. We evaluate this policy on data-driven instances calibrated to a large U.S. bank call center using publicly available data from the \citet{SEEData}. The data leads to a 26-dimensional test problem, but this problem is intractable because the associated Markov Decision Process (MDP) suffers from the curse of dimensionality. We also consider two variants of this problem. Across all three test problems, our proposed policy outperforms the best available benchmark policy we could find.

The remainder of the paper is organized as follows. Section \ref{sec:literature} reviews the related literature. Section \ref{sec:model} introduces the queueing model with customer transfers and formulates the value-based objective. Section \ref{sec:bcp} derives the diffusion approximation in the Halfin–Whitt regime and the associated HJB equation. Section \ref{sec:computational} presents the computational method. Section \ref{sec:calibration} describes the U.S. Bank call-center data, constructs the data-driven test instances, and introduces the benchmark policies used to evaluate our approach. Section \ref{sec:results} presents computational results comparing the proposed policy with the benchmark policies. The appendices provide further details on the data and the offline approximation used to evaluate the auxiliary function $F(x,v)$, along with derivations of the computational algorithm and the proposed prelimit policy.
\section{Literature Review}\label{sec:literature}
There is a large literature on call center operations; see \citet{gans2003telephone}, \citet{aksin2007modern}, and \citet{koole2023practice} for survey papers. One of the key operational problems in call centers is how to assign agents with different skills to callers with different service needs or attributes. This assignment decision affects waiting times in the system, but it also affects whether the caller's problem is resolved or further service is needed. In reality, serving a customer can remove work from one queue but may generate work at another, depending on the outcome of the first service. Therefore, in this paper, we connect skill-based routing to the control of queueing networks with Markovian routing. 

Several papers that focus on call routing also consider service outcome. \citet{zhou2005managing} consider heterogeneous agents whose service rates and call resolution probabilities differ. Unresolved callers return for service, so the routing decision affects the total time needed to resolve a customer's problem, including callbacks. \citet{mehrotra2012routing} use simulation to learn the tradeoff between waiting times and resolution in a system with multiple call types and agent groups. A similar approach is to assign a value to the match between call types and agent groups. \citet{sisselman2007value} study value-based routing policies, considering expected revenue or the likelihood of first-call resolution as the value resulting as an outcome of the matches.

Unresolved service can cause calls to be transferred or callers to repeat their contact later. Therefore, transfers and repeats have also been studied in other call center models. \citet{hathaway2023gatekeeper} study an agent's decision to continue serving a customer or transfer the customer, using a gatekeeper model and behavioral experiments. \citet{ding2015fluid} develop a fluid approximation for a call center with delayed retrials after abandonment and reconnects after service. In our setting, a caller can move directly from an original class (first service) to a rework class (second service) when their first interaction with an agent ends. The transfer probabilities depend on the service activity, that is, on both the customer class and agent pool providing the service. Consequently, assigning an agent to a call affects the immediate service outcome as well as the amount and destination class of the subsequent work. We account for both effects through a value-based objective with resolution rewards, transfer penalties, and waiting and abandonment costs.

Much of the theoretical work on dynamic scheduling problems in queueing networks involves asymptotic analysis. Two commonly considered asymptotic regimes are the conventional heavy-traffic regime and the many-server Halfin--Whitt regime proposed by \citet{halfin1981heavy}. In the conventional heavy-traffic regime, \citet{harrison1999heavy} study skill-based routing and resource pooling in parallel-server systems and propose discrete-review policies based on an approximating Brownian control problem. \citet{ata2005heavy} establish asymptotic optimality of discrete-review policies for a broader class of open processing networks with feedback routing. Feedback also changes scheduling and routing priorities in other service applications. \citet{huang2015control} study a multiclass queue motivated by emergency departments and derive asymptotically optimal policies that account for repeat service in conventional heavy traffic.

An important antecedent of our model is \citet{harrison2003broader}. This paper describes stochastic processing networks in terms of activities that consume and produce material subject to resource constraints and includes the system manager's economic objective in the definition of the nominal plan. In our model, a service activity removes a call from one class and may produce a call in a rework class. The Markovian routing probabilities then determine the expected net flow associated with each activity, and the resolution rewards and transfer penalties determine its expected value. Following \citet{harrison2003broader}, we choose a nominal plan that maximizes value subject to flow balance and capacity constraints and derive the objective of the Brownian control problem as the loss relative to the nominal plan. However, we focus on the Halfin--Whitt regime in our study.

In the Halfin--Whitt regime, several papers study dynamic scheduling under different network topologies. \citet{harrison2004dynamic} consider a multiclass queueing system with a single server pool, where a system manager chooses dynamic server allocations to minimize infinite-horizon discounted abandonment and holding costs. They characterize the approximating control problem through its HJB equation. \citet{armony2005dynamic} studies skill-based routing in an inverted-$V$ system with a single customer class served by multiple server pools and proposes a fastest-server-first (FSF) routing policy, establishing its asymptotic optimality for steady-state queue length and virtual waiting time. Focusing on tree-like networks, Atar (\citeyear{atar2005diffusion},\,\citeyear{atar2005scheduling}) considers multiple customer classes and server pools. The proposed policies are asymptotically optimal and require knowledge of the value function and its gradient for the limiting control problem. Other policies for many-server systems include the queue-and-idleness-ratio (QIR) controls of Gurvich and Whitt (\citeyear{gurvich2009queue},\,\citeyear{gurvich2009scheduling}) and the static priority policy studied by \citet{tezcan2010dynamic} for an $N$-system with pool-dependent service rates. In these parallel-server models, a customer departs from the system immediately after service completion.

A difference of our model is that the service activity used also affects how customers move through the network. A completion can simultaneously reduce the original-class population and increase a rework-class population. Thus, routing probabilities enter both the drift and the covariance of the diffusion approximation, and the allocation decision depends on congestion at the rework classes. Moreover, an activity's immediate value depends on whether it resolves the call or produces a transfer. We combine these features with multiple server pools, time-varying arrivals and staffing, and a finite planning horizon. The resulting problem requires a policy that accounts for the value of service together with its consequences for future workload throughout the network.

The heavy-traffic approach to solving dynamic control problems generally involves deriving the approximating Brownian control problem (BCP) and using the solution to the associated HJB equation to propose policies for the original (prelimit) system. Therefore, the heavy-traffic approach provides a way to study our problem, but finding an implementable policy in practice requires a further step. That step involves solving the HJB equation, which is computationally intractable in high dimensions using traditional methods such as finite-element methods. Therefore, earlier literature has considered relatively low-dimensional instances; the computational study in \citet{harrison2004dynamic} considers a two-dimensional queueing system. Our contribution is to develop a computational method for the high-dimensional network control problem and evaluate the resulting scheduling policy on data-driven test problems.

To overcome this computational challenge, we build on methods that use deep neural networks to approximate solutions of PDEs. One approach is based on physics-informed neural networks (PINNs), introduced by \citet{raissi2019physics}. A single neural network is trained to satisfy the differential equation as well as the initial and boundary conditions by minimizing a loss function evaluated at randomly sampled points within a domain and on its boundary. Derivatives of the approximation are evaluated by automatic differentiation. For a second-order PDE such as our HJB equation, this entails computing Hessians through automatic differentiation, which can be computationally expensive in high-dimensional settings. However, several techniques have been proposed to speed up these computations; see, for example, \citet{he2023learning} and \citet{hu2024hutchinson}.

A second approach is to reformulate semilinear PDEs as backward stochastic differential equations (BSDEs), building on the work of \citet{pardoux1990adapted} and \citet{ma1994solving}. The deep BSDE method introduced by \citet{e2017deep} and \citet{han2018solving} treats the value function $V$ at a given initial state as an unknown parameter and uses feedforward neural networks to approximate its spatial gradient $\nabla_x V$ at later discrete time points. A discretized BSDE propagates these approximations along simulated state paths. The initial value and the network parameters are fitted jointly by minimizing a global loss that measures the mean squared difference between the simulated value at time $T$ and the terminal condition. This stochastic formulation avoids calculating Hessians and, in our setting, allows training data to be sampled under a reference policy. For reviews of this literature, see \citet{e2021algorithms}, \citet{beck2023overview}, \citet{chessari2023numerical} and \citet{han2025brief}.

A computational challenge with the deep BSDE method is that the neural network parameters for all time points are trained jointly, creating a large optimization problem. \citet{beck2021deep} propose the deep splitting method, which divides the computation into a sequence of learning problems over shorter time intervals. For a terminal-value problem, the method proceeds backward in time and uses a Feynman--Kac representation to write the value function at each step in terms of conditional expectations. This leads to a separate optimization problem on each interval that is much smaller. \citet{germain2022approximation} provide an approximation error analysis of deep splitting for semilinear PDEs. We adapt this method to our queueing network by learning the value function and its gradient at successive time points. The resulting gradient estimates are used to implement a dynamic scheduling policy in the original (prelimit) system.

Recent research in operations management uses these neural network-based PDE methods to solve dynamic control problems. Two papers closely related to our work are \citet{ata2026dynamic} and \citet{ata2026parallel}. The former develops a neural network method for a nonstationary multiclass queue with one many-server pool over a finite horizon. The latter considers a parallel-server system with multiple pools under an infinite-horizon discounted-cost criterion. Both use Halfin--Whitt approximations to construct scheduling policies for systems with many customer classes. Our paper extends these single-pass models to a queueing network with activity-dependent Markovian routing, where service completions can generate additional work in rework classes. Consequently, service allocations affect both the immediate service outcome and the subsequent workload throughout the network. Our setting also introduces additional computational challenges. We consider a nonstationary setting in which evaluating the neural network training loss exactly requires solving optimization problems that vary over time. To avoid repeatedly solving these optimization problems during training, we approximate their solutions offline by neural networks and evaluate these networks during training for each time interval. We also consider a value-based heavy-traffic condition through a unique value-maximizing nominal plan. Our scheduling objective maximizes resolution rewards net of transfer penalties and waiting and abandonment costs. This formulation allows routing decisions to account for the practical importance of resolving callers’ needs and the economic value of service, as well as congestion costs. For other applications of similar neural network-based methods in operations management, see \citet{ata2024drift}, \citet{ata2024singular}, \citet{ata2025eviction}, \citet{ata2025matching} and \citet{ata2025replenishment}.

Finally, we draw the reference policies for our computational study from the literature. The classical $c\mu$ rule of \citet{cox1961queues} assigns priority according to holding costs and service rates. \citet{mandelbaum2004scheduling} establish heavy-traffic optimality of a generalized $c\mu$ rule for flexible-server systems with convex delay costs in the conventional heavy traffic regime. The $c\mu/\theta$ rule of \citet{atar2010cmu} incorporates abandonment. We also draw on the fastest-server-first rule of \citet{armony2005dynamic} and the resolution-based and value-based routing ideas of \citet{zhou2005managing} and \citet{sisselman2007value}; see Section \ref{sect:benchmarks} for further details.
\color{black}

\section{Model}\label{sec:model}
We consider a queueing network with \(I\) customer classes and \(K\) server pools. Let \(\mathcal I=\{1,\ldots,I\}\) and \(\mathcal K=\{1,\ldots,K\}\) denote the sets of classes and server pools, respectively. Each pool contains many servers that have identical capabilities. The system operates over a finite horizon \([0,T]\) and both the staffing levels and the external arrival rates may vary over time. We denote by $N_{k}(t)$ the number of servers in pool $k$ at time $t$.

Service is provided through activities. We represent the compatibility structure by the bipartite graph \(G=(\mathcal I\cup\mathcal K,\mathcal{E})\), where an edge \((i,k)\in \mathcal{E}\) indicates that server pool \(k\) can serve class \(i\). We refer to each edge as an activity and let \(J=|\mathcal{E}|\) be the total number of activities. Let \(\mathcal J=\{1,\ldots,J\}\) denote the activity set. For each activity \(j\in\mathcal J\), let \(b(j)\in\mathcal I\) and \(s(j)\in\mathcal K\) denote its associated customer class and server pool, respectively. Thus, activity \(j\) corresponds to the edge \((b(j),s(j))\in \mathcal{E}\) and represents the service of class \(b(j)\) by pool \(s(j)\). For later use, we define
\[ \mathcal{B}(i)=\{j\in\mathcal J:b(j)=i\},\qquad i\in\mathcal I, \]
and
\[ \mathcal{S}(k)=\{j\in\mathcal J:s(j)=k\},\qquad k\in\mathcal K. \]
Accordingly, \(\mathcal{B}(i)\) is the set of activities through which class \(i\) can be served, whereas \(\mathcal{S}(k)\) is the set of activities performed by server pool \(k\).
We partition the customer classes into the disjoint sets of original and rework classes as follows: $\mathcal I=\mathcal I_{o}\cup\mathcal I_{r}.$
Without loss of generality, we index the classes so that
\[ \mathcal I_{o}=\{1,\ldots,|\mathcal I_{o}|\}, \qquad \mathcal I_{r}=\{|\mathcal I_{o}|+1,\ldots,I\}. \]
We partition and index the activities analogously. In particular, let
\[ \mathcal J_{o}=\{j\in\mathcal J:b(j)\in\mathcal I_{o}\}, \qquad \mathcal J_{r}=\{j\in\mathcal J:b(j)\in\mathcal I_{r}\}, \]
and order them so that
\[ \mathcal J_{o}=\{1,\ldots,|\mathcal J_{o}|\}, \qquad \mathcal J_{r}=\{|\mathcal J_{o}|+1,\ldots,J\}. \]

A class \(i\in\mathcal I_{o}\) represents a caller’s first service. Calls of this class arrive externally according to a nonhomogeneous Poisson process with rate \(\lambda_i(t)>0\) at time $t$. A rework class \(i\in\mathcal I_{r}\) represents a caller’s second service. Rework may arise because the call was misclassified upon arrival or because the first service, although correctly classified, did not fully resolve the caller’s need. Rework classes receive no external arrivals, so \(\lambda_i(t)=0\) for \(i\in\mathcal I_{r}\), and receive work only through internal transfers following first-service completions.

As described above, the routing structure consists of two layers. Upon completion of an original-class activity \(j\), with \(b(j)\in\mathcal I_{o}\), the caller is routed to rework class \(l\in\mathcal I_{r}\) with probability \(P_{jl}\geq 0\). If the caller is not routed to a rework class, then he exits the system. Thus, the departure probability following activity \(j\) is
\begin{equation}
q_j=1-\sum_{l\in\mathcal I_{r}}P_{jl}.
\end{equation}
After completing a rework-class service, the caller exits the system with probability one.

\begin{assumption}[Rework service completions exit]\label{ass:rework}
  For every activity $j$ with $b(j)\in\calI_r$ we have $P_{jl}=0$ for all
  $l\in\calI$.
\end{assumption}

\begin{figure}[h]
    \centering
    \includegraphics[width=0.8\linewidth]{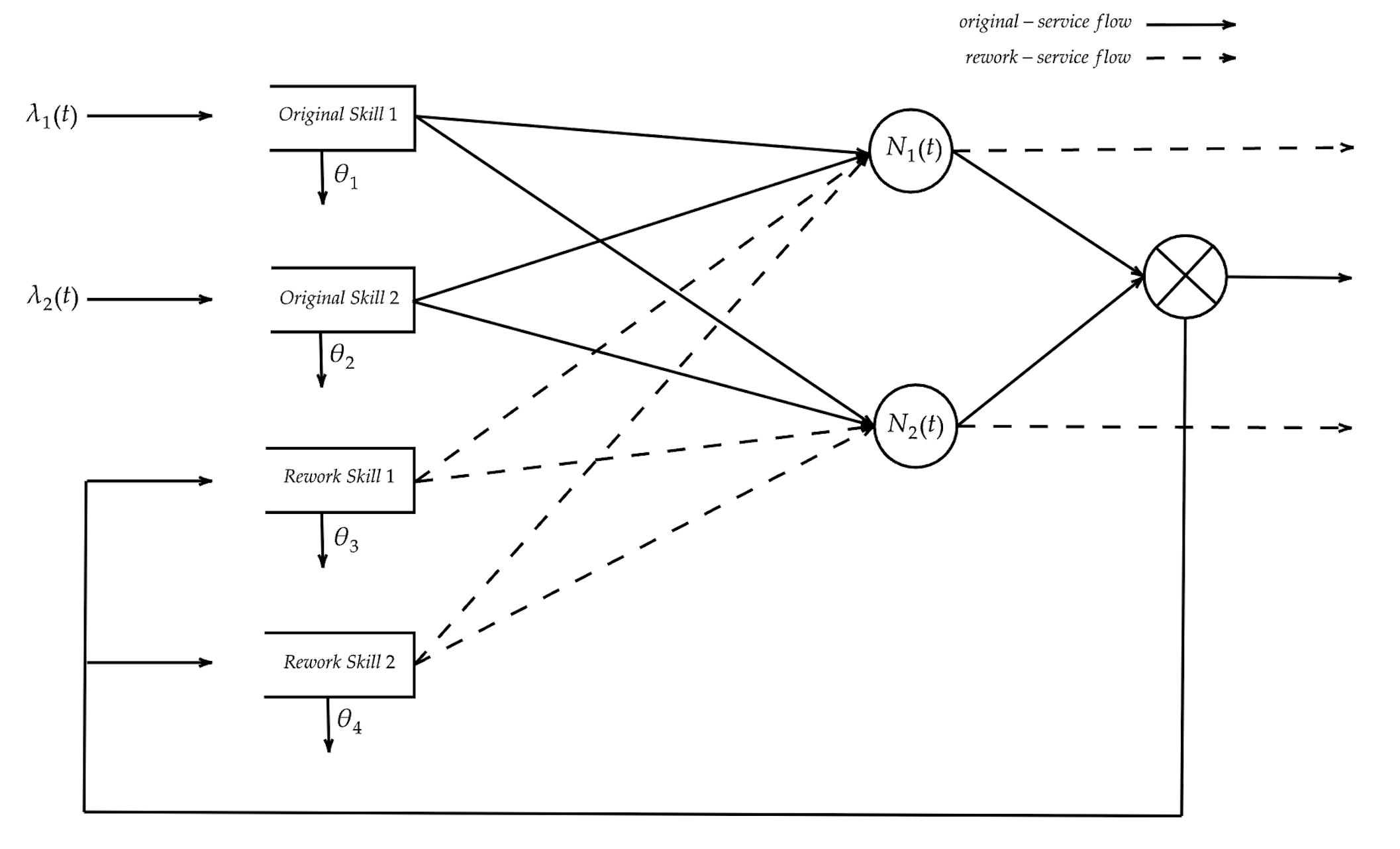}
    \caption{Queueing-network structure with four customer classes and two server pools, each of which can serve all four classes. The original classes (Original Skills 1 and 2) receive external Poisson arrivals at rates \(\lambda_1(t)\) and \(\lambda_2(t)\), respectively. The rework classes (Rework Skills 1 and 2) receive no external arrivals and are populated only through transfers following original-service completions. Solid edges represent original-service flows. After completing an original service, a caller reaches the routing node \(\otimes\) and either exits the system following resolution or is transferred to a rework class. Dashed edges represent rework-service flows, after which callers exit the system directly. Class $i$ callers abandon at rate \(\theta_i\). The system is a unitary stochastic processing network in the sense of \citet{bramson2003two}. Its structure parallels the job-shop model of \citet{kumar1999} and is depicted in the style of \citet[Fig. 3]{bramson2003two}, with \(\otimes\) representing the routing of callers from original to rework classes.}
    \label{fig:model-structure}
\end{figure}

The dataset we use supports this two-tier service structure. In our U.S. Bank data, 97.36\% of agent-handled contacts are resolved after a single service, 2.59\% require exactly two services, and therefore one transfer. Only 0.05\% involve more than two services, or equivalently, more than one transfer (Table \ref{tab:subcalls}). Motivated by this empirical pattern, we focus on a system in which each caller can receive at most one rework service. After the initial service, the caller either exits the system or is routed to a rework class and after completing the rework service, the caller exits the system.

Service times for activity~$j$ form an i.i.d.\ sequence of exponential
random variables with mean $1/\mu_j>0$, where $\mu_j$ is the corresponding
service rate, which can depend on both the class~$b(j)$ served and the server
pool~$s(j)$.  Callers may abandon while
waiting. Class~$i$ callers abandon at rate $\theta_i > 0$, i.e., their patience time is exponential with mean $1/\theta_i$.  We assume the arrival,
service, routing, and abandonment processes are mutually independent.

We describe the association between server pools and activities by the
$K\times J$ capacity-consumption matrix $A$, and the association
between classes and activities by the $I\times J$ constituency matrix
$C$.  The matrix $A$ is
\begin{equation}\label{eq:A-matrix}
  A_{kj} = \begin{cases} 1, & \text{if } s(j)=k,\\ 0, & \text{otherwise,}\end{cases}
  \qquad k\in\calK,\ j\in\calJ,
\end{equation}
and the matrix $C$ is
\begin{equation}\label{eq:C-matrix}
  C_{ij} = \begin{cases} 1, & \text{if } b(j)=i,\\ 0, & \text{otherwise,}\end{cases}
  \qquad i\in\calI,\ j\in\calJ.
\end{equation}
Both $A$ and $C$ are matrices of zeros and ones with exactly one nonzero
entry in each column, since an activity uses a single pool and serves a
single class, and at least one nonzero entry in each row. Column~$j$ equals
$e_{s(j)}$ and $e_{b(j)}$, respectively, while row~$k$ of $A$ and row~$i$ of
$C$ are supported on $\mathcal{S}(k)$ and $\mathcal{B}(i)$, where $e_{1}, \ldots, e_{I}$ are the basis vectors for $\mathbb{R}^{I}$.

A completion of activity \(j\) removes one customer from class \(b(j)\) and routes that customer to class $l$ with probability $P_{jl}$. More specifically, if \(b(j)\in\mathcal I_{o}\), a server engaged in activity \(j\) completes service at rate \(\mu_j\), thereby depleting class \(b(j)\) at rate \(\mu_j\) and generating class $l$ rework customers at rate \(\mu_jP_{jl}\), for $l\in\mathcal I_{r}$. If \(b(j)\in\mathcal I_{r}\), the server again depletes class \(b(j)\) at rate \(\mu_j\), but generates no additional work because the customer exits after completing the rework service. Thus, \(P_{jl}=0\) for every \(l\in\mathcal I\) whenever \(j\in\mathcal J_{r}\). We collect the routing probabilities into the \(J\times I\) matrix
\begin{equation}\label{eq:P-block}
P= \begin{pmatrix} 0 & P^{o}\\ 0 & 0 \end{pmatrix}
\end{equation}
where the rows are partitioned according to \(\mathcal J_{o}\) and \(\mathcal J_{r}\), and the columns according to \(\mathcal I_{o}\) and \(\mathcal I_{r}\). The only potentially nonzero block is
\[ P^{o}=\bigl(P_{jl}\bigr)_{j\in\mathcal J_{o},\,l\in\mathcal I_{r}}, \]
which contains the transfer probabilities from original work classes to rework classes. All remaining blocks are zero because no service completion routes a customer into an original class and every rework-service completion results in departure from the system.

The system state is $X(t)=(X_1(t),\ldots,X_I(t))^{'}$, where $X_i(t)$
denotes the number of class~$i$ customers in the system at time~$t$.
The system manager assigns servers to callers dynamically over time by
choosing how many servers to devote to each activity.  Thus, the system
manager's control is the $J$-dimensional process
$\psi=\{\psi(t),\,t\geq 0\}$, where $\psi_j(t)$ denotes the number of
servers working on activity~$j$ at time~$t$.  For simplicity, we assume
services can be interrupted at any time and resumed later without
efficiency loss. Given time~$t$ and system state $X(t)=x$, the action
$\psi(t)$ must belong to the set of feasible actions
\begin{equation}\label{eq:flexibility}
  \mathcal{A}(t,x) = \left\{
    \psi\in \mathbb{Z}_+^J \;:\;
    \sum_{j\in\mathcal{B}(i)}\psi_j\leq x_i,\; i\in\calI;\quad
    \sum_{j\in\mathcal{S}(k)}\psi_j\leq N_k(t),\; k\in\calK
  \right\}.
\end{equation}
The first constraint ensures that the number of class~$i$ customers in
service does not exceed the number in the system ($i\in\calI$).  The
second constraint states that the number of busy servers at pool~$k$
cannot exceed the staffing level~$N_k(t)$ ($k\in\calK$).  

In addition, we use $Y_i(t)$ to denote the number of class~$i$ customers
waiting in queue and $Z_k(t)$ to denote the number of idle servers at
pool~$k$ at time~$t$.  These processes jointly satisfy the
following constraints for $i\in\calI$, $k\in\calK$,  $j \in \mathcal{J}$ and $t\geq 0$:
\begin{align}
  &Y_i(t)+\sum_{j\in\mathcal{B}(i)}\psi_j(t) = X_i(t),
    \label{eq:queue-balance}\\
  &Z_k(t)+\sum_{j\in\mathcal{S}(k)}\psi_j(t) = N_k(t), 
    \label{eq:server-balance}\\
  &Y_{i}(t) \geq 0, \quad Z_{k}(t) \geq 0, \quad X_{i}(t) \geq 0, \quad \psi_{j}(t) \geq 0. \label{eq:positivity}
\end{align}

Next, we define the arrival, service, routing, and abandonment processes
formally.  The cumulative number of external arrivals to class $i$ until
time~$t$, denoted by $E_i(t)$, is
\begin{equation}\label{eq:arrival}
  E_i(t) = N_i^a\Big(\int_0^t\lambda_i(s)\,ds\Big),\qquad i\in\calI,\quad 0\le t\le T,
\end{equation}
where $N_i^a(\cdot)$ is a unit-rate Poisson process and $\lambda_i(\cdot)$
is the time-varying external arrival-rate function of class~$i$, with
$E_i(\cdot)\equiv 0$ for every rework class $i\in\calI_r$. That is, $\lambda_{i}(\cdot) \equiv 0$ for $i \in \mathcal{I}_{r}.$  Given a control
$\psi$, let $T_j(t)=\int_0^t\psi_j(s)\,ds$ be the cumulative server-effort spent on
activity~$j$ by time~$t$.  The cumulative number of service completions via activity~$j$ by time~$t$ is then given as follows:
\begin{equation}\label{eq:service}
  S_j(T_j(t)) = N_j^s\big(\mu_j\,T_j(t)\big),\qquad j\in\calJ,\quad t\geq 0,
\end{equation}
where $N_j^s(\cdot)$ is a unit-rate Poisson process, and we let $S(T(t))=\big(S_1(T_1(t)),\ldots,S_J(T_J(t))\big)^{\prime}$.  

To describe the routing process, we associate with each activity~$j$ an i.i.d.\ sequence
$\{\phi_j(m)\}_{m\geq 1}$ of $I$-dimensional routing vectors taking values
in $\{e_0,e_1,\ldots,e_I\}$, where $e_0$ is the zero vector and, for
$l\in\calI$, $e_l$ is the $l$th unit vector in $\mathbb{R}^{I}$.  The vector $\phi_j(m)$ records
the destination of the $m$th customer served by activity~$j$. It equals
$e_l$ when that customer is routed to class~$l$ and $e_0$ when the caller
exits the system.  We set $\Pr(\phi_j(1)=e_l)=P_{jl}$ for $l\in\calI$ and write $P=(P_{jl})$
for the $J\times I$ routing matrix, so that
$\Pr(\phi_j(1)=e_0)=1-\sum_{l\in\calI}P_{jl}$.  The cumulative routing
process associated with activity~$j$ is the $I$-dimensional process
\begin{equation}\label{eq:routing}
  \Phi_j(l) = \sum_{m=1}^{l}\phi_j(m),\qquad l=1,2,\ldots,
\end{equation}
whose $i$th component $\Phi_{j,i}(l)$ counts how many of the first $l$
customers served by activity~$j$ are routed to class~$i$.
Lastly, the cumulative number of class~$i$ abandonments until time~$t$,
denoted by $A_i(t)$, is
\begin{equation}\label{eq:abandonment}
  A_i(t) = N_i^b\!\left(\theta_i\int_0^t Y_i(s)\,ds\right),
  \qquad i\in\calI,\quad t\geq 0,
\end{equation}
where $N_i^b(\cdot)$ is a unit-rate Poisson process. We assume
that the processes $N_{i}^{a}(\cdot), N_{j}^{s}(\cdot), \{\phi_{j}(m)\}$ and $N_i^b(\cdot)$ for $i\in\calI$ and
$j\in\calJ$ are mutually independent.  One can then describe the system
dynamics by
\begin{equation}\label{eq:dynamics}
  X(t) = X(0) + E(t) + \sum_{j\in\calJ}\Phi_j\big(S_j(T_j(t))\big)
    - C\,S(T(t)) - A(t),\qquad 0\le t\le T.
\end{equation}
The last four terms on the right-hand side represent external arrivals, routing inflows, service completions, and
abandonments, respectively.

\subsection{System Manager's Objective}\label{sec:objective}
The system manager's objective combines service value with waiting and abandonment costs. For activity $j$, $v_{j} > 0$ denotes the value generated by the activity if the caller leaves the system upon completion of the activity. If an original-service activity $j \in \mathcal{J}_{o}$ instead routes the caller to rework class $l \in \mathcal{I}_{r}$, the system manager incurs a penalty $\kappa_{jl}\geq0$. Because the parameters are activity specific, they may depend on both the service class $b(j)$ served and the server pool $s(j)$ providing the service.

By time $t$, activity $j \in \mathcal{J}_{o}$ has completed $S_{j}(T_{j}(t))$ services, of which $\Phi_{j,l}(S_{j}(T_{j}(t)))$ have been routed to rework class $l$. Thus, the number of its completions after which the caller leaves the system is given by 
\begin{equation}
    S_{j}(T_{j}(t)) - \sum_{l \in \mathcal{I}_{r}} \Phi_{j,l}(S_{j}(T_{j}(t))).
\end{equation}
Thus, the realized cumulative value generated by activity $j$ by time $t$, denoted by $\mathcal{R}_{j}(t)$, is therefore
\begin{equation}\label{eq:realized-reward-activity}
  \mathcal R_j(t)
  =
  \begin{cases}
    \displaystyle
    v_{j}\Big[S_j(T_j(t))-\sum_{l\in\calI_r}\Phi_{j,l}\big(S_j(T_j(t))\big)\Big]
    -\sum_{l\in\calI_r}\kappa_{jl}\,\Phi_{j,l}\big(S_j(T_j(t))\big),
    & j\in\calJ_{o},\\[2.5ex]
    \displaystyle
    v_{j}\,S_j(T_j(t)),
    & j\in\calJ_{r}.
  \end{cases}
\end{equation}
Here, the second case uses the fact that every completed rework service ends the call.
\color{black}
The total realized service value is
\begin{equation}\label{eq:realized-reward}
  \mathcal R(t)=\sum_{j\in\calJ}\mathcal R_j(t).
\end{equation}

For class~$i$, the holding cost rate is $h_i\geq 0$ per customer per unit
of time, and the abandonment penalty is $a_i\geq 0$ per customer who
abandons.  We define the effective cost rate of class~$i$,
\begin{equation}\label{eq:effective-cost}
  c_i = h_i + \theta_i a_i \geq 0,\qquad i\in\calI,
\end{equation}
the cost incurred per unit time by a waiting class~$i$ caller in the queue.  Thus, under a
feasible control $\psi$, the cumulative net value earned by time~$t$ is
\begin{equation}\label{eq:cumulative-value}
  \Pi(t) = \mathcal R(t)
    - \int_0^t h\cdot Y(s)\,ds
    - a\cdot A(t).
\end{equation}

To facilitate the analysis that follows, we define the mean value generated per completion of activity~$j$ by  
\begin{equation}\label{eq:vbar}
  \bv_j
  =
  \begin{cases}
    \displaystyle
    q_j\,v_{j} - \sum_{l\in\calI_r} P_{jl}\,\kappa_{jl},
    & \quad j\in\calJ_{o},\\[2ex]
    v_{j},
    & \quad j\in\calJ_{r}.
  \end{cases}
\end{equation}
Also, for $j \in \mathcal{J}_{o}$, let 
\begin{equation}
    K_{j,l}(t) = \Phi_{j,l}\,(S_{j}(T_{j}(t))) - P_{jl}S_{j}(T_{j}(t))
\end{equation}
denote the centered-routing count process, which has zero mean. Then, one can verify that 
\begin{equation}\label{eq:reward-decomposition}
  \mathcal R_j(t)
  =
  \begin{cases}
    \displaystyle
    \bv_j\,S_j(T_j(t))
    - \sum_{l\in\calI_r}\big(v_{j}+\kappa_{jl}\big)\,K_{j,l}(t),
    & j\in\calJ_{o},\\[2ex]
    v_{j}\,S_j(T_j(t)),
    & j\in\calJ_{r}.
  \end{cases}
\end{equation}
Because $K_{j,l}$ has zero mean, taking expectations
in~\eqref{eq:reward-decomposition} gives
\begin{equation}\label{eq:expected-reward-reduction}
  \E_{x}^{\psi}\big[\mathcal R_{j}(t)\big]
  = \bv_{j}\,\E_{x}^{\psi}\big[S_{j}(T_{j}(t))\big]
  = \bv_{j}\,\mu_{j}\,\E_{x}^{\psi}\!\left[\int_{0}^{t}\psi_{j}(s)\,ds\right],
  \qquad j\in\calJ,
\end{equation}
where the second equality follows from Equation~\eqref{eq:service}.
Thus, every completion of activity~$j$ is valued at $\bv_j$ on average,
so the instantaneous value generation rate at time~$t$ is
$\sum_{j\in\calJ}\bv_j\mu_j\psi_j(t)$.  Similarly, by
Equation~\eqref{eq:abandonment}, class~$i$ callers abandon at rate
$\theta_i Y_i(t)$, so the instantaneous cost rate at time~$t$ is
$c\cdot Y(t)$, where $c_i$ is defined in
Equation~\eqref{eq:effective-cost}.

Callers remaining in the system at time $T$ are served during
overtime. Although these services generate value, they also incur
additional labor costs. As reported in Section~\ref{sec:calibration},
the mean hourly wage of a U.S.\ call center representative is \$17,
and each agent handles approximately $15$ calls per hour, implying a
labor cost of at least \$1.13 per call. The activity values in
Tables~\ref{tab:vj-orig} and~\ref{tab:vj-rework} range from \$1.86 to
\$8.75 and exceed this cost. Thus, crediting the system manager with
the value of overtime services net of labor costs would provide little
incentive to complete services within the planning horizon. To
encourage service completion by time $T$, we let $g(X(T))$ denote the terminal
value assigned in the objective to the callers remaining at time $T$,
and we set
\begin{equation}\label{eq:terminal-zero}
  g(X(T)) = 0 .
\end{equation}
This choice removes the credit for serving callers after the planning
horizon and discourages leaving a terminal backlog.

The system manager seeks to maximize the expected net value over the
time horizon, that is, the value generated by service completions less rework
penalties and the costs of waiting and abandonment. Therefore, the total
expected net value under policy~$\psi$ over the interval $[t,T]$, given
that $X(t) = x$, can be written as
\color{black}
\begin{equation}\label{eq:objective-vbar}
  J(t,x;\psi)
  =\E^{\psi}_{x}\!\left[\int_{t}^{T}
    \Big(\sum_{j\in\calJ}\bv_j\mu_j\psi_j(s) - \sum_{i\in\calI}c_iY_i(s)\Big)\,ds\right],
\end{equation}
where $\E_{x}^{\psi}$ denotes the conditional expectation starting in
state~$x$ under policy~$\psi$.

The problem formulated above is a Markov decision process (MDP).  Although
it can in principle be solved by standard dynamic programming, this
approach becomes computationally intractable when the state vector is
high-dimensional due to the curse of dimensionality.  To address this,
we derive a diffusion approximation of the control problem and solve it
using deep neural network approximations.

\section{Approximating Brownian Control Problem}\label{sec:bcp}

In order to derive a tractable approximation to the control problem
described in Section~\ref{sec:model}, we consider a sequence of systems
indexed by $n=1,2,\ldots$, each having the structure described in
Section~\ref{sec:model}.\footnote{Our treatment is purely formal and is
meant to motivate our computational method. A rigorous weak convergence
result justifying the approximation is left for future research.}  A superscript of $n$ is attached to various stochastic
processes to emphasize their dependence on $n$.  We assume that
the arrival, service, and abandonment rates vary with $n$ as follows:  For
$i\in\calI$, $j\in\calJ$, and $0\le t\le T$,
\vspace{-2mm}
\begin{align}
  &\lambda_i^n(t) = n\lambda_i(t) + \sqrt{n}\,\zeta_i(t) + o(\sqrt n),
    \label{eq:scaling}\\
  &\mu_j^n = \mu_j,\qquad \theta_i^n = \theta_i,\qquad P^n = P,
    \label{eq:scaling-fixed}
\end{align}
where $\lambda_{i}(\cdot)$ and $\zeta_{i}(\cdot)$ are given functions. We estimate them from data in Section \ref{sec:calibration} for our test problems. Similarly,
the number of agents varies with $n$ as follows:
\vspace{-3mm}
\begin{equation}\label{eq:scaling-staff}
  N_k^n(t) = n\nu_k(t),\qquad k\in\calK,\ 0\le t\le T.
\end{equation}
Our approximation assumes a critically loaded sequence of systems. Consequently, following \citet{harrison1999heavy}, \citet{harrison2003broader} and \citet{atar2005scheduling}, we
introduce a \emph{planning problem} that chooses a nominal processing plan
$\xi(\cdot)$ over the planning horizon $[0,T]$. Following \citep{harrison1988brownian, harrison2000brownian}, we define the input-output
matrix $R$ as the $I\times J$ matrix with entries
\begin{equation}\label{eq:R}
  R_{ij} =
  \begin{cases}
    \mu_j\,, & b(j)=i,\\[2pt]
    -\,P_{ji}\,\mu_j\,,  & b(j)\neq i,
  \end{cases}
\end{equation}
and the mean activity value rate $\tilde v_j(t) = \bv_j\mu_j\nu_{s(j)}(t)$,
where $\bv_j$ is the mean value of one completion of activity~$j$
defined in Equation \eqref{eq:vbar}. To interpret $\tilde v_j(t)$, recall that server 
pool $s(j)$ has $n\nu_{s(j)}(t)$ agents in the $n$th system from Equation \eqref{eq:scaling-staff}.  If the
server pool devoted all of its capacity to activity~$j$, service completions
would occur at rate $n\mu_j\nu_{s(j)}(t)$, each generating an expected
value of $\bv_j$. Thus, $n\tilde v_j(t)$ is the first-order value generation rate when server pool~$s(j)$ devotes all of its
capacity to activity~$j$. The staffing levels enter the planning problem through the diagonal $J \times J$ staffing matrix
\begin{equation}\label{eq:staffing-matrix}             D(t) = \operatorname{diag}\!\big(\nu_{s(1)}(t),\ldots,\nu_{s(J)}(t)\big)                           = \operatorname{diag}\!\big(A^\top\nu(t)\big),        \end{equation}  
whose $j$th diagonal entry is the limiting staffing level of the server pool $s(j)$ that performs activity $j$.
We choose a nominal plan $\xi(\cdot)$ so as to
  \begin{align}
  \label{eq:spp}
    \text{maximize}
    \quad & \int_0^{T}\tilde v(t)\cdot \xi(t)\,dt\\
    \text{subject to}
    \quad & R\,D(t)\,\xi(t) = \lambda(t),\quad t\in[0,T], \label{eq:flow_balance}\\
    & A\,\xi(t) \leq e,\qquad\ \qquad t\in[0,T],\\
    & \xi(t)\geq 0,\qquad\qquad \quad \,t\in[0,T], \label{eq:spp-const3}
  \end{align}
where $e$ is the $K$-vector of ones and $A$ is the capacity-consumption
matrix of Section~\ref{sec:model}.  The decision variable $\xi_j(t)$ is the
fraction of pool~$s(j)$'s capacity allocated to activity~$j$ at time~$t$,
and the objective is the cumulative expected value generated over the time horizon.
The constraint $R\,D(t)\,\xi(t)=\lambda(t)$ is the flow-balance condition. For each class, the net rate at which work is processed equals the
external arrival rate.  The constraint $A\xi(t)\le e$ reflects the capacity constraint of each server pool, i.e., $\sum_{j\in\mathcal{S}(k)}\xi_j(t)\leq 1$.

\begin{assumption}[Heavy Traffic Assumption]\label{ass:ht}
  The planning problem~\eqref{eq:spp}-\eqref{eq:spp-const3} admits a unique
  optimal solution with $A\xi^*(t)=e$ for every $t \in [0,T]$.
\end{assumption}

At time $t$ activity~$j$ is \emph{basic} if $\xi_j^*(t)>0$ and
\emph{nonbasic} if $\xi_j^*(t)=0$.  We write
\begin{equation*}
\calB(t)=\{j\in\calJ:\xi_j^*(t)>0\} \quad \text{and} \quad \calN(t)=\{j\in\calJ:\xi_j^*(t)=0\}
\end{equation*}
for the sets of basic and nonbasic activities, respectively.  The nominal number of
servers engaged in activity~$j$ is 
\begin{equation}
\psi_j^*(t) = \nu_{s(j)}(t)\,\xi_j^*(t),    
\end{equation} 
so that
\begin{equation}
x_i^*(t) = \sum_{j\in\mathcal{B}(i)}\psi_j^*(t)   \label{eq:nominal_service} 
\end{equation}
is the nominal number of class~$i$ customers in service. The flow balance condition \eqref{eq:flow_balance} reads
\begin{equation}\label{eq:flow-balance-psi}
  \sum_{j\in\mathcal{B}(i)}\mu_j\psi_j^*(t) \;-\; \sum_{j\in\calJ}P_{ji}\,\mu_j\psi_j^*(t)
    \;=\; \lambda_i(t),\qquad i\in\calI .
\end{equation}
Next, we introduce the scaled state, queue-length, and idleness
processes $\hat X^n$, $\hat Y^n$, and $\hat Z^n$ respectively, as follows:  For
$0\le t\le T$ and $n\geq 1$, let
\begin{align}
  \hat{X}_i^n(t) &= \frac{X_i^n(t)-n x_i^*(t)}{\sqrt n},
    \quad i\in\calI,
    \label{eq:scaled-state}\\
  \hat{Y}_i^n(t) &= \frac{Y_i^n(t)}{\sqrt n},
    \quad \quad \quad \quad \,\,\,\,\, i\in\calI,
    \label{eq:scaled-queue-def}\\
  \hat{Z}_k^n(t) &= \frac{Z_k^n(t)}{\sqrt n},
    \quad \quad \quad \quad \,\,\,\,  k\in\calK.
    \label{eq:scaled-idle-def}
\end{align}
Following \citet{harrison2003broader}, we
restrict attention to control policies that deviate from the nominal
plan by order $\sqrt n$. That is, for the $n^{\text{th}}$ system, we consider policies of the following form:
\begin{align}
  &\psi_j^n(t) = n\psi_j^*(t) - \sqrt n\,\hat\psi_j^n(t) + o(\sqrt n),\quad j\in\calJ,\label{eq:control-prelimit}
\end{align}
where $\hat\psi^n$ is the scaled deviation control, measuring the
shortfall of servers on activity~$j$ relative to the nominal
allocation. Recall that $\psi^{*}_{j}(t) > 0$ for $j \in \calB(t)$ and $\psi^{*}_{j}(t) = 0$ for $j \in \calN(t)$. Because we must have $\psi_{j}^{n}(t) \geq 0$ for all $j$, $t$, the restrictions below follow from \eqref{eq:control-prelimit}:
\begin{align}
    \hat{\psi}_{j}^{n}(t) &\leq \sqrt{n}\psi^{*}_{j}(t), \quad j \in \calB(t), \label{eq:basic}\\
    \hat{\psi}_{j}^{n}(t) &\leq 0, \quad \quad \quad \quad j \in \calN(t), \label{eq:nonbasic}
\end{align}
where the restriction \eqref{eq:basic} gets weaker as $n$ gets large. Using the scaled processes and suppressing the $o(1)$
terms arising from \eqref{eq:control-prelimit}, one can
derive from
Equations~\eqref{eq:queue-balance}--\eqref{eq:positivity} the following: For
$0\le t\le T$, $i \in \mathcal{I}$ and $k \in \mathcal{K}$, we have
\begin{align}
  &\hat{Y}_i^n(t) = \hat{X}_i^n(t)+\sum_{j\in\mathcal{B}(i)}\hat\psi_j^n(t),
    \label{eq:scaled-queue}\\
  &\hat{Z}_k^n(t) = \sum_{j\in\mathcal{S}(k)}\hat\psi_j^n(t),
    \label{eq:scaled-server}\\
  &\hat{Y}_{i}^{n}(t) \geq 0, \quad \hat{Z}^{n}_{k}(t) \geq 0. \label{eq:scaled-positivity}
\end{align}
For such policies, we now derive the infinitesimal drift and covariance of
the scaled state process $\hat X^n$ to facilitate the formal derivation of
its diffusion limit.  In particular, for $t\geq 0$, distinct classes
$i,i'\in\calI$, and small $h>0$, writing
$\Delta\hat X_i^n = \hat X_i^n(t+h)-\hat X_i^n(t)$, the following holds:
\begin{align}
  &\E\big[\Delta\hat X_i^n \mid \hat X^n(t)=x\big]
    = \Big(\zeta_i(t) + \sum_{j\in\mathcal{B}(i)}\mu_j\hat\psi_j^n(t)
      - \sum_{j\in\calJ}P_{ji}\mu_j\hat\psi_j^n(t)
      - \theta_i\hat Y_i^n(t)\Big)h + o(h),
    \label{eq:inf-drift}\\
  &\E\big[(\Delta\hat X_i^n)^2 \mid \hat X^n(t)=x\big]
    = 2\sum_{j\in\mathcal{B}(i)}\mu_j\psi_j^*(t)\,h + o(h),
    \label{eq:inf-var}\\
  &\E\big[\Delta\hat X_i^n\,\Delta\hat X_{i'}^n \mid \hat X^n(t)=x\big]
    = -\Big(\sum_{j\in\mathcal{B}(i')}P_{ji}\,\mu_j\psi_j^*(t)
      + \sum_{j\in\mathcal{B}(i)}P_{ji'}\,\mu_j\psi_j^*(t)\Big)h + o(h).
    \label{eq:inf-cov}
\end{align}
The variance and covariance rates
in~\eqref{eq:inf-var}--\eqref{eq:inf-cov} are the entries of the
time-dependent covariance rate matrix $\Gamma(t)=(\gamma_{ii'}(t))_{i,i'\in\calI}$,
whose diagonal and off-diagonal entries are, for distinct $i,i'\in\calI$,
\begin{align}
  \gamma_{ii}(t)
    &= 2\sum_{j\in\mathcal{B}(i)}\mu_j\psi_j^*(t) = \lambda_i(t)
       + \sum_{j\in\mathcal{B}(i)}\mu_j\psi_j^*(t)
       + \sum_{j\in\calJ}P_{ji}\,\mu_j\psi_j^*(t),
    \label{eq:Sigma-diag}\\
  \gamma_{ii'}(t)
    &= -\Big(\sum_{j\in\mathcal{B}(i)}P_{ji'}\,\mu_j\psi_j^*(t)
             + \sum_{j\in\mathcal{B}(i')}P_{ji}\,\mu_j\psi_j^*(t)\Big),
    \label{eq:Sigma-offdiag}
\end{align}
where the second equality in~\eqref{eq:Sigma-diag} uses the flow
balance equation~\eqref{eq:flow-balance-psi}.  The diagonal entry $\gamma_{ii}(t)$ is the infinitesimal
variance parameter for class~$i$. It collects the external arrivals $\lambda_i(t)$,
the class $i$ service completions $\sum_{j\in\mathcal{B}(i)}\mu_j\psi_j^*(t)$,
and the completions of other classes routed into class~$i$,
$\sum_{j\in\calJ}P_{ji}\mu_j\psi_j^*(t)$. Transfers induce negative
correlation between classes.  Completions of an activity
$j \in \mathcal{B}(i)$ occur at rate $\mu_{j}\psi^{*}_{j}(t)$ and route
the caller into class~$i'$ with probability $P_{ji'}$, so each such
event simultaneously decreases the number of class~$i$ callers in the
system, $X_i$, and increases the number of class~$i'$ callers,
$X_{i'}$.  The two sums in~\eqref{eq:Sigma-offdiag} account for the two
directions of transfer between the classes. Equations \eqref{eq:Sigma-diag}--\eqref{eq:Sigma-offdiag} can be written in matrix notation as follows:
\begin{align}
  \Gamma(t) = \operatorname{diag}(\lambda(t)) + \sum_{j\in\calJ}
    \mu_j\psi_j^*(t)\,Q_j, \label{eq:Sigma-compact}
\end{align}
where $Q_j = q_j\,e_{b(j)}e_{b(j)}^\top + \sum_{l\in\calI}P_{jl}\,(e_l-e_{b(j)})(e_l-e_{b(j)})^\top.$ We let
$\sigma(t)$ be any $I\times I$ matrix satisfying
$\sigma(t)\sigma(t)^\top=\Gamma(t)$, for instance the symmetric square root
$\Gamma(t)^{1/2}$.

Taking the formal limit as $n\to\infty$, denoting the weak limit of
$(\hat X^n,\hat\psi^n,\hat Y^n,\hat Z^n)$ by $(\hat X,\hat\psi,\hat Y,\hat Z)$
and using the flow balance equation~\eqref{eq:flow-balance-psi}, we deduce
from Equations~\eqref{eq:inf-drift}--\eqref{eq:inf-cov} that the limiting state process
$\hat X=(\hat X_i)_{i\in\calI}$ satisfies, for $t\geq 0$,
\begin{equation}\label{eq:sde}
  d\hat{X}(t) = b\big(t,\hat{X}(t),\hat\psi(t)\big)\,dt + \sigma(t)\,dB(t),
\end{equation}
where $B(t) = (B_{i}(t))$ is an $I$-dimensional standard Brownian motion and the
drift function $b$ is given as follows:
\begin{equation}\label{eq:drift}
  b_i(t,x,\hat\psi)
  = \zeta_i(t) - \theta_{i}x_{i} + \sum_{j \in \mathcal{B}(i)} (\mu_{j} - \theta_{i})\,\hat{\psi}_{j}-\sum_{j \in \mathcal{J}} P_{ji}\mu_{j}\hat{\psi}_{j},
    \qquad i\in\calI.
\end{equation}

Also, it follows from Equations \eqref{eq:scaled-queue}--\eqref{eq:scaled-positivity} that the limiting control, queue-length, and idleness processes, $\hat{\psi}$, $\hat{Y}$, and $\hat{Z}$, respectively, satisfy the following: For $i \in \mathcal{I}$, $j \in \mathcal{J}$ and $k \in \mathcal{K}$
\begin{align}
    &\hat{Y}_{i}(t) = \hat{X}_{i}(t) + \sum_{j \in \mathcal{B}(i)}\hat{\psi}_{j}(t), \quad t \geq 0,\\
    &\hat{Z}_{k}(t) = \sum_{j \in \mathcal{S}(k)} \hat{\psi}_{j}(t), \quad t \geq 0,\\
    &\hat{Y}(t) \geq 0, \quad \hat{Z}(t) \geq 0, \quad t \geq 0.
\end{align}

To minimize technical complexity, we restrict attention to Markov controls,
i.e., $\hat{\psi}(t) = \psi(t, \hat{X}(t))$ for $t \geq 0$, where
$\psi : [0,T]\times\mathbb{R}^{I} \to \mathbb{R}^{J}$ is a measurable
control policy. We let $\Psi(t,x)$ denote the set of feasible actions at time $t$ in state $x$ and call a control policy $\psi$ admissible if
$\psi(t,x) \in \Psi(t,x)$ for every $(t,x)\in[0,T]\times\mathbb{R}^{I}$, where
\begin{equation}
\label{eq:set_feasible_policies}
\Psi(t,x) = \left\{ \psi \in \mathbb{R}^{J} :
\begin{array}{ll}
\sum_{j \in \mathcal{B}(i)} \psi_{j} \geq -x_{i}, & i \in \mathcal{I}, \\[2pt]
\sum_{j \in \mathcal{S}(k)} \psi_{j} \geq 0, & k \in \mathcal{K}, \\[2pt]
\psi_{j} \leq 0, & j \in \calN(t)
\end{array}
\right\}.
\end{equation}
The constraints defining $\Psi(t,x)$ follow from \eqref{eq:scaled-queue}--\eqref{eq:scaled-positivity} and the constraints on the scaled controls of basic and nonbasic activities in \eqref{eq:basic}--\eqref{eq:nonbasic}.
For the $n$th system, we define the centered and scaled process
\begin{equation}\label{eq:scaled-reward-process}
  \hat\Pi^n(t) = \frac{n\int_0^t \tilde v(s)\cdot\xi^*(s)\,ds - \Pi^n(t)}{\sqrt n},
  \qquad 0\le t\le T,
\end{equation}
where the cumulative net value of the $n$th system from Equation \eqref{eq:cumulative-value} is as follows:
\begin{equation}\label{eq:cumulative-value-n}
  \Pi^n(t) = \sum_{j\in\calJ}\mathcal R_j^n(t)
    - \int_0^t h\cdot Y^n(s)\,ds - a\cdot A^n(t),
\end{equation}
and $\mathcal R_j^n$ denotes the realized value
\eqref{eq:realized-reward-activity} of activity~$j$ in the $n$th system. The term $n\int_0^t \tilde v(s)\cdot\xi^*(s)\,ds$ is the
cumulative value earned by the nominal plan. One can show that it is an upper
bound on the value achievable under any admissible policy, because
$\xi^*(s)$ maximizes the instantaneous value rate subject to the
capacity and flow-balance constraints.  Following
\citet{harrison2003broader}, we refer to $\hat\Pi^n$ as the system
manager's scaled cumulative cost process.

One can then formally derive an approximating Brownian control problem by taking
the limit $n \rightarrow \infty$ of the scaled total cost.
We state the resulting problem directly and give the formal derivation
in Appendix \ref{app:cost}. The resulting total expected cost under an admissible policy $\psi$,
starting in state $x$ at time~$t$, is
\begin{equation}\label{eqn:cost_definition}
  \hat{J}(t,x;\psi)
  =\E^{\psi}_{x}\!\left[\int_{t}^{T}
    \Big(\sum_{i\in\calI}c_i\hat Y_{i}(s)
    +\sum_{j\in\calJ}\bv_j\mu_j\hat{\psi}_j(s)\Big)\,ds\right],
\end{equation}
where $\mathbb{E}_{x}^{\psi}$ denotes the conditional expectation
starting in state $x$ under policy $\psi$. We define the optimal value function as
\begin{equation}\label{eqn:value_function_definition}
  V(t,x) = \inf\, \hat J(t,x;\psi),
  \qquad (t,x)\in[0,T]\times\R^I,
\end{equation}
where the infimum is taken over the set of admissible policies. Since the terminal value is zero by
Equation~\eqref{eq:terminal-zero}, it satisfies $V(T,x) = 0$ for $x \in \mathbb{R}^{I}$.

\subsection{HJB Equation}\label{sec:hjb}

As a preliminary to introducing the HJB equation, we define the second-order
differential operator
\begin{equation}\label{eq:L}
  \mathcal{L}_t V(t,x) = \frac{1}{2}\sum_{i\in\calI}\sum_{i'\in\calI}\gamma_{ii'}(t)\,
    \frac{\partial^2 V}{\partial x_i\,\partial x_{i'}}(t,x)
\end{equation}
and the Hamiltonian
\begin{equation}\label{eq:H}
  \mathcal{H}(t,x,p) = \inf_{\hat\psi\in\Psi(t,x)}\!
    \left[\sum_{i\in\calI}b_i(t,x,\hat\psi)\,p_i
    + \sum_{i\in\calI}c_i\Big(x_i+\sum_{j\in\mathcal{B}(i)}\hat\psi_j\Big)
    + \sum_{j\in\calJ}\bv_j\mu_j\,\hat\psi_j\right].
\end{equation}
Then the HJB equation involves finding a sufficiently smooth function $V (t, x)$ that solves the following partial differential equation: For $(t, x) \in [0, T] \times \R^{I}$,
\begin{equation}\label{eq:hjb}
  \frac{\partial V}{\partial t}(t,x)
    + \mathcal{L}_t V(t,x)
    + \mathcal{H}\big(t,x,\nabla_x V(t,x)\big)
    = 0,\qquad V(T,x)=\hat g(x).
\end{equation}
Substituting the drift~\eqref{eq:drift} into~\eqref{eq:H} and collecting
the coefficient of $\hat\psi_j$ yields
\begin{equation}\label{eq:H-explicit}
  \mathcal{H}(t,x,p)
    = \sum_{i\in\calI}(\zeta_i(t)-\theta_i x_i)\,p_i + \sum_{i\in\calI}c_i\,x_i
      + \inf_{\hat\psi\in\Psi(t,x)}\!\left[\sum_{j\in\calJ}w_j(p)\,\hat\psi_j\right],
\end{equation}
where the activity weight function for $j\in\calJ$ is
\begin{equation}\label{eq:weight}
  w_j(p) = (\mu_j-\theta_{b(j)})\,p_{b(j)}
    - \mu_j\sum_{i\in\calI}P_{ji}\,p_i
    + \bv_j\mu_j + c_{b(j)} .
\end{equation}
Then the HJB equation can be written equivalently as follows: For $(t,x) \in [0,T] \times \R^{I}$
\begin{align}\label{eq:hjb-full}
  &\frac{\partial V}{\partial t}(t,x)
  + \frac{1}{2}\sum_{i\in\calI}\sum_{i'\in\calI}\gamma_{ii'}(t)\,\frac{\partial^2 V}{\partial x_i\,\partial x_{i'}}(t,x)
  + \sum_{i\in\calI}\bigl(\zeta_i(t)-\theta_i x_i\bigr)\frac{\partial V}{\partial x_i}(t,x)
  \notag\\
  &\quad
  \qquad \,\,\,\, + \sum_{i\in\calI}c_i\,x_i
  + \inf_{\hat\psi\in\Psi(t,x)}\!\left[\sum_{j\in\calJ}w_j(\nabla_x V(t,x))\,\hat\psi_j\right]
  = 0,\\
  &V(T,x) = \hat g(x).
\end{align}
\paragraph{Proposed policy for the Brownian control problem.}
The minimization in the HJB equation~\eqref{eq:hjb-full} is a
linear program over the polyhedron $\Psi(t,x)$. Given the value
function $V$, an optimal control at each time $t$ and state $x$
is obtained by solving this linear program,
\begin{equation}\label{eq:opt-policy}
  \hat\psi^*(t,x) \in \operatorname*{arg\,min}_{\hat\psi\in\Psi(t,x)}
    \sum_{j\in\calJ} w_j\big(\nabla_x V(t,x)\big)\,\hat{\psi}_{j}.
\end{equation}

\section{Computational Method}\label{sec:computational}
Our computational method builds on the deep splitting method proposed by \citet{beck2021deep} for solving semilinear parabolic PDEs. We begin by choosing a \textit{reference policy} to simulate the sample paths of the
system state visited during the training process. Ideally, the reference policy should generate sample paths that concentrate in the regions of the state space most frequently visited under the optimal policy. We denote the reference policy by $\tilde\psi$, which is a
measurable function $\tilde\psi:[0,T]\times\R^{I}\to\R^{J}$ with
$\tilde\psi(t,x)\in\Psi(t,x)$ for all $(t,x)$.  Then, the corresponding
reference process $\tilde X$ solves the following stochastic differential equation
\begin{equation}\label{eq:reference-process}
  d\tilde X(t)
  = b\big(t,\tilde X(t),\tilde\psi(t,\tilde X(t))\big)\,dt
  + \sigma(t)\,dB(t),
  \qquad t\in[0,T],
\end{equation}
with diffusion coefficient $\sigma$ and drift function $b$ as defined in~\eqref{eq:Sigma-compact} and~\eqref{eq:drift}, respectively.  We use the resulting
sample paths as the training data for our computational method.

Next, we derive a stochastic identity that the value function satisfies along the reference process, which will serve as the basis for the loss function of our neural-network-based computational method. Evaluating the HJB
equation~\eqref{eq:hjb-full} at $(t,\tilde X(t))$ for $t \in [0,T]$ and adding and
subtracting the term
\begin{equation*}
b\big(t,\tilde X(t),\tilde\psi(t,\tilde X(t))\big)\cdot\nabla_x V(t,\tilde X(t))  
\end{equation*}
yields
\begin{align}\label{eq:hjb-identity}
  \frac{\partial V}{\partial t}\big(t,\tilde X(t)\big)
  + \mathcal{L}_t V\big(t,\tilde X(t)\big)
  &+ b\big(t,\tilde X(t),\tilde\psi(t,\tilde X(t))\big)
    \cdot\nabla_x V\big(t,\tilde X(t)\big) \nonumber \\
  &- F\big(t,\tilde X(t),\nabla_x V(t,\tilde X(t))\big)
  = 0, \quad t \in [0,T],
\end{align}
where, for $(t,x,p)\in[0,T]\times\R^I\times\R^I$, the
auxiliary function $F$ is
\begin{equation}\label{eq:F-def}
  F(t,x,p)
  := b\big(t,x,\tilde\psi(t,x)\big)\cdot p
  - \mathcal{H}(t,x,p),
\end{equation}
and $\mathcal H$ is the Hamiltonian~\eqref{eq:H}.  Using the expressions for 
the drift function~\eqref{eq:drift} and the Hamiltonian~\eqref{eq:H-explicit}, the terms
$\big(\zeta_i(t)-\theta_i x_i\big)\,p_i$ cancel and $F$ admits the form
\begin{equation}\label{eq:F-explicit}
  F(t,x,p)
  = \sum_{j\in\calJ}\Big[(\mu_j-\theta_{b(j)})\,p_{b(j)}
    - \mu_j\sum_{i\in\calI}P_{ji}\,p_i\Big]\,\tilde\psi_j(t,x)
  - \sum_{i\in\calI}c_i x_i
  - \min_{\hat\psi\in\Psi(t,x)}\sum_{j\in\calJ} w_j(p)\,\hat\psi_j,
\end{equation}
with the activity weight functions $w_j(\cdot)$ given in Equation~\eqref{eq:weight}.

To discretize the time horizon, we take a partition
$0=t_0<t_1<\cdots<t_N=T$ that is fine enough for the approximation
below to be accurate, and we set $\Delta t_n=t_{n+1}-t_n$.  The
training data consist of state pairs
$\big(\tilde X(t_n),\tilde X(t_{n+1})\big)$ for $n=0,1,\ldots,N-1$, and
we obtain each pair in two steps.  The first state $\tilde X(t_n)$ is
drawn from an empirical distribution that
Subroutine~\ref{sub:distribution} estimates by simulating the reference
process from time zero, and the second state $\tilde X(t_{n+1})$
follows from a single Euler--Maruyama step
of~\eqref{eq:reference-process}, as described in
Subroutine~\ref{sub:sampling}.

The stochastic identity~\eqref{eq:hjb-identity} consists of terms that
are linear in $V$ and a nonlinear term $F$. In our setting, the
nonlinearity arises from the minimization over the feasible set
$\Psi(t,x)$. On each subinterval $[t_n,t_{n+1}]$, the deep splitting
method approximates $F$ by its value at the left endpoint $t_n$, which
yields a linear PDE whose solution approximates $V$ on that
subinterval. The Feynman--Kac representation of this linear PDE along
the reference process then yields the following relation for
$n=0,1,\ldots,N-1$,
\begin{equation}\label{eq:fk-identity}
  \E\Big[V\big(t_{n+1},\tilde X(t_{n+1})\big)\,\Big|\,\tilde X(t_n)\Big]
  = V\big(t_n,\tilde X(t_n)\big)
  + F\big(t_n,\tilde X(t_n),\nabla_xV(t_n,\tilde X(t_n))\big)\,\Delta t_n.
\end{equation}
Appendix~\ref{app:deep-splitting} derives this relationship formally.

Evaluating $F$ defined in Equation~\eqref{eq:F-explicit} at a given $(t,x,p)$
requires solving two linear programs.  To make them
explicit, we rewrite the auxiliary function $F(t,x,p)$ as
\begin{equation}\label{eq:F-decomp}
  F(t,x,p) = \mathcal{D}(t,x)\cdot p - \mathcal{M}(t,x,p)
  - \sum_{i\in\calI}c_i x_i,
\end{equation}
\vspace{-2mm}
where
\vspace{-2mm}
\begin{align}
  \mathcal{M}(t,x,p)
  &= \min_{\hat\psi\in\Psi(t,x)}\sum_{j\in\calJ}w_j(p)\,\hat\psi_j,
    \label{eq:M-def}\\
  \mathcal{D}_i(t,x)
  &= \sum_{j\in\mathcal{B}(i)}(\mu_j-\theta_i)\,\tilde\psi_j(t,x)
   - \sum_{j\in\calJ}P_{ji}\,\mu_j\,\tilde\psi_j(t,x),
   \qquad i\in\calI,
    \label{eq:D-def}
\end{align}
and $\mathcal{D}(t,x)=\big(\mathcal{D}_1(t,x),\ldots,\mathcal{D}_I(t,x)\big)^{\top}$.

The constraints defining $\Psi(t,x)$ follow from the scaled queue and
idleness relations \eqref{eq:scaled-queue}--\eqref{eq:scaled-positivity}
and from the sign constraint \eqref{eq:nonbasic} on the nonbasic
activities. Our approach approximates the $n$th system by the Brownian
control problem. With this in mind, Equation~\eqref{eq:control-prelimit}
and the nonnegativity requirement $\psi_j^n(t)\ge 0$ imply that the
scaled controls of the $n$th system satisfy
$\hat\psi_j^n(t)\le\sqrt n\,\psi_j^*(t)$ for $j\in\calB(t)$, as stated
in the constraint \eqref{eq:basic}. Therefore, we impose the
constraint \eqref{eq:basic} on the limiting controls of the basic
activities and solve each linear program over
\begin{equation}\label{eq:Psi-bar}
  \bar\Psi(t,x) = \big\{\hat\psi\in\Psi(t,x) :
    \hat\psi_j \le \sqrt n\,\psi_j^*(t),\ j\in\calB(t)\big\},
\end{equation}
where $n$ is the size of the system under study. In particular, we
replace $\Psi(t,x)$ by $\bar\Psi(t,x)$ in Equation~\eqref{eq:M-def} and
in the reference policies of Section~\ref{sec:reference-policies}. The neural
networks $\hat{\mathcal M}$ and $\hat{\mathcal D}$ introduced next
approximate the resulting optimization problems.

From Equation~\eqref{eq:M-def} and \eqref{eq:Psi-bar}, one sees that computing $\mathcal{M}(t,x,p)$ requires solving a linear program over the feasible set $\bar\Psi(t,x)$. In addition, computing $\mathcal{D}(t,x)$ requires solving a linear program, because the reference policies we consider are defined through linear programs with the same feasible set $\bar\Psi(t,x)$, as explained below in Section~\ref{sec:reference-policies}. Since these linear programs must be solved at every state visited along the sample paths and at every iteration of the neural network training algorithm, evaluating $F(t,x,p)$ online is computationally expensive. Consequently, we approximate $\mathcal{M}$ and $\mathcal{D}$ with neural networks $\hat{\mathcal{M}}$ and $\hat{\mathcal{D}}$, respectively, training the pair of networks $(\hat{\mathcal{M}}, \hat{\mathcal{D}})$ for each grid point $t_{n}$ and writing $\hat{\mathcal{M}}_n(\cdot,\cdot)=\hat{\mathcal{M}}(t_n,\cdot,\cdot)$ and $\hat{\mathcal{D}}_n(\cdot)=\hat{\mathcal{D}}(t_n,\cdot)$. We substitute $\hat{\mathcal{M}}_{n}$ and $\hat{\mathcal{D}}_{n}$ into~\eqref{eq:F-decomp}, so that the approximation of the auxiliary function $F(t,x,p)$ used in our computational method is
\begin{equation}\label{eq:F-hat}
  \hat F(t,x,p) = \hat{\mathcal{D}}(t,x)\cdot p
  - \hat{\mathcal{M}}(t,x,p) - \sum_{i\in\calI}c_i x_i.
\end{equation}
Appendix~\ref{app:offline} describes the offline approximation of
$\mathcal{M}$ and $\mathcal{D}$ in detail.

We also represent the value function and its gradient by neural
networks, one pair for each grid point $t_{n}$.  For each $n=0,1,\ldots,N-1$, let
$H^{n}(\,\cdot\,;\omega_n)$ denote the approximation of the value function $V(t_n,\cdot)$
and let $G^{n}(\,\cdot\,;\nu_n)$ denote the approximation of the gradient function
$\nabla_xV(t_n,\cdot)$, where $\omega_n$ and $\nu_n$ are the
parameter vectors. The deep splitting method computes the neural network pairs $\big(H^{n}(\cdot;\omega_n), G^{n}(\cdot;\nu_n)\big)$ for $n = 0,1,\ldots,N-1$, recursively backward in time. The recursion begins at the terminal time step $n = N$, where we use the terminal cost function $\hat{g} (\cdot) \equiv 0$ to set $V(t_{N},\tilde{X}(t_{N})) = \hat{g}(\tilde{X}(t_{N}))$.

The conditional expectation on the left-hand side of~\eqref{eq:fk-identity}
is the $L^{2}$ projection of
$V\big(t_{n+1},\tilde X(t_{n+1})\big)$ onto the $\sigma$-algebra
generated by $\tilde X(t_n)$, so the two functions on the right-hand side, the value function $V(t_n,\cdot)$ and its gradient $\nabla_x V(t_n,\cdot)$, solve a least-squares problem. As a result, for our computational method, we define the loss function at $n = N-1$ as follows:
\begin{equation}\label{eq:loss-terminal}
\begin{aligned}
  \ell_{N-1}(\omega_{N-1},\nu_{N-1})
  = \E\Big[\Big(
    &\hat g\big(\tilde X(t_N)\big)
    - H^{N-1}\big(\tilde X(t_{N-1});\omega_{N-1}\big)\\
    &- \hat F\big(t_{N-1},\tilde X(t_{N-1}),
        G^{N-1}(\tilde X(t_{N-1});\nu_{N-1})\big)\,\Delta t_{N-1}
  \Big)^{2}\Big],
\end{aligned}
\end{equation} 
and we set
$(\omega_{N-1}^{\star},\nu_{N-1}^{\star})
\in\operatorname{argmin}\ell_{N-1}(\omega_{N-1},\nu_{N-1})$. Similarly, for $n=N-2,\ldots,1,0$, fixing the trained network $H^{n+1}(\,\cdot\,;\omega_{n+1}^{\star})$ in place, we define the loss function $\ell_{n}(\omega_{n}, \nu_{n})$ as follows:
\begin{equation}\label{eq:loss-recursive}
\begin{aligned}
  \ell_{n}(\omega_{n},\nu_{n})
  = \E\Big[\Big(
    &H^{n+1}\big(\tilde X(t_{n+1});\omega_{n+1}^{\star}\big)
    - H^{n}\big(\tilde X(t_{n});\omega_{n}\big) - \hat F\big(t_{n},\tilde X(t_{n}),
        G^{n}(\tilde X(t_{n});\nu_{n})\big)\,\Delta t_{n}
  \Big)^{2}\Big],
\end{aligned}
\end{equation}
and we set $(\omega_{n}^{\star}, \nu_{n}^{\star}) \in \operatorname{argmin} \ell_{n}(\omega_{n}, \nu_{n}).$ Algorithm~\ref{alg:ds} describes the resulting neural network training algorithm, with the loss functions replaced by their empirical estimates as shown in Equations \eqref{eq:empirical-loss-n-1}--\eqref{eq:empirical-loss-n-2} and the auxiliary function $F$ replaced by the offline approximation
$\hat F$ as shown in Equation \eqref{eq:F-hat}.
\begin{subroutine}[H]
\caption{Distribution generator}
\label{sub:distribution}
\begin{minipage}{\textwidth}
\begin{algorithmic}[1]
  \Statex \textbf{Input:} The offline networks
    $\hat{\mathcal{D}}_n(\cdot)$ for $n=0,1,\ldots,N-1$ from Subroutine~\ref{sub:offline-D} in Appendix \ref{app:offline}, the
    diffusion coefficient $\sigma(\cdot)$, the time horizon $T$, the
    number of intervals $N$, a discretization step-size $\Delta t_n$
    (for simplicity, we assume $\Delta t_n\triangleq T/N$), an initial distribution $\pi_{0}$,
    and the number of sample path replications $S_0$.
  \Statex \textbf{Output:} Empirical distributions $\pi_n$ for
    $n=1,2,\ldots,N-1$.
  \Function{Generator}{$T,\Delta t_n,\pi_0,S_0$}
    \State For the time interval $[0,T]$ and $N=T/\Delta t_n$, construct
      the partition $0=t_0<\cdots<t_N=T$,
    \Statex \quad\, where $\Delta t_n=t_{n+1}-t_n$ for
      $n=0,\ldots,N-1$.
    \For{$r\gets 1$ to $S_0$}
      \State Draw $x_0^{(r)}\sim\pi_{0}$ independently across replications
        and set $\tilde X^{(r)}(t_0)=x_0^{(r)}$.
      \State Generate $N$ i.i.d.\ $I$-dimensional Gaussian random
        vectors $\Delta B^{(r)}(t_n)=(\Delta B_i^{(r)}(t_n))$
      \Statex \quad\quad\, with mean zero and covariance matrix
        $\Delta t_n\,\mathrm{Id}_I$ for $n=0,1,\ldots,N-1$.
      \For{$n\gets 0$ to $N-1$}
        \For{$i\gets 1$ to $I$}
        \vspace{4mm}
          \State $\begin{aligned}[t]
            \tilde X_i^{(r)}(t_{n+1}) \gets{}
              \tilde X_i^{(r)}(t_n)
              + \big(\zeta_i(t_n) - \theta_i\,\tilde X_i^{(r)}(t_n)
                &+ \hat{\mathcal{D}}_{n,i}(\tilde X^{(r)}(t_n))\big)\,\Delta t_n\\
            &+ \sum_{i'\in\calI}\sigma_{ii'}(t_n)\,\Delta B_{i'}^{(r)}(t_n)
          \end{aligned}$
        \EndFor
      \EndFor
    \EndFor
    \For{$n\gets 1$ to $N-1$}
      \State Estimate the mean vector
        $m(t_n) = \frac{1}{S_0}\sum_{r=1}^{S_0}\tilde X^{(r)}(t_n)$.
      \State Estimate the covariance matrix
        $C(t_n) = \frac{1}{S_0}\sum_{r=1}^{S_0}
          \big(\tilde X^{(r)}(t_n)-m(t_n)\big)
          \big(\tilde X^{(r)}(t_n)-m(t_n)\big)^{\top}$.
      \State Fit the empirical distribution as
        $\pi_n := \mathscr{N}\big(m(t_n),C(t_n)\big)$.
    \EndFor
    \State \textbf{return} $\pi_n$ for $n=1,2,\ldots,N-1$.
  \EndFunction
\end{algorithmic}
\end{minipage}
\end{subroutine}

\begin{subroutine}[H]
\caption{Sampling process}
\label{sub:sampling}
\begin{minipage}{\textwidth}
\begin{algorithmic}[1]
  \Statex \textbf{Input:} The offline networks
    $\hat{\mathcal{D}}_n(\cdot)$ for $n=0,1,\ldots,N-1$ from Subroutine~\ref{sub:offline-D} in Appendix \ref{app:offline}, the
    diffusion coefficient $\sigma(\cdot)$, a batch size $S$, a
    discretization step-size $\Delta t_n$ (for simplicity, we assume
    $\Delta t_n\triangleq T/N$), and a distribution $\pi_n$ (see
    Subroutine~\ref{sub:distribution} for the \textsc{Generator} of
    $\pi_n$).
  \Statex \textbf{Output:} State pair samples
    $\{\tilde X^{(s)}(t_n),\tilde X^{(s)}(t_{n+1})\}_{s=1}^{S}$.
  \Function{Sampling}{$t_n,\Delta t_n,\pi_n$}
    \For{$s\gets 1$ to $S$}
      \State Initialize a random state
        $\tilde X^{(s)}(t_n)\sim\pi_n$.
      \State Generate one $I$-dimensional Gaussian random vector
        $\Delta B^{(s)}(t_n)$ with mean zero and
      \Statex \quad\quad\, covariance matrix $\Delta t_n\,\mathrm{Id}_I$.
      \For{$i\gets 1$ to $I$}
        \State $\begin{aligned}[t]
          \tilde X_i^{(s)}(t_{n+1}) \gets{}
            \tilde X_i^{(s)}(t_n)
            + \big(\zeta_i(t_n) - \theta_i\,\tilde X_i^{(s)}(t_n)
              &+ \hat{\mathcal{D}}_{n,i}(\tilde X^{(s)}(t_n))\big)\,\Delta t_n\\
          &+ \sum_{i'\in\calI}\sigma_{ii'}(t_n)\,\Delta B_{i'}^{(s)}(t_n)
        \end{aligned}$
      \EndFor
    \EndFor
    \State \textbf{return}
      $\{\tilde X^{(s)}(t_n),\,\tilde X^{(s)}(t_{n+1})\}_{s=1}^{S}$
  \EndFunction
\end{algorithmic}
\end{minipage}
\end{subroutine}
\vspace{-5mm}
\begin{algorithm}[H]
\caption{Deep Splitting}
\label{alg:ds}
\begin{minipage}{\textwidth}
\begin{algorithmic}[1]
  \Statex \textbf{Input:} A batch size $S$, a time horizon $T$, the
    number of intervals $N$, the number of subnetworks $2N$, a
    discretization step-size $\Delta t_n$ (for simplicity, we assume
    $\Delta t_n\triangleq T/N$), the empirical distributions $\pi_n$
    from Subroutine~\ref{sub:distribution}, the offline networks
    $\hat{\mathcal{M}}_n(\cdot,\cdot)$ and $\hat{\mathcal{D}}_n(\cdot)$
    from Subroutines \ref{sub:offline-M} and~\ref{sub:offline-D} in Appendix \ref{app:offline}, and an
    optimization solver (SGD, ADAM, RMSProp, etc.).
  \Statex \textbf{Output:} The approximation of the value function
    $H^n(\,\cdot\,;\omega_n^\star)$, the approximation of the gradient
    function $G^n(\,\cdot\,;\nu_n^\star)$, and the neural network
    weights $(\omega_n^\star,\nu_n^\star)$ for $n=0,\ldots,N-1$.
  \State Set $V(t_N,\cdot)=\hat g(\cdot)$.
  \State Generate $S$ state pairs
    $\{\tilde X^{(s)}(t_{N-1}),\tilde X^{(s)}(t_N)\}_{s=1}^{S}$ by
    invoking \textsc{Sampling}$(t_{N-1},\Delta t_{N-1},\pi_{N-1})$.
  \State Compute the minimizer of the empirical loss
    $\hat\ell_{N-1}(\omega_{N-1},\nu_{N-1})$
  \vspace{-4mm}
  \Statex \begin{align} \label{eq:empirical-loss-n-1}
    \hat\ell_{N-1}(\omega_{N-1},\nu_{N-1})
    = \frac{1}{S}&\sum_{s=1}^{S}\Big(
      \hat g\big(\tilde X^{(s)}(t_N)\big)
      - H^{N-1}\big(\tilde X^{(s)}(t_{N-1});\omega_{N-1}\big) \nonumber\\
      &- \hat F\big(t_{N-1},\tilde X^{(s)}(t_{N-1}),
        G^{N-1}(\tilde X^{(s)}(t_{N-1});\nu_{N-1})\big)\,
        \Delta t_{N-1}\Big)^2.
  \end{align}
  \vspace{-4mm}
  \State Update
    $(\omega_{N-1}^\star,\nu_{N-1}^\star)
    \in\operatorname{argmin}\hat\ell_{N-1}(\omega_{N-1},\nu_{N-1})$.
  \For{$n\gets N-2$ to $0$}
    \State Generate $S$ state pairs
      $\{\tilde X^{(s)}(t_n),\tilde X^{(s)}(t_{n+1})\}_{s=1}^{S}$ by
      invoking \textsc{Sampling}$(t_n,\Delta t_n,\pi_n)$.
    \State Initialize the neural network weights as
      $(\omega_n,\nu_n)=(\omega_{n+1}^\star,\nu_{n+1}^\star)$.
    \State Compute the minimizer of the empirical loss
      $\hat\ell_n(\omega_n,\nu_n)$
    \vspace{-4mm}
    \Statex \begin{align} \label{eq:empirical-loss-n-2}
      \hat\ell_n(\omega_n,\nu_n)
      = \frac{1}{S}\sum_{s=1}^{S}\Big(
        H^{n+1}&\big(\tilde X^{(s)}(t_{n+1});\omega_{n+1}^\star\big)
        - H^n\big(\tilde X^{(s)}(t_n);\omega_n\big) \nonumber\\
        &- \hat F\big(t_n,\tilde X^{(s)}(t_n),
          G^n(\tilde X^{(s)}(t_n);\nu_n)\big)\,\Delta t_n\Big)^2.
    \end{align}
    \State Update
      $(\omega_n^\star,\nu_n^\star)
      \in\operatorname{argmin}\hat\ell_n(\omega_n,\nu_n)$.
  \EndFor
  \State \textbf{return} Functions $H^n(\,\cdot\,;\omega_n^\star)$ and
    $G^n(\,\cdot\,;\nu_n^\star)$ for $n=0,1,\ldots,N-1$.
\end{algorithmic}
\end{minipage}
\end{algorithm}

\paragraph{Proposed policy for the prelimit system.}
Using the trained gradient networks $G^{n}(\cdot;\nu_{n}^{\star})$, which approximate $\nabla_x V(t_n,\cdot)$, we construct a policy for the prelimit system as follows. At time $t\in[t_n,t_{n+1})$, we observe the state $X^n(t)$, form the scaled state $\hat X^n(t)$ defined in \eqref{eq:scaled-state} and the gradient estimate $p=G^{n}\big(\hat X^n(t);\nu_{n}^{\star}\big)$, and choose the server assignments $\psi_j^n(t)$, $j\in\calJ$, by solving the linear program
\begin{align}
  \text{Maximize}\quad
  & \sum_{j\in\calJ}\Big[c_{b(j)} + \bv_j\mu_j
    + (\mu_j-\theta_{b(j)})\,p_{b(j)}
    - \mu_j\sum_{i\in\calI}P_{ji}\,p_i\Big]\,\psi_j^n(t)
    \label{eq:prelimit-policy}\\
  \text{subject to}\quad
  & \sum_{j\in\mathcal{B}(i)}\psi_j^n(t) \le X_i^n(t),\qquad \, i\in\calI,
    \label{eq:pp-class}\\
  & \sum_{j\in\mathcal{S}(k)}\psi_j^n(t) \le N_k^n(t),\qquad k\in\calK,
    \label{eq:pp-pool}\\
  & \psi_j^n(t) \ge 0, \qquad \qquad \qquad \,\,\,\, j\in\calJ. \label{eq:pp-pos}
\end{align}
\begin{remark}
This policy is motivated by the minimization in the HJB equation \eqref{eq:hjb-full}, which characterizes the optimal control $\hat{\psi}^{\star}$ of the limiting problem. Specifically, at time $t \in [t_{n}, t_{n+1})$ and in state $x$, the control $\hat\psi$ is obtained by solving
\begin{equation}\label{eq:limit-policy}
\begin{aligned}
  \text{Minimize}\quad
  & \sum_{j\in\calJ}\Big[c_{b(j)} + \bv_j\mu_j
    + (\mu_j-\theta_{b(j)})\,p_{b(j)}
    - \mu_j\sum_{i\in\calI}P_{ji}\,p_i\Big]\,\hat\psi_j\\
  \text{subject to}\quad
  & \hat\psi\in\bar\Psi(t_n,x),
\end{aligned}
\end{equation}
where $p=\nabla_xV(t_n,x)$.  To translate this solution into a policy
recommendation for the prelimit system, one sets $\psi^n$ using
$\hat\psi$ and the scaling relation~\eqref{eq:control-prelimit}.
Because that relation carries a minus sign, the minimization
in~\eqref{eq:limit-policy} turns into the maximization
in~\eqref{eq:prelimit-policy}.  Also using the scaling relations, one
can show that the constraints defining $\bar\Psi(t_n,x)$
correspond to the constraints~\eqref{eq:pp-class}--\eqref{eq:pp-pos};
see Appendix~\ref{app:prelimit-policy} for details.
\end{remark}
\color{black}
Next, we introduce the reference policies considered in our
computational study.

\subsection{Reference Policies}\label{sec:reference-policies}
In our computational study, we consider seven reference policies.  The
first six are index rules.  Each index rule assigns a weight
$\tilde w_j$ to activity~$j$.  At each grid point $t_n$ and state $x$,
the control $\tilde\psi(t_n,x)$ is obtained by solving the linear
program
\begin{equation}\label{eq:ref-lp}
\begin{aligned}
  \text{Minimize}\quad & \sum_{j\in\calJ} \tilde w_j\,\hat\psi_j\\
  \text{subject to}\quad & \hat\psi\in\bar\Psi(t_n,x),
\end{aligned}
\end{equation}
\color{black}
where the vector of activity weights
$\tilde w=(\tilde w_1,\ldots,\tilde w_J)$ depends on the rule, and it
sets $\tilde\psi(t,x)=\tilde\psi(t_n,x)$ for $t\in[t_n,t_{n+1})$.  The
seventh policy requires no optimization.
\color{black}
\begin{itemize}
\item \emph{The $c\mu$ rule}, with weights
  $\tilde w_j = c_{b(j)}\mu_j$, following \citet{cox1961queues}. 
\item \emph{The $c\mu/\theta$ rule}, with weights
  $\tilde w_j = c_{b(j)}\mu_j/\theta_{b(j)}$, following
  \citet{atar2010cmu}.
\item \emph{The fastest-server-first (FSF) rule}, with weights
  $\tilde w_j=\mu_j$, following \citet{armony2005dynamic}.
\item \emph{The myopic rule}, with weights
  $\tilde w_j = c_{b(j)}+\bv_j\mu_j$.  This rule maximizes the
  instantaneous rate of value generation and holding cost relief.  Its
  weights are given by the activity weights in~\eqref{eq:weight}
  evaluated at $p=0$.  Thus, it coincides with the optimal control
  in~\eqref{eq:opt-policy} when the gradient of the value function is
  zero.
\item \emph{The abandonment-adjusted myopic rule}, with weights
  $\tilde w_j=(c_{b(j)}+\bv_j\mu_j)/\theta_{b(j)}$, paralleling the
  $c\mu/\theta$ rule of \citet{atar2010cmu}.
\item \emph{The value-rate rule}, with weights
  $\tilde w_j=\bv_j\mu_j$.  This rule maximizes the instantaneous rate
  of value generation.
\item \emph{The nominal policy}, which sets $\tilde\psi(t,x)=0$ and
  requires no linear program.  Under this policy, the reference
  process fluctuates around the nominal plan with no second-order
  control, and the training data concentrate around the trajectory on
  which the diffusion approximation is centered.
\end{itemize}

The reference policy enters the computational method in two places. It enters the drift of the reference process $\tilde X$ in
Equation~\eqref{eq:reference-process}, which generates the training
data, and it enters the auxiliary function $F$ in
Equation~\eqref{eq:F-explicit}, which appears in the loss function.  In
both expressions, the reference policy enters through the term
\begin{equation*}
\sum_{j\in\mathcal{B}(i)}(\mu_j-\theta_i)\,\tilde\psi_j(t,x)
-\sum_{j\in\calJ}P_{ji}\,\mu_j\,\tilde\psi_j(t,x)    
\end{equation*}
This term is the mapping $\mathcal{D}$ defined
in Equation~\eqref{eq:D-def}, so that the drift under the reference policy is
\begin{equation*}
b_i\big(t,x,\tilde\psi(t,x)\big) = \zeta_i(t) - \theta_i x_i
+ \mathcal{D}_i(t,x) \quad \text{for} \quad i\in\calI.  
\end{equation*}
Under an index rule, evaluating
$\tilde\psi(t,x)$ requires solving~\eqref{eq:ref-lp} at every state
visited along the sample paths, so we approximate $\mathcal{D}$
offline, as described in Appendix~\ref{app:offline}.

\section{Data, Test Problems, and Benchmark Policies}\label{sec:calibration}

\subsection{Data}\label{sec:data}

Our data come from the call center of a large U.S.\ bank and are publicly
available through the Service Enterprise Engineering (SEE) Laboratory at
the Technion \citep{SEEData}.  The records span March~2001 through
October~2003 and document both individual calls and agent activity.  The
center operates around the clock, seven days a week, and spans four
physical locations in New York, Pennsylvania, Rhode Island, and Massachusetts. Its workforce is organized into groups called \emph{nodes} that do not coincide with the physical locations, as a node may draw agents from several sites, and a single site may host agents from several nodes.  On a typical weekday the center receives between 330{,}000 and 350{,}000 calls and staffs more than a thousand agents, who are assigned calls according to
their skills.

An arriving call first reaches the voice response unit (VRU), an
automated system that lets the customer carry out transactions on their
own.  Most callers complete their service there, but roughly
50{,}000--65{,}000 callers per weekday
request a live agent.  These are the calls we focus on.  Because the menu
of services changed over the life of the data set, with some services
discontinued after 2001 and several new ones launched in November~2002,
we work with the $84$ business days from July through October~2002, a
period with a stable set of services and agent groups.  These days
comprise $4{,}548{,}264$ agent service completions.  Every call is
stored as one or more sub-calls that trace its path through the system. To
remove outliers, we keep only the calls whose outcome is a normal
termination, a transfer, a short abandonment, or an abandonment, which
together account for over $99\%$ of all observations, and we focus on calls with 
service times below $30$ minutes and waiting times below $15$ minutes.

There are eleven service types named Retail, Premier, Business,
Platinum, Consumer Loans, Online Banking, EBO, Telesales, Subanco, Case
Quality, and Priority Service.  Retail
alone generates roughly $71\%$ of the calls, which the system routes to
three service nodes, labeled 1, 2, and~3.  We therefore treat the Retail
calls handled at each node as a separate class, for a total of thirteen
customer types.  In the tables that follow we abbreviate the thirteen
classes as R1, R2, and R3 for the three Retail node classes, Pre for
Premier, Bus for Business, Plt for Platinum, CL for Consumer Loans, OB
for Online Banking, EBO, Tel for Telesales, Sub for Subanco, CQ for Case
Quality, and PS for Priority Service.  Following Section~\ref{sec:model}, each service type gives rise
to an \emph{original} class and a \emph{rework} class, according to
whether the caller receives a first or a second agent service, so that
$|\calI_o|=|\calI_r|=13$ and $I=26$.  A call is stored as an ordered
sequence of sub-calls, one per agent service, and each transfer opens a
new sub-call. 

Table~\ref{tab:subcalls} shows the distribution of the number of agent
services per contact.  Among the $4{,}429{,}080$ contacts that reach at
least one agent, $97.36\%$ end after a single service and $2.59\%$
require a second one.  Only $0.05\%$ of contacts go on to a third
service, so nearly all rework consists of exactly one additional
service.  This pattern supports the two-layer structure of
Section~\ref{sec:model} and its rework-exit assumption. Accordingly, we drop the observations of the third and later agent services of these contacts and treat the completion of the second service as the departure of the caller.

\begin{table}[ht]
\centering
\begin{tabular}{rrrr}
\toprule
Agent services & Contacts & \% of contacts & Cumulative \%\\
\midrule
$1$      & 4{,}312{,}297 & 97.363\% & 97.363\%\\
$2$      & 114{,}479     & 2.585\%  & 99.948\%\\
$3$      & 2{,}221       & 0.050\%  & 99.998\%\\
$4$      & 72            & 0.002\%  & 100.000\%\\
$\ge 5$  & 11            & 0.000\%  & 100.000\%\\
\bottomrule
\end{tabular}
\caption{Distribution of the number of agent services (sub-calls) per
agent-handled contact.}
\label{tab:subcalls}
\end{table}

\begin{figure}[ht]
\centering
\includegraphics[width=\linewidth]{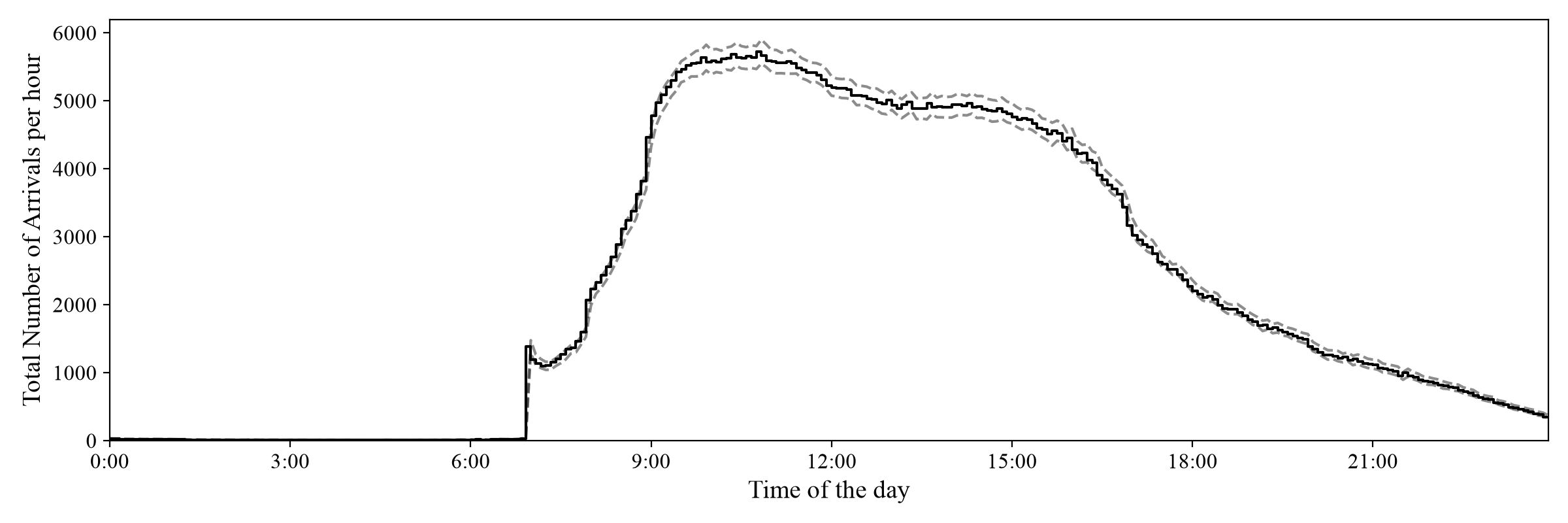}
\caption{Hourly arrival rate of callers on weekdays during July--October
2002, aggregated over all classes.  The resolution of the horizontal
axis is five minutes.  The solid curve is the across-day average and the
dashed curves give $99\%$ confidence intervals based on the day-to-day
variation of each five-minute count. }
\label{fig:arrivals-total}
\end{figure}

A class~$i$ arrival occurs when a caller whose original class is $i$
joins the agent queue.  We count these arrivals over five-minute
intervals and average the counts across the $84$ days to obtain the
time-varying arrival-rate functions $\lambda_i(\cdot)$.  Figure~\ref{fig:arrivals-total} displays the aggregate
pattern. Appendix \ref{app:data} contains the
corresponding graph for every original class.  Rework classes have no external arrivals because their arrivals consist entirely of first-service completions routed to them.

A caller who cannot be served immediately waits in a queue and may hang
up before reaching an agent.  Abandonment times are observed only for
the callers who abandon, while for everyone else they are censored at
the moment service begins.  Fewer than $3\%$ of the calls end in
abandonment, so abandonment times are heavily censored, which can lead to biased estimates.  We therefore estimate mean patience times with the
bias-corrected Kaplan--Meier estimator of \citet{stute1994jackknife}.  For an
original class, the patience observations come from first sub-calls
waiting in the agent queue, and for a rework class, from second sub-calls
waiting for the second agent service. Tables~\ref{tab:stats}
and~\ref{tab:stats-rework} summarize the two layers of the network, respectively.  Table~\ref{tab:stats} presents the number of contacts, the
arrival and rework percentages, and the abandonment rate of each
original class, and Table~\ref{tab:stats-rework} presents the number of
rework arrivals and the abandonment rate of each rework class.

\begin{table}[ht]
\centering
\setlength{\tabcolsep}{5pt}
{\small
\begin{tabular}{lcccc}
\toprule
Class & Number of & Arrival & Rework & $\theta$\\
\rule{0pt}{3.5ex} & contacts & percentage (\%) & percentage (\%) & (per hr.)\\
\midrule
Retail (Node: 1) & 1{,}138{,}786 & 25.71 & 3.02 & 7.61\\
Retail (Node: 2) & 1{,}164{,}545 & 26.29 & 1.57 & 6.63\\
Retail (Node: 3) & 841{,}566 & 19.00 & 4.18 & 6.92\\
Premier & 146{,}557 & 3.31 & 2.12 & 13.50\\
Business & 266{,}492 & 6.02 & 1.41 & 7.52\\
Platinum & 21{,}756 & 0.49 & 1.75 & 7.91\\
Consumer Loans & 292{,}142 & 6.60 & 3.29 & 4.79\\
Online Banking & 144{,}637 & 3.27 & 2.97 & 7.01\\
EBO & 37{,}009 & 0.84 & 0.71 & 5.05\\
Telesales & 309{,}243 & 6.98 & 0.72 & 7.94\\
Subanco & 25{,}370 & 0.57 & 7.94 & 7.15\\
Case Quality & 17{,}460 & 0.39 & 7.95 & 9.20\\
Priority Service & 23{,}517 & 0.53 & 8.05 & 15.77\\
\bottomrule
\end{tabular}
}
\caption{Summary statistics for the thirteen \emph{original} classes.  The second column gives the first-service volume of each
class and the fourth column its probability of generating a second agent
service.}
\label{tab:stats}
\end{table}

\begin{table}[ht]
\centering
\setlength{\tabcolsep}{5pt}
{\small
\begin{tabular}{lcc}
\toprule
Class & Number of & $\theta$\\
\rule{0pt}{3.5ex} & rework arrivals & (per hr.)\\
\midrule
Retail (Node: 1) & 5{,}789 & 7.66\\
Retail (Node: 2) & 5{,}612 & 8.99\\
Retail (Node: 3) & 4{,}804 & 4.83\\
Premier & 2{,}830 & 10.54\\
Business & 11{,}079 & 5.86\\
Platinum & 344 & 7.91$^{\dagger}$\\
Consumer Loans & 27{,}258 & 5.33\\
Online Banking & 16{,}758 & 7.94\\
EBO & 17{,}528 & 6.15\\
Telesales & 20{,}579 & 4.73\\
Subanco & 922 & 5.00\\
Case Quality & 1{,}345 & 9.20$^{\dagger}$\\
Priority Service & 1{,}932 & 15.77$^{\dagger}$\\
\bottomrule
\end{tabular}
}
\caption{Summary statistics for the thirteen \emph{rework} classes. The dagger indicates the rework classes whose queues produce almost no abandonments in the data. Their abandonment rates are taken from the corresponding original classes.}
\label{tab:stats-rework}
\end{table}

Skill-based routing assigns each call to an agent qualified for the
requested service. In the dataset, every agent carries a group code that identifies the
set of skills shared within the group.  Sixteen group codes, forming fourteen agent groups, appear in the dataset between July and October 2002, namely 1, 5, 9, 15/16, 19/20, 26, 28, 30, 31, 33,
34, 38, 39, and~40.  The paired codes 15/16 and 19/20 each denote a
single group whose code alternates between days.  Within each pair the
two codes never operate on the same day and always handle the same
services.

\begin{figure}[ht]
\centering
\includegraphics[width=\linewidth]{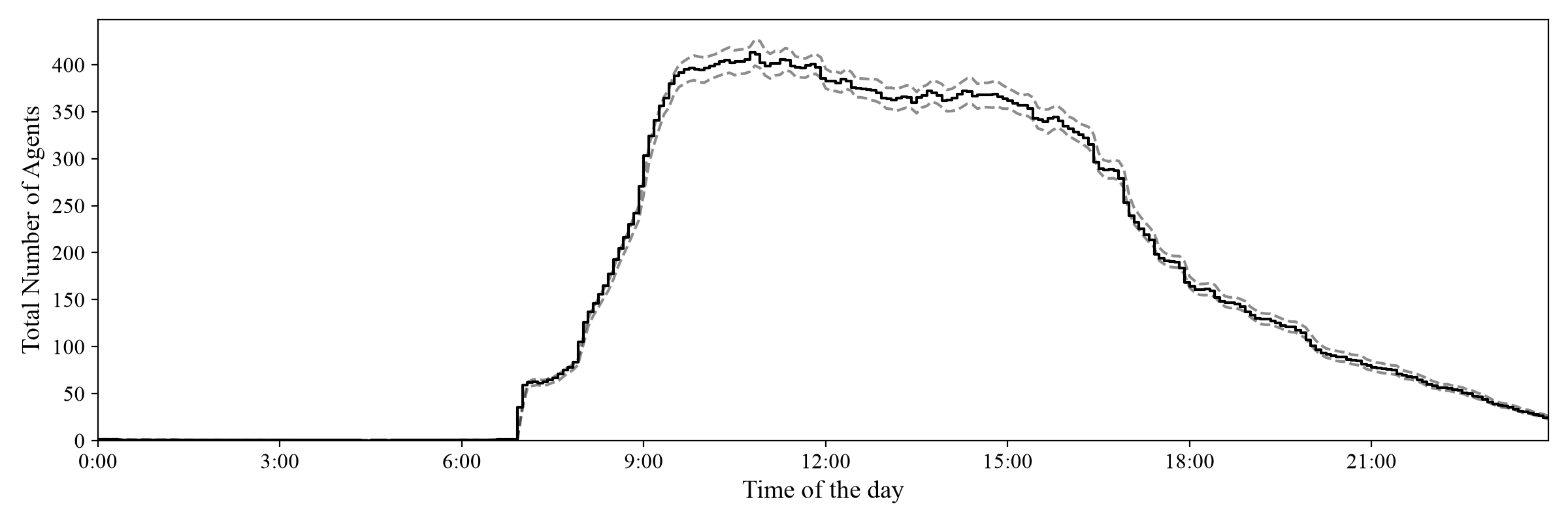}
\caption{Number of agents on duty on weekdays during July--October 2002,
aggregated over the nine server pools.  The resolution of the horizontal
axis is five minutes.  The solid curve is the across-day average and the
dashed curves give $99\%$ confidence intervals based on the day-to-day
variation of each five-minute count.}
\label{fig:staffing-total}
\end{figure}

The data also record the status of every agent second by second.  We
treat an agent as on duty from sign-in to sign-out, excluding breaks,
and count the agents on duty in each group over five-minute intervals.
Averaging these counts across days produces the time-varying staffing
levels $N_k(\cdot)$.  Figure~\ref{fig:staffing-total} shows the aggregate
staffing pattern over the day.

We construct skill-based server pools by combining agent group codes
with similar service capabilities, ensuring that each pool contains a
substantial number of agents.  Codes 26, 28, and~30 primarily serve
Business callers and form one pool.  Codes 15/16, 38, 39, and~40 form
another pool that collectively serves Retail, Premier, Subanco, Case
Quality, and Priority Service callers.  Each remaining code forms a separate
pool, yielding $K=9$ server pools.  Table~\ref{tab:pools} lists the
corresponding codes and classes served by each pool.
Appendix~\ref{app:data} shows the staffing levels for each server pool
throughout the day.

\begin{table}[ht]
\centering
{\small
\begin{tabular}{ccl}
\toprule
Pool & Agent group codes & Classes served\\
\midrule
1 & 1 & R1, R2, R3, Bus, Tel\\
2 & 5 & R2, R3, Pre, Bus\\
3 & 9 & R1, R2, Pre, Bus, EBO\\
4 & 26, 28, 30 & R2, R3, Bus, Plt, Tel\\
5 & 19/20 & R2, R3, Pre, Bus, CL\\
6 & 31 & R1, R3, CL\\
7 & 33 & R1, R3, Pre, Bus, CL, OB\\
8 & 34 & R3, Tel\\
9 & 15/16, 38, 39, 40 & R1, R3, Pre, Sub, CQ, PS\\
\bottomrule
\end{tabular}
}
\caption{The nine server pools, their corresponding agent group codes, and
the original classes they serve.}
\label{tab:pools}
\end{table}

An activity $(i,k)\in\mathcal E$ of the model exists whenever server pool~$k$ serves class~$i$ in the data. To eliminate outliers and rare assignments made under unforeseen increases in the workload, we exclude the (class, server pool) pairs with very few services. The two layers differ in volume, so we use a different threshold in each. In the original layer, we exclude a pair if the pool performs less than $0.1\%$ of its services in that layer on the class. In the rework layer, pool volumes are two orders of magnitude smaller. Therefore, we exclude a pair in the rework layer if the pool performs fewer than $100$ services on the class. This leaves $41$ activities serving the original classes and $21$ serving the rework classes, so the underlying network has $I+K=35$ nodes and $J=62$ activities. For each activity~$j$,
we estimate the hourly service rate $\mu_j$ from the mean handling time
of the services performed through that activity, separately for the two
layers.  Tables~\ref{tab:mu-orig}
and~\ref{tab:mu-rework} in Appendix \ref{app:data} list the estimated service rates for the original and rework classes respectively.
Comparing Tables~\ref{tab:mu-orig} and~\ref{tab:mu-rework} shows that
rework services run generally longer than original services of the
same service type.

Under Assumption~\ref{ass:rework}, routing moves callers only from
original classes into rework classes.  We estimate the routing
probabilities for each activity.  For an activity $j\in\calJ_o$, let
$n_j$ denote the number of first agent services delivered through $j$
and $n_{jl}$ the number of them followed by a second agent service of
class~$l$.  We estimate
\begin{equation}\label{eq:phat}
  \hat P_{jl} = \frac{n_{jl}}{n_j},\qquad
  \hat{q}_{j} = 1 - \sum_{l\in \mathcal{I}_{r}}\hat P_{jl}.
\end{equation}
Appendix~\ref{app:data} shows the estimated routing
probabilities~\eqref{eq:phat} for every activity.

In total, our analysis covers $4{,}429{,}080$ contacts across the eleven
service types received on weekdays between July and October~2002,
together with the $116{,}783$ second sub-calls they generate.  With the
Retail class split across service nodes and the original and rework layers, the resulting system has $26$ customer classes and $9$ server pools.

We use secondary data sources to estimate the economic model parameters of Section~\ref{sec:objective}. We take the opportunity cost of an hour of waiting to be the caller's
foregone hourly wage.  The U.S.\ Bureau of Labor Statistics reports an
average hourly wage of \$25 in the retail
industry\footnote{Available at
https://www.bls.gov/news.release/empsit.t19.htm.}, which we adopt as the
holding cost rate of the Retail classes.  We group the remaining service
types by their perceived priority relative to the Retail class.
Premier, Business, Platinum, and Priority Service form the
higher-priority group, with rates between \$27 and \$33 per hour.  The
lower-priority group is further graded by call volume.  Service types
drawing less than $1\%$ of total arrivals, namely EBO, Subanco, and Case
Quality, receive the lowest rate of \$20 per hour, and the others,
Online Banking, Consumer Loans, and Telesales, receive \$23 per hour.
This assignment keeps the weighted average holding cost rate close to
the \$25 used for the Retail classes.  Table~\ref{tab:cost-rates}
reports the resulting holding cost rates together with the arrival percentage of each class.

For the abandonment penalties, we approximate the caller's value for the service using the value of time spent during service. Assuming servers handle approximately 15 calls per hour, we set the abandonment penalty of class $i$ as $a_{i} = h_{i}/15$. While this is a crude approach, it aligns the abandonment penalties with the structure of the holding cost rates, as shown in Table \ref{tab:cost-rates}. A service type's economic parameters are the same for both its original and rework classes. 

According to
ZipRecruiter\footnote{Available at
https://www.ziprecruiter.com/Salaries/Call-Center-Agent-Salary}, the
average hourly wage of a call center representative in the U.S.\ is
approximately \$17. At a service rate of $15$ calls per hour, this
implies a labor cost of approximately \$1.13 per call. The activity
values in Tables~\ref{tab:vj-orig} and \ref{tab:vj-rework} range from
\$1.86 to \$8.75 and therefore exceed this estimated cost. Crediting
the net value of services completed after the planning horizon could
thus weaken the incentive to complete them within the planning
horizon, motivating our choice of zero terminal value in
Section~\ref{sec:objective}.
\color{black}
\begin{table}[ht]
\centering
{\small
\begin{tabular}{lccc}
\toprule
Service type & Arrival & $h$ & $a$\\
\rule{0pt}{3.5ex} & percentage (\%) & (per hr.) & (per call)\\
\midrule
Case Quality           & 0.39  & \$20 & \$1.33\\
Subanco                & 0.57  & \$20 & \$1.33\\
EBO                    & 0.84  & \$20 & \$1.33\\
\midrule
Online Banking         & 3.27  & \$23 & \$1.53\\
Consumer Loans         & 6.60  & \$23 & \$1.53\\
Telesales              & 6.98  & \$23 & \$1.53\\
\midrule
Retail (Node: 1, 2, 3) & 71.00 & \$25 & \$1.67\\
\midrule
Premier                & 3.31  & \$27 & \$1.80\\
Business               & 6.02  & \$30 & \$2.00\\
Platinum               & 0.49  & \$33 & \$2.20\\
Priority Service       & 0.53  & \$33 & \$2.20\\
\bottomrule
\end{tabular}
}
\caption{Holding cost rates and abandonment penalties per service type.  Rows are grouped
by holding cost rate. A service type's parameters apply to both its original and its rework
class.}
\label{tab:cost-rates}
\end{table}

Next, we calibrate the values $v_j$ and the transfer penalties
$\kappa_{jl}$ defined in Equation \eqref{eq:vbar}. Every customer contacts the call center with a request they would like resolved. As a result, we assume an activity generates value only if it resolves the request of the caller. We let $s_{j}$ denote the probability that the request is resolved, given that the caller leaves the system after activity $j$. We assume that this
probability differs between original and rework activities and set $s_{j} = s_{o}$ for $j \in \mathcal{J}_{o}$ and $s_{j} = s_{r}$ for $j \in \mathcal{J}_{r}$. The value of a resolution depends also on the class $b(j)$
served and on the server pool $s(j)$ providing the service. As a
result, we define
\begin{equation}\label{eq:vj-def}
  v_j = s_j\, u_{b(j)}\, \gamma_j,
  \qquad j\in\calJ,
\end{equation}
where $u_{i}$ is the value of resolving a class $i$ request when service is provided by an agent from its specialty pool. The matching factor $\gamma_{j} \in (0,1]$ discounts this value when pool $s(j)$ is not the specialty pool for class $b(j)$. We calibrate the resolution probabilities $s_{o}$ and $s_{r}$ first, then the customer values $u_{i}$ and then the matching factors $\gamma_{j}$. Finally, we calibrate the transfer penalties $\kappa_{jl}$.

We first specify the resolution probabilities. Let $\bar{s}$ denote the average first-call resolution probability, and let $s_{o}$ and $s_{r}$ denote the resolution probabilities for contacts that leave after their original service and those routed to a rework class, respectively. We assume
\begin{equation}\label{eq:s-mixture}
  s_o\,\mathbb{P}(\text{contact leaves after original service})
  + s_r\,\mathbb{P}(\text{contact is routed to a rework class})
  = \bar s .
\end{equation}
We calibrate these probabilities using findings from SQM Group, a
provider of call center quality assurance software that measures
first-call resolution through post-call surveys at more than 500 North
American call centers.\footnote{Available at
https://www.linkedin.com/company/sqm-group-inc-, accessed September 19, 2026.}
SQM Group reports an average first-call resolution rate of
$69\%$,\footnote{Available at
https://www.sqmgroup.com/resources/library/blog/call-center-fcr-benchmark-2024-results-by-industry, accessed September 19, 2026.}
with a rate $14\%$ lower for transferred contacts, which tend to be
more complex.\footnote{Available at
https://www.sqmgroup.com/resources/library/blog/call-transfer-hold-performance-impact-csat-and-fcr
and
https://www.sqmgroup.com/resources/library/blog/agent-call-handling-impact-fcr-and-call-resolution, accessed September 19, 2026.}
Accordingly, we set $\bar s = 0.69$ and $s_r = 0.86\, s_o$. In our
data, $97.4\%$ of contacts leave after their original service, while
$2.6\%$ are routed to a rework class (Table~\ref{tab:subcalls}). Thus,
we solve
\begin{equation}\label{eq:s-solve}
  0.974\, s_o + 0.026\, s_r = 0.69,
  \qquad
  s_r = 0.86\, s_o,
\end{equation}
which yields $s_o = 0.693$ and $s_r = 0.596$.

We translate successful resolutions into monetary value through their
effect on customer retention. Specifically, we define $u_i$ as the
difference in expected customer lifetime value (CLV) between a class $i$
customer whose request is resolved and one whose request remains
unresolved after the call. Let $m_i$ denote the annual margin generated
by a class $i$ customer, received at the end of each year the customer
is retained. Let $\rho$ denote the average annual retention probability
and $\delta$ the annual discount rate. Following
\citet[Equation~3]{gupta2003customers}, the expected customer lifetime
value is
\begin{equation}\label{eq:margin-multiple}
  \mathrm{CLV}_i
  = \sum_{k=1}^{\infty}\frac{m_i\,\rho^{k}}{(1+\delta)^{k}}
  = \frac{\rho\, m_i}{1+\delta-\rho},
  \qquad i\in\calI.
\end{equation}
We assume that resolving a request affects retention only in the
current year, and that the annual retention probability is $\rho$ in
every later year. A request is resolved with probability $\bar s$, in which
case the current-year retention probability is $\rho+\Delta_1$.
Otherwise, the request remains unresolved, and the current-year
retention probability is $\rho-\Delta_2$. The corresponding expected
lifetime values are
\begin{equation}\label{eq:clv-unresolved}
  \mathrm{CLV}_i^{U}
  = \sum_{k=1}^{\infty}\frac{m_i\,(\rho-\Delta_2)\,\rho^{k-1}}{(1+\delta)^{k}}
  = \frac{(\rho-\Delta_2)\, m_i}{1+\delta-\rho},
  \qquad i\in\calI,
\end{equation}
and
\begin{equation}\label{eq:clv-resolved}
  \mathrm{CLV}_i^{R}
  = \sum_{k=1}^{\infty}\frac{m_i\,(\rho+\Delta_1)\,\rho^{k-1}}{(1+\delta)^{k}}
  = \frac{(\rho+\Delta_1)\, m_i}{1+\delta-\rho},
  \qquad i\in\calI.
\end{equation}
The value of resolving a class $i$ request is therefore
\begin{equation}\label{eq:ui-def}
  u_i
  = \mathrm{CLV}_i^{R} - \mathrm{CLV}_i^{U}
  = \frac{\Delta\rho\; m_i}{1+\delta-\rho},
  \qquad i\in\calI,
\end{equation}
where $\Delta\rho = \Delta_1+\Delta_2$ is the average difference in
retention probability between resolved and unresolved requests. We refer to
$\Delta\rho$ as the retention lift. We take $\rho$, $\delta$, and
$m_i$ from the literature and calibrate $\Delta\rho$.

We take estimates of $\rho$, $\delta$, and $m_i$ from
\citet{gupta2004valuing}, who report them for the financial
institutions in their study. Their estimates of the retention
probability $\rho$ range from 85\% for the credit-card company Capital
One to 95\% for the brokerage accounts of Ameritrade, and we set
$\rho = 90\%$, the midpoint of this range. For the discount rate, \citet{gupta2004valuing} adopt
$\delta = 12\%$ as the midpoint of the $8$--$16\%$ cost-of-capital
range suggested by the finance literature. For the annual margins $m_i$,
\citet[Table~2]{gupta2004valuing} report quarterly per customer
margins of \$13.71 for Capital One, \$43.02 for E*Trade, and \$50.39
for Ameritrade, which annualize to approximately \$55, \$172, and
\$200 per customer. We assume that the service types with higher
holding costs also have higher margins. Therefore, we assign margins
between \$55 and \$200 across the holding cost groups of
Table~\ref{tab:cost-rates}.
Table~\ref{tab:econ-params} reports the margins.

We calibrate the retention lift $\Delta\rho$ by distributing the
annual attrition attributable to service failures across the
agent-handled calls a customer makes each year,
\begin{equation}\label{eq:delta-rho}
  \Delta\rho
  = \frac{(1-\rho)\,\chi}{L},
\end{equation}
where $1-\rho$ is the annual attrition rate, $\chi$ is the share of
attrition attributable to service failures, and $L$ is the average
number of agent-handled calls per customer per year. With an annual
retention probability of $\rho = 0.90$, the bank loses $10\%$ of its
customers each year. \citet{keaveney1995customer} identifies failed
service encounters as a leading reason for customer switching in
service industries, reported by one-third of survey respondents. Based
on this finding, we set $\chi = 1/3$. \citet{campbell2010cost} report
an average of $0.2$ call center transactions per customer per month at
a large U.S.\ bank, giving $L = 2.4$ calls per year. Combining these
inputs yields
\begin{equation}\label{eq:delta-rho-value}
  \Delta\rho
  = \frac{0.10\times(1/3)}{2.4}
  \approx 0.0139 .
\end{equation}
Thus, we set the retention lift to $1.39\%$. Table~\ref{tab:econ-params} reports
the resulting customer values. A customer whose request is resolved
after rework is the same customer with the same request. Consequently, the
rework class of a service type has the same customer value $u_i$ as its
original class.
\color{black}

\begin{table}[ht]
\centering
\begin{tabular}{lrr}
\toprule
Service type & $m_i$ (\$ per year) & $u_i$ (\$ per resolved request)\\
\midrule
EBO, Subanco, Case Quality                  & 55  & 3.48\\
Online Banking, Consumer Loans, Telesales   & 65  & 4.11\\
Retail                                     & 75  & 4.74\\
Premier                                    & 125 & 7.90\\
Business                                   & 150 & 9.48\\
Platinum, Priority Service                  & 200 & 12.64\\
\bottomrule
\end{tabular}
\caption{Annual margins and the value of resolving a request, by
service type. The annual margins $m_i$ span the range of per-customer
margins reported for financial institutions by
\citet[Table~2]{gupta2004valuing}. Customer values $u_i$ are calculated
using Equation~\eqref{eq:ui-def} with $\rho=90\%$, $\delta=12\%$, and
$\Delta\rho=1.39\%$.}
\label{tab:econ-params}
\end{table}
We assign the matching factors $\gamma_j$ using the observed service
shares of each server pool. For each class, we rank the pools in
descending order of the percentage of its services they perform. We
set the highest-ranked pool as the specialty pool of the class and
assign its activities a matching factor of $\gamma_j = 1$, so that the
customer value is not discounted. For the remaining pools, we assume
larger discounts as their rank decreases,
\begin{equation}\label{eq:gamma}
  \gamma_j =
  \begin{cases}
    1,   & r(j) = 1,\\
    0.9, & r(j) = 2,\\
    0.8, & r(j) = 3,\\
    0.7, & r(j) \ge 4,
  \end{cases}
\end{equation}
where $r(j)$ is the rank of pool $s(j)$ among the pools serving class
$b(j)$. Tables~\ref{tab:shares-orig} and~\ref{tab:shares-rework} in
Appendix~\ref{app:data-shares} report the service shares for each class
and pool from the data. Tables~\ref{tab:vj-orig} and~\ref{tab:vj-rework} report the
resulting activity values, calculated using
Equation~\eqref{eq:vj-def}.
\color{black}

\begin{table}[ht]
\centering
{\small
\setlength{\tabcolsep}{4pt}
\begin{tabular}{lrrrrrrrrrrr}
\toprule
Class & $s_o$ & $u_i$ & \multicolumn{9}{c}{Server pool}\\
\cmidrule(lr){4-12}
 & & & 1 & 2 & 3 & 4 & 5 & 6 & 7 & 8 & 9\\
\midrule
Retail (Node: 1) & 0.693 & 4.74 & \textbf{3.28} &  & 2.95 &  &  & 2.30 & 2.30 &  & 2.63\\
Retail (Node: 2) & 0.693 & 4.74 & 2.95 & \textbf{3.28} & 2.63 & 2.30 & 2.30 &  &  &  & \\
Retail (Node: 3) & 0.693 & 4.74 & 2.95 & \textbf{3.28} &  & 2.30 & 2.30 & 2.30 & 2.30 & 2.30 & 2.63\\
Premier & 0.693 & 7.90 &  & 4.92 & 3.83 &  & \textbf{5.47} &  & 3.83 &  & 4.38\\
Business & 0.693 & 9.48 & 4.59 & 5.91 & 4.59 & \textbf{6.56} & 5.25 &  & 4.59 &  & \\
Platinum & 0.693 & 12.64 &  &  &  & \textbf{8.75} &  &  &  &  & \\
Consumer Loans & 0.693 & 4.11 &  &  &  &  & 2.56 & \textbf{2.84} & 2.28 &  & \\
Online Banking & 0.693 & 4.11 &  &  &  &  &  &  & \textbf{2.84} &  & \\
EBO & 0.693 & 3.48 &  &  & \textbf{2.41} &  &  &  &  &  & \\
Telesales & 0.693 & 4.11 & 2.28 &  &  & 2.56 &  &  &  & \textbf{2.84} & \\
Subanco & 0.693 & 3.48 &  &  &  &  &  &  &  &  & \textbf{2.41}\\
Case Quality & 0.693 & 3.48 &  &  &  &  &  &  &  &  & \textbf{2.41}\\
Priority Service & 0.693 & 12.64 &  &  &  &  &  &  &  &  & \textbf{8.75}\\
\bottomrule
\end{tabular}
}
\caption{Activity values $v_j$ for the $41$ activities
serving the original classes, from Equation~\eqref{eq:vj-def}. Columns are server pools and bold marks the specialty
pool of the class, where $\gamma_j = 1$.}
\label{tab:vj-orig}
\end{table}

\begin{table}[ht]
\centering
{\small
\setlength{\tabcolsep}{4pt}
\begin{tabular}{lrrrrrrrrrrr}
\toprule
Class & $s_r$ & $u_i$ & \multicolumn{9}{c}{Server pool}\\
\cmidrule(lr){4-12}
 & & & 1 & 2 & 3 & 4 & 5 & 6 & 7 & 8 & 9\\
\midrule
Retail (Node: 1) & 0.596 & 4.74 & \textbf{2.82} &  & 2.54 &  &  &  &  &  & \\
Retail (Node: 2) & 0.596 & 4.74 & 2.54 & \textbf{2.82} & 2.26 &  &  &  &  &  & 1.98\\
Retail (Node: 3) & 0.596 & 4.74 & 2.54 & \textbf{2.82} &  &  &  & 1.98 &  &  & 2.26\\
Premier & 0.596 & 7.90 &  &  &  &  & \textbf{4.70} &  &  &  & \\
Business & 0.596 & 9.48 &  &  &  & \textbf{5.64} &  &  &  &  & \\
Platinum & 0.596 & 12.64 &  &  &  & \textbf{7.53} &  &  &  &  & \\
Consumer Loans & 0.596 & 4.11 &  &  &  &  &  & \textbf{2.45} &  &  & \\
Online Banking & 0.596 & 4.11 &  &  &  &  &  &  & \textbf{2.45} &  & \\
EBO & 0.596 & 3.48 &  &  & \textbf{2.07} &  &  &  &  &  & \\
Telesales & 0.596 & 4.11 &  &  &  &  &  &  &  & \textbf{2.45} & \\
Subanco & 0.596 & 3.48 &  & 1.86 &  &  &  &  &  &  & \textbf{2.07}\\
Case Quality & 0.596 & 3.48 &  &  &  &  &  &  &  &  & \textbf{2.07}\\
Priority Service & 0.596 & 12.64 &  &  &  &  &  &  &  &  & \textbf{7.53}\\
\bottomrule
\end{tabular}
}
\caption{Activity values $v_j$ for the $21$ activities
serving the rework classes, from Equation~\eqref{eq:vj-def}. Columns are server pools and bold marks the specialty
pool of the class, where $\gamma_j = 1$.}
\label{tab:vj-rework}
\end{table}
We calibrate the transfer penalties $\kappa_{jl}$ using customer
satisfaction scores (CSAT), which measure the satisfaction of customers
with the service received. SQM Group reports that CSAT scores are
$12\%$ lower for transferred customers.\footnote{Available at
https://www.sqmgroup.com/resources/library/blog/call-transfer-hold-performance-impact-csat-and-fcr, accessed September 19, 2026.}
We use this finding to motivate a proportional discount on the value
generated by a transferred contact. Under Equation~\eqref{eq:vbar}, an
original-service activity $j$ that transfers a caller to rework class
$l$ incurs a penalty $\kappa_{jl}$. If the caller subsequently receives
service through activity $j'\in\calJ_r$, with $b(j') = l$, the original
service generates no value, the transfer incurs the penalty
$\kappa_{jl}$, and the rework service generates $v_{j'} = s_{r}u_{l}\gamma_{j'}$. The total
value generated by the contact is therefore $v_{j'} - \kappa_{jl}$.
We set the transfer penalty to a fraction $\kappa_0 = 0.12$ of the
expected value of the rework service,
\begin{equation}\label{eq:kappa-recipe}
  \kappa_{jl} = \kappa_0\, s_r\, u_l\, \bar\gamma_l,
  \qquad j\in\calJ_o,\ l\in\calI_r,
\end{equation}
where $s_r\, u_l$ is the expected value of a class $l$ request served
by its specialty pool, and
\begin{equation}\label{eq:gamma-bar}
  \bar\gamma_l = \sum_{j'\in\mathcal{B}(l)} \alpha_{lj'}\,\gamma_{j'},
  \qquad l\in\calI_r,
\end{equation}
is the average matching factor for class $l$. Here, $\alpha_{lj'}$ is
the observed share of class $l$ services performed by activity $j'$ from the data. Because the pool that will provide rework service is unknown at the
time of transfer, we use these observed shares to calculate the
expected value of the rework service. Table~\ref{tab:kappa} reports the resulting transfer
penalties.
\color{black}

\begin{table}[ht]
\centering
{\small
\begin{tabular}{lrrrr}
\toprule
Rework class $l$ & $s_r$ & $u_l$ & $\bar\gamma_l$ & $\kappa_{jl}$\\
\midrule
Retail (Node: 1)                          & 0.596 & 4.74  & 0.98 & 0.33\\
Retail (Node: 2, 3)                       & 0.596 & 4.74  & 0.95 & 0.32\\
Premier                                   & 0.596 & 7.90  & 1    & 0.56\\
Business                                  & 0.596 & 9.48  & 1    & 0.68\\
Platinum, Priority Service                & 0.596 & 12.64 & 1    & 0.90\\
Consumer Loans, Online Banking, Telesales & 0.596 & 4.11  & 1    & 0.29\\
EBO, Case Quality                         & 0.596 & 3.48  & 1    & 0.25\\
Subanco                                   & 0.596 & 3.48  & 0.99 & 0.25\\
\bottomrule
\end{tabular}
}
\caption{Transfer penalties by rework class, from
Equation~\eqref{eq:kappa-recipe} with $\kappa_0 = 12\%$ and
$u_l$ and $\bar\gamma_l$.}
\label{tab:kappa}
\end{table}
\color{black}
To summarize, the data give us the following primitives at a resolution
of five minutes over the day.  For each original class $i$, we observe
the arrival rate $\lambda_i(\cdot)$, and for each server
pool $k$ the staffing level $N_k(\cdot)$.  The service rates $\mu_j$,
the abandonment rates $\theta_i$ and the routing probabilities $P_{jl}$
are estimated once for the planning horizon, and the economic parameters
$h_i$, $a_i$, $v_j$ and $\kappa_{jl}$ come from the secondary sources
above.  The resulting network has $13$ original classes, $13$ rework
classes, $9$ server pools and $62$ activities.  The next section
turns these primitives into the test problems.

\subsection{Main test problem and its variants}\label{section:main_test}

We calibrate the main test problem using the U.S.\ Bank call center
data described in Section~\ref{sec:data}. We also consider two variants
of this problem that differ only in their prelimit arrival rates. All
three problems have the same first-order primitives but different
second-order arrival-rate terms $\zeta$, as defined in
Equation~\eqref{eq:scaling}.

We consider the 9~A.M.\ to 5~P.M.\ working hours, with $t = 0$
corresponding to 9~A.M.\ and the length of the planning horizon
$T = 8$ hours. We partition the time horizon $[0,T]$ as
$0 = \tau_0 < \tau_1 < \cdots < \tau_M = T$ into $M = 96$ five-minute
intervals. For each interval $[\tau_m,\tau_{m+1})$, let
$\lambda^n_i(\tau_m)$ denote the arrival rate of class $i$, estimated
as shown in Appendix~\ref{app:data-arrivals}, and let $N_k(\tau_m)$
denote the number of agents on duty in server pool $k$, estimated as
shown in Appendix~\ref{app:data-staffing}. In addition, we set the
system parameter to $n = 100$, reflecting the scale of the staffing
levels observed in the data, and we use this value to calculate the
limiting staffing levels as
\begin{equation}\label{eq:nu-calib}
  \nu_k(\tau_m) = \frac{N_k(\tau_m)}{n}, \qquad k\in\calK,\
  m = 0,1,\ldots,M-1 .
\end{equation}

\begin{figure}[ht]
\centering
\includegraphics[width=0.8\linewidth]{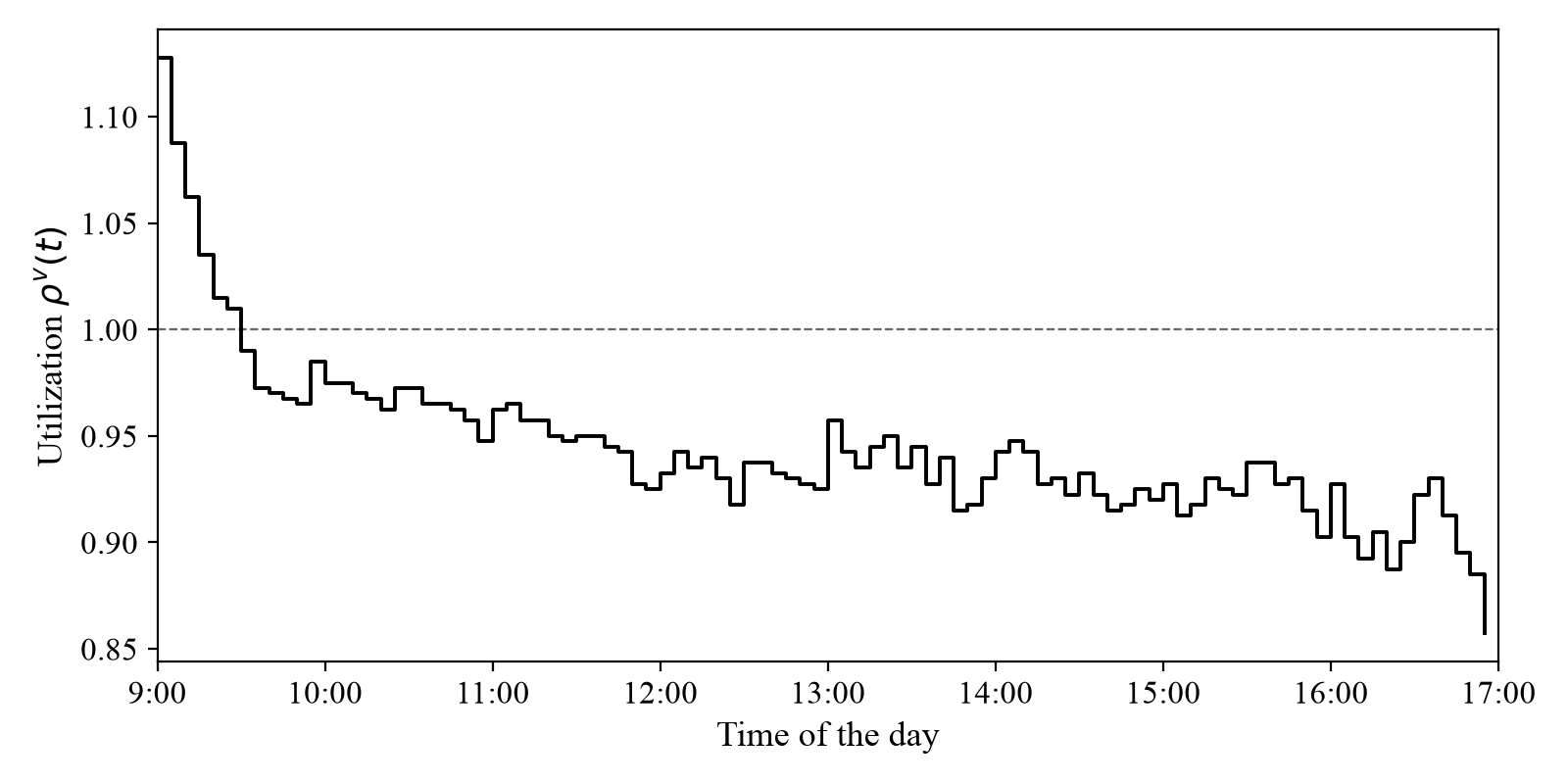}
\caption{Utilization $\rho(\cdot)$ under the value-maximizing plan at
the estimated arrival rates between 9~A.M.\ and 5~P.M.  The resolution
of the horizontal axis is five minutes and the dashed line marks $\rho = 1$.}
\label{fig:utilization}
\end{figure}

We assume that the arrival rates and the staffing levels are constant
in each five-minute interval and consider each interval separately. Recall from Equation \eqref{eq:scaling} that the arrival rate function satisfies 
\begin{equation}\label{eq:scaling-recall}
    \lambda^{n}(t) = n\lambda(t) + \sqrt{n}\,\zeta(t) + o(\sqrt{n}),
\end{equation}
and that the first-order term $\lambda(t)$ serves as an input to the planning problem \eqref{eq:spp}-\eqref{eq:spp-const3}. In order to determine $\lambda(t)$, we proceed as follows: First, we solve the following auxiliary problem numerically for different values of $\rho$:
\begin{align}
    \text{maximize}\quad & \sum_{j\in\calJ}\bv_j\,\mu_j\,\psi^n_j \label{eq:rho-v}\\
    \text{subject to}\quad & R\,\psi^n = \lambda^n(\tau_m), \label{eq:rho-v-const1}\\
    & A\,\psi^n \le \rho\,N(\tau_m), \label{eq:rho-v-const2}\\
    & \psi^n \ge 0, \label{eq:rho-v-const3}
\end{align}
where $\psi^n_j$ is the number of agents working on activity $j$. We first consider a relatively large value of $\rho$ ($\rho = 2$) and observe numerically that as $\rho$ decreases, we eventually reach a solution that satisfies\footnote{It appears that this is an artifact of the overlapping capabilities of the server pools; see Table \ref{tab:shares-orig}. Similar properties have also been observed in earlier work; see for example \citet{harrison1999heavy} and \citet{ata2005heavy}.}:
\begin{equation}\label{eq:exhaust}
    A\psi^{n*}(\rho) = \rho N(\tau_m).
\end{equation}
That is, all server pools have the same utilization $\rho$. We denote this value by $\rho(\tau_m)$, and Figure~\ref{fig:utilization} shows $\rho(\cdot)$ over the time horizon. Then we set
\begin{equation}\label{eq:lambda-calib}
  \lambda(\tau_m) = \frac{\lambda^n(\tau_m)}{\rho(\tau_m)\,n},
  \qquad m = 0,1,\ldots,M-1 ,
\end{equation}
and solve the following planning problem to determine the nominal plan $\xi^*(\tau_m)$ on the interval $[\tau_m,\tau_{m+1})$:
\begin{align}
    \text{maximize}\quad & \tilde v(\tau_m)\cdot\xi \label{eq:spp-lp}\\
    \text{subject to}\quad & R\,D(\tau_m)\,\xi = \lambda(\tau_m), \label{eq:spp-lp-flow}\\
    & A\,\xi \le e, \label{eq:spp-lp-cap}\\
    & \xi \ge 0, \label{eq:spp-lp-pos}
\end{align}
where $\tilde v_j(\tau_m) = \bv_j\,\mu_j\,\nu_{s(j)}(\tau_m)$. One can verify the following connection between problems \eqref{eq:rho-v}--\eqref{eq:rho-v-const3} and \eqref{eq:spp-lp}--\eqref{eq:spp-lp-pos}:
\begin{equation}\label{eq:psi-xi-connection}
    \psi^{n*}\big(\rho(\tau_m)\big) = n\,\rho(\tau_m)\,D(\tau_m)\,\xi^{*}(\tau_m),
\end{equation}
from which it follows that $A\xi^{*}(\tau_{m}) = e$ on every interval. We also verify that the solution is unique on every interval (see Appendix \ref{app:uniqueness} for details). Hence, the heavy traffic assumption (in Assumption \ref{ass:ht}) holds.

The second-order terms then follow from Equation~\eqref{eq:scaling},
\begin{equation}\label{eq:zeta-calib}
  \zeta(\tau_m)
  = \frac{\lambda^n(\tau_m)-n\,\lambda(\tau_m)}{\sqrt n},
  \qquad m = 0,1,\ldots,M-1 .
\end{equation}
\vspace{-10mm}

To study the policy under heavier arrival loads, we consider two
variants of the main test problem, with $5\%$ and $10\%$ more arrivals
than the main test problem, respectively. The staffing levels, the
service rates, the abandonment rates, the routing probabilities, the
economic parameters, the limiting arrival rates $\lambda(\cdot)$, the
limiting staffing levels $\nu(\cdot)$, and the nominal plan
$\xi^*(\cdot)$ are the same across the three problems. The prelimit
arrival rates of a variant are $\beta\,\lambda^n(\tau_m)$ for
$m = 0,1,\ldots,M-1$, with $\beta = 1.05$ for the first variant and
$\beta = 1.10$ for the second variant. Then, for these test problems,
we set the second-order terms as
\begin{equation}\label{eq:zeta-variant}
  \zeta(\tau_m)
  = \frac{\beta\,\lambda^n(\tau_m)-n\,\lambda(\tau_m)}{\sqrt n},
  \qquad m = 0,1,\ldots,M-1 .
\end{equation}

\begin{figure}[ht]
\centering
\includegraphics[width=0.8\linewidth]{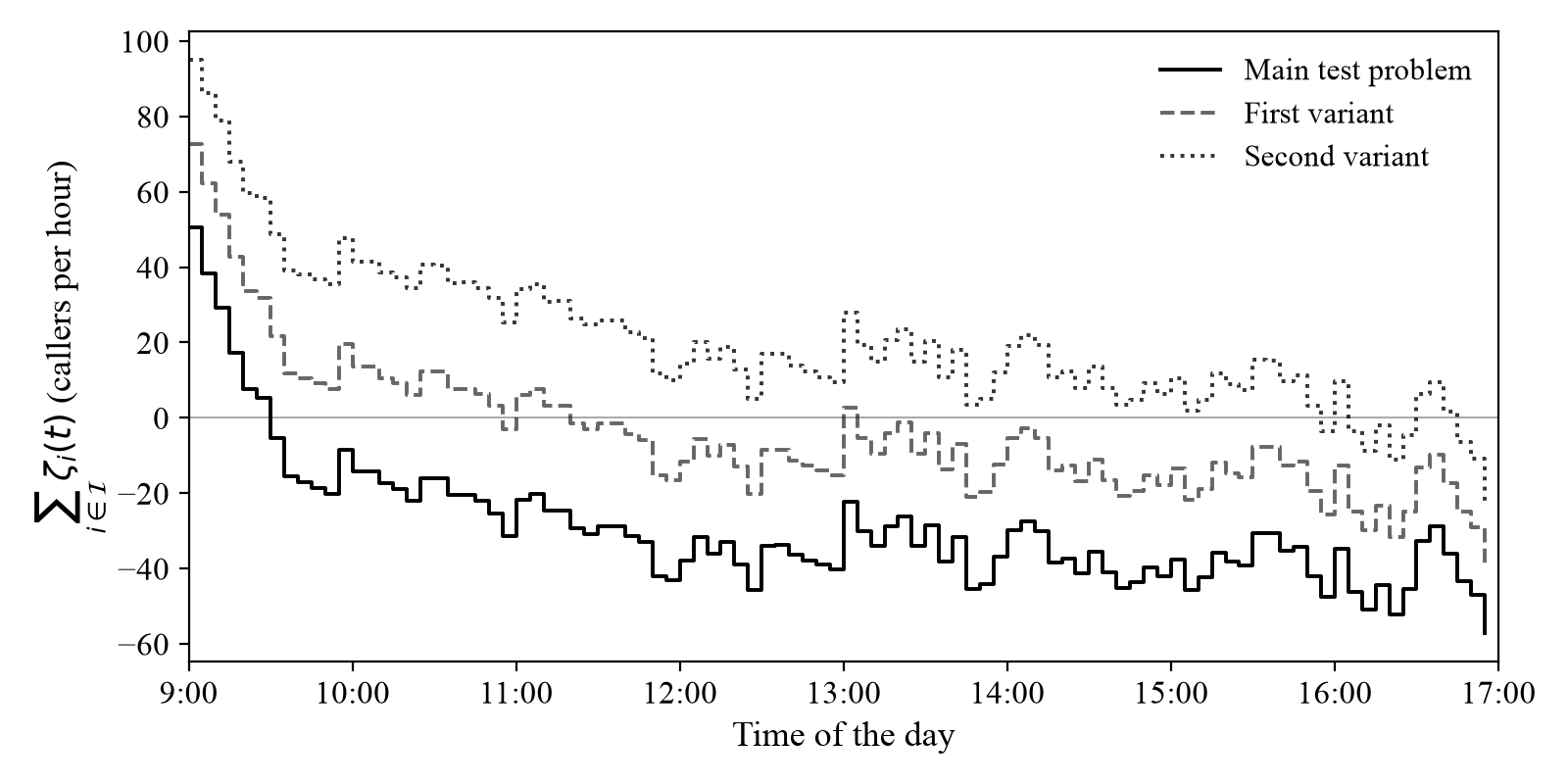}
\caption{The sum of the second-order terms over the classes,
$\sum_{i\in\calI}\zeta_i(\cdot)$, for the main test problem and its two
variants.  The resolution of the horizontal axis is five minutes.}
\label{fig:zeta-total}
\end{figure}

Figure~\ref{fig:zeta-total} shows the sum of the second-order terms
over the classes for the main test problem and its two variants, and
Figure~\ref{fig:zeta-classes} in Appendix~\ref{app:data-arrivals}
shows the term of each original class. For the main test problem the
sum is negative for most of the time horizon, so the main test problem
is underloaded for most of the time horizon. The first variant is
critically loaded and the second variant is overloaded for most of the
time horizon.

\color{black}
\vspace{-5mm}
\subsection{Benchmark policies}\label{sect:benchmarks}

This section introduces the benchmark policies used to evaluate the
proposed policy in our three test problems. Section~\ref{sec:results}
reports the computational results. Each benchmark policy considered is
an index rule that assigns a weight $w_j$ to
activity $j$. At time $t$, the policy is determined by solving the linear
program~\eqref{eq:prelimit-policy}--\eqref{eq:pp-pos} with the activity
weights in~\eqref{eq:prelimit-policy} replaced by $w_j$. Specifically,
in state $X^n(t)$, the benchmark policy chooses $\psi^n(t)$ that
maximizes
\begin{align}
  \text{Maximize}\quad
  & \sum_{j\in\calJ} w_j\,\psi_j^n(t)
    \label{eq:bench-lp}\\
  \text{subject to}\quad
  & \sum_{j\in\mathcal{B}(i)}\psi_j^n(t) \le X_i^n(t),\qquad \, i\in\calI,
    \notag\\
  & \sum_{j\in\mathcal{S}(k)}\psi_j^n(t) \le N_k^n(t),\qquad k\in\calK,
    \notag\\
  & \psi_j^n(t) \ge 0, \qquad \qquad \qquad \,\,\,\, j\in\calJ .
    \notag
\end{align}
We consider seven benchmark policies. The first six benchmark policies
use the weights of the reference policies introduced in
Section~\ref{sec:reference-policies}.
\begin{itemize}
\item The $c\mu$ rule, following \citet{cox1961queues}, uses
  $w_j = c_{b(j)}\mu_j$.
\item The $c\mu/\theta$ rule, following \citet{atar2010cmu}, uses
  $w_j = c_{b(j)}\mu_j/\theta_{b(j)}$.
\item The fastest-server-first (FSF) rule, following
  \citet{armony2005dynamic}, uses $w_j=\mu_j$.
\item The myopic rule uses $w_j = c_{b(j)}+\bv_j\mu_j$, setting $p = 0$
  in~\eqref{eq:prelimit-policy}.
\item The abandonment-adjusted myopic rule uses
  $w_j=(c_{b(j)}+\bv_j\mu_j)/\theta_{b(j)}$ and adjusts the myopic
  weights by abandonment rates in the same manner as the $c\mu/\theta$
  rule.
\item The value-rate rule uses $w_j=\bv_j\mu_j$.
\item The nominal-plan rule uses the optimal dual variables of the
  planning problem in~\eqref{eq:spp}--\eqref{eq:spp-const3} in place
  of the gradient of the value function in~\eqref{eq:prelimit-policy}.
\end{itemize}
For the nominal-plan rule, let $y(\tau_m)\in\R^{I}$ denote the vector
of optimal dual variables associated with the flow balance
constraints~\eqref{eq:flow_balance} of the planning problem solved at
grid point $\tau_m$. Here $y_i(\tau_m)$ measures the increase in the
optimal value rate per unit increase in the limiting arrival rate of
class $i$. Since the objective of the Brownian control problem is to
minimize cost, we set $p_i(\tau_m) = -y_i(\tau_m)$. For
$t\in[\tau_m,\tau_{m+1})$, the rule uses the weights
\begin{equation}\label{eq:price-weight}
  w_j(t) = c_{b(j)} + \bv_j\mu_j
    - (\mu_j-\theta_{b(j)})\,y_{b(j)}(t)
    + \mu_j\sum_{i\in\calI}P_{ji}\,y_i(t),
  \qquad j\in\calJ .
\end{equation}
\vspace{-10mm}
\section{Computational results}\label{sec:results}

This section compares our proposed policy, derived using the
computational method in Section~\ref{sec:computational}, against the
benchmark policies introduced in Section~\ref{sect:benchmarks}. We
consider three test problems\footnote{The code, the calibrated inputs of the three test problems, and the call
center data are available at
\url{https://github.com/ekasikaralar/transfer_network}.} with the same first-order primitives, but
different prelimit arrival rates, representing underloaded, critically
loaded, and overloaded traffic regimes. The proposed policy outperforms
all benchmarks in each test problem. We evaluate the policies using
$10{,}000$ simulation replications with common random seeds across
policies. All reported performance figures are subject to simulation
and discretization errors.

Tables~\ref{tab:results-main}, \ref{tab:results-first},
and~\ref{tab:results-second} report the results for the main test
problem and its two variants, respectively. Each table reports the
expected net value, the value generated by service completions,
holding and abandonment costs, the number of transfers, and agent
utilization. Each metric is averaged over the simulation replications
and reported using $99\%$ confidence intervals.

In this section, we define agent utilization as the fraction of total
staffed agent time spent serving calls. It is calculated by dividing
the cumulative number of agents in service by the cumulative number of
agents on duty over the time horizon. Since staffing levels are the
same across all policies, the differences in utilization under
different policies reflect different service assignments. For the same
number of arrivals, assigning callers to activities with lower service
rates keeps agents occupied longer, while choosing faster activities
can create more idle time. The last row of each table reports the
percentage gap between the best benchmark and the proposed policy.

\begin{table}[ht]
\centering
\setlength{\tabcolsep}{8pt}
\resizebox{0.94\textwidth}{!}{%
\begin{tabular}{lcccccc}
\toprule
Policy & Net value & Value of service & Holding & Abandonment & Transfers & Utilization\\
 & & completions & cost & cost & &\\
\midrule
Our policy & 135{,}279.7 $\pm$ 15.3 & 136{,}233.2 $\pm$ 15.9 & 419.9 $\pm$ 3.4 & 165.8 $\pm$ 1.3 & 1{,}041.4 $\pm$ 0.8 & 0.944\\
Nominal-plan rule & 135{,}141.6 $\pm$ 15.9 & 136{,}051.0 $\pm$ 16.5 & 396.2 $\pm$ 3.5 & 157.1 $\pm$ 1.3 & 1{,}010.3 $\pm$ 0.8 & 0.932\\
Ab.-adj.\ myopic & 131{,}215.7 $\pm$ 17.1 & 131{,}625.0 $\pm$ 17.2 & 85.1 $\pm$ 0.9 & 42.4 $\pm$ 0.5 & 807.8 $\pm$ 0.8 & 0.837\\
Myopic & 131{,}204.2 $\pm$ 16.9 & 131{,}611.4 $\pm$ 17.0 & 85.1 $\pm$ 0.9 & 40.8 $\pm$ 0.4 & 806.8 $\pm$ 0.7 & 0.837\\
Value-rate & 131{,}172.3 $\pm$ 16.8 & 131{,}597.9 $\pm$ 17.0 & 99.3 $\pm$ 0.9 & 45.2 $\pm$ 0.4 & 806.5 $\pm$ 0.7 & 0.837\\
FSF & 119{,}652.6 $\pm$ 19.1 & 120{,}132.5 $\pm$ 19.4 & 112.0 $\pm$ 0.9 & 50.6 $\pm$ 0.5 & 889.5 $\pm$ 0.9 & 0.827\\
$c\mu/\theta$ & 119{,}635.2 $\pm$ 19.1 & 120{,}112.0 $\pm$ 19.3 & 111.6 $\pm$ 0.9 & 51.1 $\pm$ 0.5 & 880.1 $\pm$ 0.8 & 0.827\\
$c\mu$ & 119{,}168.9 $\pm$ 19.3 & 119{,}696.1 $\pm$ 19.5 & 118.6 $\pm$ 1.0 & 52.5 $\pm$ 0.5 & 1{,}026.2 $\pm$ 0.9 & 0.832\\
\midrule
Performance gap & $-0.10\%$ $\pm$ 0.02\% & & & & &\\
\bottomrule
\end{tabular}}
\caption{Simulation results for the main test problem, based on
$10{,}000$ replications per policy. Each value is the average over the
replications of the total over the eight-hour day, reported with the
half-width of its $99\%$ confidence interval. The first four numerical
columns are measured in dollars and the fifth in callers. Utilization confidence interval half-widths are below
$0.0002$ for all policies. The last row reports the percentage gap in
expected net value between the best benchmark (the nominal-plan rule)
and the proposed policy.}
\label{tab:results-main}
\end{table}

\begin{table}[ht]
\centering
\setlength{\tabcolsep}{8pt}
\resizebox{0.94\textwidth}{!}{%
\begin{tabular}{lcccccc}
\toprule
Policy & Net value & Value of service & Holding & Abandonment & Transfers & Utilization\\
 & & completions & cost & cost & &\\
\midrule
Our policy & 139{,}679.6 $\pm$ 15.4 & 141{,}246.1 $\pm$ 15.3 & 856.9 $\pm$ 5.5 & 340.2 $\pm$ 2.2 & 1{,}042.8 $\pm$ 0.8 & 0.963\\
Nominal-plan rule & 139{,}013.2 $\pm$ 18.3 & 141{,}479.4 $\pm$ 15.6 & 1{,}525.0 $\pm$ 9.4 & 567.6 $\pm$ 3.3 & 1{,}053.1 $\pm$ 0.9 & 0.969\\
Myopic & 137{,}152.6 $\pm$ 16.8 & 137{,}681.0 $\pm$ 17.0 & 157.0 $\pm$ 1.4 & 72.8 $\pm$ 0.7 & 851.8 $\pm$ 0.8 & 0.879\\
Ab.-adj.\ myopic & 137{,}142.0 $\pm$ 16.8 & 137{,}677.2 $\pm$ 17.1 & 157.4 $\pm$ 1.4 & 78.3 $\pm$ 0.7 & 853.1 $\pm$ 0.8 & 0.879\\
Value-rate & 137{,}120.5 $\pm$ 16.7 & 137{,}681.7 $\pm$ 16.9 & 182.0 $\pm$ 1.5 & 81.1 $\pm$ 0.7 & 850.2 $\pm$ 0.8 & 0.878\\
FSF & 126{,}382.5 $\pm$ 19.1 & 127{,}018.3 $\pm$ 19.6 & 196.7 $\pm$ 1.4 & 88.9 $\pm$ 0.7 & 978.8 $\pm$ 0.9 & 0.870\\
$c\mu/\theta$ & 126{,}302.1 $\pm$ 19.0 & 126{,}929.7 $\pm$ 19.4 & 193.1 $\pm$ 1.4 & 90.1 $\pm$ 0.7 & 963.5 $\pm$ 0.9 & 0.869\\
$c\mu$ & 125{,}916.6 $\pm$ 19.0 & 126{,}593.4 $\pm$ 19.5 & 210.1 $\pm$ 1.5 & 91.6 $\pm$ 0.7 & 1{,}075.0 $\pm$ 0.9 & 0.873\\
\midrule
Performance gap & $-0.48\%$ $\pm$ 0.02\% & & & & &\\
\bottomrule
\end{tabular}}
\caption{Simulation results for the first variant, based on $10{,}000$
replications per policy. The columns are as in
Table~\ref{tab:results-main}. The last row reports the percentage gap
in expected net value between the best benchmark (the nominal-plan
rule) and the proposed policy.}
\label{tab:results-first}
\end{table}

\begin{table}[ht]
\centering
\setlength{\tabcolsep}{8pt}
\resizebox{0.94\textwidth}{!}{%
\begin{tabular}{lcccccc}
\toprule
Policy & Net value & Value of service & Holding & Abandonment & Transfers & Utilization\\
 & & completions & cost & cost & &\\
\midrule
Our policy & 144{,}177.0 $\pm$ 15.1 & 145{,}299.4 $\pm$ 15.2 & 541.8 $\pm$ 3.2 & 234.6 $\pm$ 1.4 & 975.5 $\pm$ 0.8 & 0.958\\
Myopic & 142{,}663.9 $\pm$ 16.4 & 143{,}524.5 $\pm$ 16.7 & 377.9 $\pm$ 2.9 & 166.0 $\pm$ 1.2 & 897.7 $\pm$ 0.8 & 0.919\\
Value-rate & 142{,}633.5 $\pm$ 16.3 & 143{,}538.6 $\pm$ 16.7 & 411.2 $\pm$ 2.8 & 178.7 $\pm$ 1.3 & 894.0 $\pm$ 0.8 & 0.919\\
Ab.-adj.\ myopic & 142{,}605.6 $\pm$ 16.5 & 143{,}493.6 $\pm$ 16.9 & 380.1 $\pm$ 2.9 & 189.5 $\pm$ 1.4 & 900.2 $\pm$ 0.8 & 0.920\\
Nominal-plan rule & 138{,}878.5 $\pm$ 32.1 & 145{,}570.1 $\pm$ 15.1 & 4{,}639.0 $\pm$ 19.4 & 1{,}665.2 $\pm$ 6.7 & 1{,}082.5 $\pm$ 0.9 & 0.991\\
FSF & 132{,}669.6 $\pm$ 18.1 & 133{,}670.8 $\pm$ 18.9 & 422.1 $\pm$ 2.6 & 191.0 $\pm$ 1.2 & 1{,}080.1 $\pm$ 1.0 & 0.912\\
$c\mu/\theta$ & 132{,}466.5 $\pm$ 17.8 & 133{,}460.6 $\pm$ 18.6 & 414.2 $\pm$ 2.5 & 200.7 $\pm$ 1.3 & 1{,}059.3 $\pm$ 1.0 & 0.911\\
$c\mu$ & 132{,}203.2 $\pm$ 17.9 & 133{,}271.3 $\pm$ 19.0 & 468.9 $\pm$ 2.9 & 199.9 $\pm$ 1.2 & 1{,}135.9 $\pm$ 0.9 & 0.913\\
\midrule
Performance gap & $-1.06\%$ $\pm$ 0.02\% & & & & &\\
\bottomrule
\end{tabular}}
\caption{Simulation results for the second variant, based on $10{,}000$
replications per policy. The columns are as in
Table~\ref{tab:results-main}. The last row reports the percentage gap
in expected net value between the best benchmark (the myopic rule) and
the proposed policy.}
\label{tab:results-second}
\end{table}

Tables~\ref{tab:results-main}--\ref{tab:results-second} show that the
proposed policy achieves the highest expected net value in all three
test problems. As summarized in Table~\ref{tab:results-summary}, the
proposed policy outperforms the best benchmark by $0.10\%$, $0.48\%$,
and $1.06\%$ for the underloaded, critically loaded, and overloaded
regimes, respectively. As expected, congestion eases as system utilization decreases, leaving less room for our proposed policy to outperform the benchmarks. Our results are consistent with this intuition. The nominal-plan rule, which uses the optimal dual variables of the
planning problem in~\eqref{eq:spp}--\eqref{eq:spp-const3}, is the best
benchmark in the underloaded
and critically loaded regimes, while the myopic rule is the best
benchmark for the overloaded regime.

\begin{table}[ht]
\centering
{\small
\begin{tabular}{lccc}
\toprule
Method & Main test problem & First variant & Second variant\\
\midrule
Our policy & 135{,}279.7 $\pm$ 15.3 & 139{,}679.6 $\pm$ 15.4 & 144{,}177.0 $\pm$ 15.1\\
Best benchmark & 135{,}141.6 $\pm$ 15.9 & 139{,}013.2 $\pm$ 18.3 & 142{,}663.9 $\pm$ 16.4\\
\midrule
Performance gap & $-0.10\%$ $\pm$ 0.02\% & $-0.48\%$ $\pm$ 0.02\% & $-1.06\%$ $\pm$ 0.02\%\\
\bottomrule
\end{tabular}
}
\caption{The expected net value of our policy and of the best benchmark
policy in the three test problems, with the half-length of the $99\%$
confidence interval, and the percentage gap between the best benchmark
and our policy, relative to the best benchmark.}
\label{tab:results-summary}
\end{table}

For the main test problem, the system is underloaded for most of the
time horizon and the congestion costs (holding and abandonment costs)
are small. Since the nominal-plan rule maximizes value generation, it
outperforms all other benchmark policies. Both our policy and the nominal-plan rule generate substantially higher utilization and service value than the
other benchmarks.

For the critically loaded (first variant) test problem, policies that
prioritize only high-value classes create more congestion in the
system, making holding and abandonment costs more significant. The
nominal-plan rule is a static rule that depends on the value
maximizing nominal plan and does not depend on the current congestion
levels in the system. The proposed policy is a state-dependent policy
that adjusts the service assignments based on the congestion levels.
The proposed policy obtains almost the same service value as the
nominal-plan rule, yet it reduces the combined holding and abandonment
costs from \$2{,}093 to \$1{,}197 (a $43\%$ reduction).

This distinction becomes more evident in the overloaded (second
variant) test problem. The nominal-plan rule generates the highest
service value, but its combined holding and abandonment costs increase
to \$6{,}304 and it is no longer the best benchmark policy. The
proposed policy obtains almost the same service value, yet its
congestion cost is only \$776. It also generates more service value
than the myopic rule, which is the best benchmark in this regime. These
results show that the proposed policy is able to find service
assignments based on both the value generated from service completions
and the congestion costs. The computational framework produces
policies that outperform the best benchmark in each traffic regime.

The small percentage gaps in expected net value partly reflect the
large first-order contribution of the nominal plan, which is of order
$n$. Our policy is derived from the Brownian control problem, whose
objective captures performance on the second-order $\sqrt n$ scale.
Appendix~\ref{app:scaled-objective} reports the centered and scaled
costs and compares the policies on this second-order scale, where the
improvement of our policy over the best benchmark is substantially
larger.
\color{black}
\begingroup
\setlength{\bibsep}{0pt}
\bibliographystyle{apalike}
\bibliography{references}
\endgroup
\newpage
\appendix
\section{A Formal Derivation of the Brownian Control Problem}
\label{app:strong}

Note by the strong approximations \citep{horvath1993weighted} that for a
rate-one Poisson process, denoted by $N(t)$, we have as
$n \rightarrow \infty$
\begin{equation}
    N(nt) = nt + \sqrt{n}\,B(t) + o(\sqrt{n}),
    \label{eq:strong-approx}
\end{equation}
where $B(\cdot)$ is a standard Brownian motion. We will apply
\eqref{eq:strong-approx} to derive the approximating Brownian
control problem formally. Throughout, 
$T_j^n(t)=\int_0^t\psi_j^n(s)\,ds$.

We consider the arrival, abandonment, and service completion processes of the $n$th system one at a time. Using \eqref{eq:strong-approx} and \eqref{eq:scaling}, we can write
\begin{align}
    E_{i}^{n}(t)
    & = N_{i}^{a}\Big(\int_0^t \lambda_{i}^{n}(s)\,ds\Big) \nonumber\\
    & = N_{i}^{a}\Big(n\!\int_0^t\!\lambda_{i}(s)\,ds
        + \sqrt{n}\!\int_0^t\!\zeta_{i}(s)\,ds + o(\sqrt n)\Big) \nonumber\\
    & \approx n\!\int_0^t\!\lambda_{i}(s)\,ds
        + \sqrt{n}\!\int_0^t\!\zeta_{i}(s)\,ds
        + \sqrt{n}\,B_{i}^{a}\Big(\int_{0}^{t}\lambda_{i}(s)ds\Big)
        + o(\sqrt{n}),
    \qquad i\in\calI,
    \label{eq:sa-arrival}
\end{align}
where $B_{i}^{a}$ is a standard Brownian motion. Similarly, for the
abandonment process, from \eqref{eq:strong-approx},
\eqref{eq:abandonment}, and $Y_i^n=\sqrt n\,\hat Y_i^n$, we can write
\begin{align}
    A_{i}^{n}(t)
    &= N_{i}^{b}\Big(\theta_{i}\!\int_{0}^{t}\!Y_{i}^{n}(s)\,ds\Big)
    \nonumber\\
    &= N_{i}^{b}\Big(\sqrt{n}\,\theta_{i}\!\int_{0}^{t}\!
        \hat{Y}_{i}^{n}(s)\,ds\Big) \nonumber\\
    &\approx \sqrt{n}\,\theta_{i}\!\int_{0}^{t}\!\hat{Y}_{i}^{n}(s)\,ds
        + o(\sqrt{n}),
    \qquad i\in\calI.
    \label{eq:sa-abandon}
\end{align}
Lastly, we consider the service completions and their routing jointly. We
abbreviate
$\Phi_{j,l}^n(t):=\Phi_{j,l}\big(S_j(T_j^n(t))\big)$ for the number of
completions of activity~$j$ routed to class~$l$ by time~$t$, and write
$\Phi_{j,\mathrm{out}}^n(t)$ for the number that exit.  By Poisson
thinning, the completions of activity~$j$ split into mutually independent
Poisson substreams, one per destination class and one for exits, with
unit-rate Poisson processes
$\{N_{jl}^\phi, N_{j,\mathrm{out}}^\phi\}$ mutually independent,
\begin{equation}\label{eq:split}
  \Phi_{j,l}^n(t) = N_{jl}^\phi\big(P_{jl}\,\mu_j T_j^n(t)\big),\quad l\in\calI,
  \qquad
  \Phi_{j,\mathrm{out}}^n(t) = N_{j,\mathrm{out}}^\phi\big(q_j\,\mu_j T_j^n(t)\big),
\end{equation}
which together make up $S_j(T_j^n(t))$. In applying
\eqref{eq:strong-approx} to the substreams, we first consider nonbasic
activities, i.e., $j\in\calN(t)$. Using \eqref{eq:control-prelimit}, it
follows that for $l\in\calI_r$,
\begin{align}
    \Phi_{j,l}^{n}(t)
    &= N_{jl}^{\phi}\big(P_{jl}\,\mu_{j}T_j^n(t)\big) \nonumber\\
    &= N_{jl}^{\phi}\Big(-\sqrt{n}\,P_{jl}\,\mu_{j}\!\int_{0}^{t}\!
        \hat{\psi}_{j}^{n}(s)\,ds\Big) \nonumber\\
    &\approx -\sqrt{n}\,P_{jl}\,\mu_{j}\!\int_{0}^{t}\!
        \hat{\psi}_{j}^{n}(s)\,ds + o(\sqrt{n}).
    \label{eq:sa-phi-nonbasic}
\end{align}
Now we consider basic activities, i.e., $j\in\calB(t)$. Using
\eqref{eq:strong-approx} and \eqref{eq:control-prelimit}, we have
\begin{align}
    \Phi_{j,l}^{n}(t)
    &= N_{jl}^{\phi}\big(P_{jl}\,\mu_{j}T_j^n(t)\big) \nonumber\\
    &= N_{jl}^{\phi}\Big(P_{jl}\,\mu_{j}
        \Big(n\!\int_0^t\!\psi_{j}^{*}(s)\,ds
        - \sqrt{n}\!\int_0^t\!\hat{\psi}_{j}^{n}(s)\,ds\Big)
        + o(\sqrt n)\Big) \nonumber\\
    &\approx n\,P_{jl}\,\mu_{j}\!\int_0^t\!\psi_{j}^{*}(s)\,ds
        - \sqrt{n}\,P_{jl}\,\mu_{j}\!\int_{0}^{t}\!
          \hat{\psi}_{j}^{n}(s)\,ds
        + \sqrt{n}\,B_{jl}^{\phi}
          \Big(P_{jl}\,\mu_{j}\!\int_0^t\!\psi_{j}^{*}(s)\,ds\Big)
        + o(\sqrt{n}),
    \label{eq:sa-phi-basic}
\end{align}
where $B_{jl}^{\phi}$ is a standard Brownian motion. Because
$\psi_{j}^{*}(t)=0$ for $j\in\calN(t)$, the two cases combine. For all
$j\in\calJ$ and $l\in\calI_r$,
\begin{equation}
  \Phi_{j,l}^n(t)
  \approx n\,P_{jl}\,\mu_{j}\!\int_0^t\!\psi_{j}^{*}(s)\,ds
  - \sqrt{n}\,P_{jl}\,\mu_{j}\!\int_{0}^{t}\!\hat{\psi}_{j}^{n}(s)\,ds
  + \sqrt{n}\,B_{jl}^{\phi}
    \Big(P_{jl}\,\mu_{j}\!\int_0^t\!\psi_{j}^{*}(s)\,ds\Big)
  + o(\sqrt{n}),
  \label{eq:sa-phi}
\end{equation}
and analogously
\begin{equation}
  \Phi_{j,\mathrm{out}}^n(t)
  \approx n\,q_{j}\,\mu_{j}\!\int_0^t\!\psi_{j}^{*}(s)\,ds
  - \sqrt{n}\,q_{j}\,\mu_{j}\!\int_{0}^{t}\!\hat{\psi}_{j}^{n}(s)\,ds
  + \sqrt{n}\,B_{j,\mathrm{out}}^{\phi}
    \Big(q_{j}\,\mu_{j}\!\int_0^t\!\psi_{j}^{*}(s)\,ds\Big)
  + o(\sqrt{n}),
  \label{eq:sa-phi-out}
\end{equation}
with all Brownian motions mutually independent and independent of
$\{B_i^a\}$.

By the system dynamics~\eqref{eq:dynamics}, the state of the $n$th system
evolves as
\begin{equation}\label{eq:sa-state-n}
  X_i^n(t)
  = X_i^n(0) + E_i^n(t)
  + \sum_{j\in\calJ}\Phi_{j,i}^n(t)
  - \sum_{j\in\mathcal{B}(i)}
    \Big[\Phi_{j,\mathrm{out}}^n(t)
    + \sum_{l\in\calI_r}\Phi_{j,l}^n(t)\Big]
  - A_i^n(t),
  \qquad i\in\calI,
\end{equation}
where we write the completions of each activity $j\in\mathcal B(i)$
through its thinned substreams,
$S_j(T_j^n(t))
=\Phi_{j,\mathrm{out}}^n(t)+\sum_{l\in\calI_r}\Phi_{j,l}^n(t)$.
Recall that for $t\geq0$,
\begin{equation}
  X_i^n(t) = \sqrt{n}\,\hat X_i^n(t) + n\,x_i^*(t).
  \label{eq:sa-scaled-relation}
\end{equation}
Therefore, using \eqref{eq:sa-arrival}, \eqref{eq:sa-abandon},
\eqref{eq:sa-phi}, \eqref{eq:sa-phi-out}, and
\eqref{eq:sa-scaled-relation}, together with
$q_j+\sum_{l\in\calI_r}P_{jl}=1$ for $j\in\mathcal B(i)$, we can write
Equation~\eqref{eq:sa-state-n} as
\begin{align}
  \hat X_i^n(t) = \hat X_i^n(0)
  &+ \sqrt n\!\int_0^t\!\Big[
      \lambda_i(s)
      + \sum_{j\in\calJ}P_{ji}\,\mu_j\psi_j^*(s)
      - \sum_{j\in\mathcal{B}(i)}\mu_j\psi_j^*(s)
    \Big]ds
  - \sqrt n\,\big(x_i^*(t)-x_i^*(0)\big)
  \notag\\
  &+ \int_0^t\!\Big[
      \zeta_i(s)
      + \sum_{j\in\mathcal{B}(i)}\mu_j\hat\psi_j^n(s)
      - \sum_{j\in\calJ}P_{ji}\,\mu_j\hat\psi_j^n(s)
      - \theta_i\hat Y_i^n(s)
    \Big]ds
  + B_i^a\Big(\int_{0}^{t}\lambda_i(s)ds\Big)
  \notag\\
  &+ \sum_{j\in\calJ}
      B_{ji}^\phi\Big(P_{ji}\,\mu_j\!\int_0^t\!\psi_j^*(s)\,ds\Big)
  - \sum_{j\in\mathcal{B}(i)}\Big[
      B_{j,\mathrm{out}}^\phi
        \Big(q_j\,\mu_j\!\int_0^t\!\psi_j^*(s)\,ds\Big)
  \notag\\
  &+ \sum_{l\in\calI_r}
      B_{jl}^\phi\Big(P_{jl}\,\mu_j\!\int_0^t\!\psi_j^*(s)\,ds\Big)
    \Big]
  + o(1).
  \label{eq:sa-scaled-state-pre}
\end{align}
Recall the flow-balance condition~\eqref{eq:flow-balance-psi}:
\begin{equation}
  \sum_{j\in\mathcal{B}(i)}\mu_j\psi_j^*(t)
  - \sum_{j\in\calJ}P_{ji}\,\mu_j\psi_j^*(t)
  = \lambda_i(t),
  \qquad i\in\calI.
  \label{eq:sa-flow-balance}
\end{equation}
Substituting \eqref{eq:sa-flow-balance} into
\eqref{eq:sa-scaled-state-pre}, the $O(\sqrt n)$ integral vanishes.
Suppressing the centering term $\sqrt n\,(x_i^*(t)-x_i^*(0))$ arising
from the time variation of the nominal plan, consistent with our formal
treatment, we obtain
\begin{align}
  \hat X_i^n(t) = \hat X_i^n(0)
  &+ \int_0^t\!\Big[
      \zeta_i(s)
      + \sum_{j\in\mathcal{B}(i)}\mu_j\hat\psi_j^n(s)
      - \sum_{j\in\calJ}P_{ji}\,\mu_j\hat\psi_j^n(s)
      - \theta_i\hat Y_i^n(s)
    \Big]ds
  + B_i^a\Big(\int_{0}^{t}\lambda_i(s)ds\Big)
  \notag\\
  &+ \sum_{j\in\calJ}
      B_{ji}^\phi\Big(P_{ji}\,\mu_j\!\int_0^t\!\psi_j^*(s)\,ds\Big)
  - \sum_{j\in\mathcal{B}(i)}\Big[
      B_{j,\mathrm{out}}^\phi
        \Big(q_j\,\mu_j\!\int_0^t\!\psi_j^*(s)\,ds\Big)
  \notag\\
  &+ \sum_{l\in\calI_r}
      B_{jl}^\phi\Big(P_{jl}\,\mu_j\!\int_0^t\!\psi_j^*(s)\,ds\Big)
    \Big]
  + o(1).
  \label{eq:sa-scaled-state}
\end{align}
Given that
\begin{align}
  B_i^a\Big(\int_{0}^{t}\lambda_i(s)ds\Big)
  &\sim N\Big(0,\int_{0}^{t}\lambda_i(s)ds\Big), \nonumber\\
  B_{jl}^\phi\Big(P_{jl}\,\mu_j\!\int_0^t\!\psi_j^*(s)\,ds\Big)
  &\sim N\Big(0,P_{jl}\,\mu_j\!\int_0^t\!\psi_j^*(s)\,ds\Big),
  \nonumber
\end{align}
that the Brownian motions are mutually independent, and that
$P_{j,b(j)}=0$, the sum of the Brownian terms in
\eqref{eq:sa-scaled-state} is a driftless Brownian motion whose variance
rate is
\begin{equation*}
  \lambda_i(t)
  + \sum_{j\in\calJ}P_{ji}\,\mu_j\psi_j^*(t)
  + \sum_{j\in\mathcal{B}(i)}\mu_j\psi_j^*(t)
  = \gamma_{ii}(t).
\end{equation*}
For $i\neq i'$, the only Brownian motions shared by the class-$i$ and
class-$i'$ components are those of the substreams routed from one class
to the other, each entering the two components with opposite signs, so
that the covariance rate is
\begin{equation*}
  -\Big(\sum_{j\in\mathcal{B}(i')}P_{ji}\,\mu_j\psi_j^*(t)
  + \sum_{j\in\mathcal{B}(i)}P_{ji'}\,\mu_j\psi_j^*(t)\Big)
  = \gamma_{ii'}(t).
\end{equation*}
Thus the Brownian terms have covariance rate $\Gamma(t)$ as
in~\eqref{eq:Sigma-compact}, and they are equal in distribution to
$\int_0^\cdot\sigma(s)\,dB(s)$, for an $I$-dimensional
standard Brownian motion $B$. Denoting the limit of the processes
$(\hat X^n,\hat\psi^n,\hat Y^n,\hat Z^n)$ by
$(\hat X,\hat\psi,\hat Y,\hat Z)$ as $n\to\infty$,
Equation~\eqref{eq:sa-scaled-state} can be written as the limiting state
process~\eqref{eq:sde} with drift function~\eqref{eq:drift}.
\vspace{-3mm}
\section{Derivation of the Limiting Cost}\label{app:cost}

We derive the limit of the scaled cost
process~\eqref{eq:scaled-reward-process}, and thereby the cost
objective~\eqref{eqn:cost_definition}, by evaluating the
three terms of the net value~\eqref{eq:cumulative-value-n} one at a time.

For the value term, recall from~\eqref{eq:split} that the exit count and
the routed counts of activity~$j$ are the thinned substreams
$\Phi_{j,\mathrm{out}}^n(t)$ and $\Phi_{j,l}^n(t)$, so the realized value
\eqref{eq:realized-reward-activity} in the $n$th system is
\vspace{-3mm}
\begin{equation}\label{eq:value-term-thinned}
  \mathcal R_j^n(t)
  = v_{j}\,\Phi_{j,\mathrm{out}}^n(t)
  - \sum_{l\in\calI_r} \kappa_{jl}\,\Phi_{j,l}^n(t).
\end{equation}
Substituting the expansions~\eqref{eq:sa-phi} and \eqref{eq:sa-phi-out}
into \eqref{eq:value-term-thinned} and using
$q_jv_{j}-\sum_{l\in\calI_r}P_{jl}\kappa_{jl}=\bv_j$ from~\eqref{eq:vbar},
we have
\vspace{-3mm}
\begin{align}
  \mathcal R_j^n(t)
  \approx{}& n\,\bv_j\mu_j\!\int_0^t\!\psi_j^*(s)\,ds
  - \sqrt n\,\bv_j\mu_j\!\int_0^t\!\hat\psi_j^n(s)\,ds
  + \sqrt n\,v_{j}\,B_{j,\mathrm{out}}^\phi
      \Big(q_j\,\mu_j\!\int_0^t\!\psi_j^*(s)\,ds\Big)
  \notag\\
  &- \sqrt n\sum_{l\in\calI_r} \kappa_{jl}\,
      B_{jl}^\phi\Big(P_{jl}\,\mu_j\!\int_0^t\!\psi_j^*(s)\,ds\Big)
  + o(\sqrt n),
  \qquad j\in\calJ.
  \label{eq:value-term}
\end{align}
Since $\psi_j^*(s)=\nu_{s(j)}(s)\,\xi_j^*(s)$, we have
$\sum_{j\in\calJ}\bv_j\mu_j\psi_j^*(s)=\tilde v(s)\cdot\xi^*(s)$, so the
leading $O(n)$ term in \eqref{eq:value-term}, summed over $j\in\calJ$, is
exactly the centering in~\eqref{eq:scaled-reward-process}. Therefore,
\begin{align}
  &\frac{1}{\sqrt n}\left(
    n\!\int_0^t\!\sum_{j\in\calJ}\bv_j\mu_j\psi_j^*(s)\,ds
    - \sum_{j\in\calJ}\mathcal R_j^n(t)\right)
  \notag\\
  &\qquad=
  \int_0^t\!\sum_{j\in\calJ}\bv_j\mu_j\,\hat\psi_j^n(s)\,ds
  - \sum_{j\in\calJ}v_{j}\,B_{j,\mathrm{out}}^\phi
      \Big(q_j\,\mu_j\!\int_0^t\!\psi_j^*(s)\,ds\Big)
  \notag\\
  &\qquad\phantom{=}
  + \sum_{j\in\calJ}\sum_{l\in\calI_r}\kappa_{jl}\,
      B_{jl}^\phi\Big(P_{jl}\,\mu_j\!\int_0^t\!\psi_j^*(s)\,ds\Big)
  + o(1).
  \label{eq:od-reward-expansion}
\end{align}
The Brownian terms above are mutually independent, time-changed standard
Brownian motions, and their sum has quadratic variation
\begin{equation}\label{eq:outcome-reward-qv}
  \sum_{j\in\calJ}
  \Big(q_j\,v_{j}^2
  + \sum_{l\in\calI_r}P_{jl}\,\kappa_{jl}^2\Big)
  \int_0^t\mu_j\psi_j^*(s)\,ds.
\end{equation}
The holding cost term, with $Y_i^n=\sqrt n\,\hat Y_i^n$, is
\begin{equation}\label{eq:holding-term}
  \int_0^t h\cdot Y^n(s)\,ds = \sqrt n\int_0^t h\cdot\hat Y^n(s)\,ds ,
\end{equation}
and the abandonment term is
\begin{equation}\label{eq:abandon-term}
  a\cdot A^n(t) \approx \sqrt n\sum_{i\in\calI} a_i\theta_i\!\int_0^t\!\hat Y_i^n(s)\,ds
   + o(\sqrt n).
\end{equation}
Adding the holding and abandonment
terms~\eqref{eq:holding-term}--\eqref{eq:abandon-term}
to~\eqref{eq:od-reward-expansion},
by~\eqref{eq:scaled-reward-process}, and letting $n\to\infty$, with
$c_i=h_i+\theta_i a_i$, gives the limiting cost process
\begin{align}
  \hat\Pi(t)
  ={}& \int_0^t\!\Big[\sum_{i\in\calI} c_i\,\hat Y_i(s)
    + \sum_{j\in\calJ} \bv_j\mu_j\,\hat\psi_j(s)\Big]\,ds
  \notag\\
  &- \sum_{j\in\calJ}v_{j}\,B_{j,\mathrm{out}}^\phi
      \Big(q_j\,\mu_j\!\int_0^t\!\psi_j^*(s)\,ds\Big)
  + \sum_{j\in\calJ}\sum_{l\in\calI_r}\kappa_{jl}\,
      B_{jl}^\phi\Big(P_{jl}\,\mu_j\!\int_0^t\!\psi_j^*(s)\,ds\Big).
  \label{eq:scaled-reward-limit}
\end{align}
The Brownian terms in \eqref{eq:scaled-reward-limit} are sums of Brownian
increments under the deterministic time changes determined by the nominal
plan $\psi^*$. Each increment has mean zero, so these terms disappear after taking expectations.

\section{Derivation of the Deep Splitting Algorithm}
\label{app:deep-splitting}

In what follows, we provide a purely formal derivation that conveys the
intuition behind the algorithm.  We suppress certain technical details
and refer the reader to \citet{beck2021deep} for those.

The derivation starts from the equivalent characterization of the HJB
equation obtained in Section~\ref{sec:computational}.  For
$(t,x)\in[0,T]\times\R^{I}$,
\begin{equation}\label{eq:app-ds-hjb}
  \frac{\partial V}{\partial t}(t,x)
  + \mathcal{L}_t V(t,x)
  + b\big(t,x,\tilde\psi(t,x)\big)\cdot\nabla_x V(t,x)
  - F\big(t,x,\nabla_x V(t,x)\big) = 0,
  \qquad V(T,x) = \hat g(x),
\end{equation}
where $b$, $\mathcal{L}_t$, and $F$ are given
in~\eqref{eq:drift},~\eqref{eq:L}, and~\eqref{eq:F-def}, respectively,
and $\hat g\equiv 0$ by Equation~\eqref{eq:terminal-zero}.
Fix a partition $0=t_0<t_1<\cdots<t_N=T$ and let $\Delta t_n = t_{n+1}-t_n$.
Under suitable integrability
assumptions, for $n=0,1,\ldots,N-1$, $t\in[t_n,t_{n+1}]$ and
$x\in\R^{I}$, integrating~\eqref{eq:app-ds-hjb} in time gives
\begin{equation}\label{eq:app-ds-integrated}
  V(t,x) = V(t_n,x)
  - \int_{t_n}^{t}\mathcal{L}_s V(s,x)\,ds
  - \int_{t_n}^{t} b\big(s,x,\tilde\psi(s,x)\big)\cdot\nabla_x V(s,x)\,ds
  + \int_{t_n}^{t} F\big(s,x,\nabla_x V(s,x)\big)\,ds.
\end{equation}
For a sufficiently fine partition, we replace the last integral by the
value of its integrand at the left endpoint $t_n$ times the step length
$\Delta t_n$, so that
\begin{equation}\label{eq:app-ds-frozen}
  V(t,x) \approx V(t_n,x)
  - \int_{t_n}^{t}\mathcal{L}_s V(s,x)\,ds
  - \int_{t_n}^{t} b\big(s,x,\tilde\psi(s,x)\big)\cdot\nabla_x V(s,x)\,ds
  + F\big(t_n,x,\nabla_x V(t_n,x)\big)\,\Delta t_n .
\end{equation}
Accordingly, let $\mathcal{V}:[0,T]\times\R^{I}\to\R$ be a function
with the following three properties.  First, $\mathcal{V}(T,x)=\hat g(x)$
for all $x\in\R^{I}$.  Second, for each $n\in\{0,1,\ldots,N-1\}$,
$\mathcal{V}\in C^{1,2}\big((t_n,t_{n+1}]\times\R^{I};\R\big)$ and
its partial derivatives grow at most polynomially in $x$.  Third, for each
$n\in\{0,1,\ldots,N-1\}$, $t\in(t_n,t_{n+1}]$, and $x\in\R^{I}$,
\begin{equation}\label{eq:app-ds-V}
  \mathcal{V}(t,x) = \mathcal{V}(t_n,x)
  - \int_{t_n}^{t}\mathcal{L}_s \mathcal{V}(s,x)\,ds
  - \int_{t_n}^{t} b\big(s,x,\tilde\psi(s,x)\big)\cdot\nabla_x \mathcal{V}(s,x)\,ds
  + F\big(t_n,x,\nabla_x \mathcal{V}(t_n,x)\big)\,\Delta t_n .
\end{equation}
Equation~\eqref{eq:app-ds-V} is the linear approximation of the HJB
equation on the subinterval $(t_n,t_{n+1}]$, and
comparing~\eqref{eq:app-ds-V} with~\eqref{eq:app-ds-frozen} one expects
that
\begin{equation}\label{eq:app-ds-approx}
  \mathcal{V}(t_n,x)\approx V(t_n,x)
  \qquad\text{for } n=0,1,\ldots,N .
\end{equation}
The last term of~\eqref{eq:app-ds-V} does not depend on $t$, so
differentiating~\eqref{eq:app-ds-V} in $t$ shows that, on
$(t_n,t_{n+1}]\times\R^{I}$,
\begin{equation}\label{eq:app-ds-linear}
  \frac{\partial \mathcal{V}}{\partial t}(t,x)
  + \mathcal{L}_t \mathcal{V}(t,x)
  + b\big(t,x,\tilde\psi(t,x)\big)\cdot\nabla_x \mathcal{V}(t,x) = 0 .
\end{equation}
Applying It\^o's formula to $\mathcal{V}\big(t,\tilde X(t)\big)$ and
using the dynamics~\eqref{eq:reference-process} of the reference
process,
\begin{equation}\label{eq:app-ds-ito}
  d\mathcal{V}\big(t,\tilde X(t)\big)
  = \Big[\frac{\partial \mathcal{V}}{\partial t}
    + \mathcal{L}_t \mathcal{V}
    + b\big(t,\tilde X(t),\tilde\psi(t,\tilde X(t))\big)\cdot\nabla_x \mathcal{V}\Big]
    \big(t,\tilde X(t)\big)\,dt
  + \nabla_x \mathcal{V}\big(t,\tilde X(t)\big)\cdot\sigma(t)\,dB(t).
\end{equation}
Substituting~\eqref{eq:app-ds-linear} into~\eqref{eq:app-ds-ito}
eliminates the drift term, so that
$d\mathcal{V}(t,\tilde X(t))
= \nabla_x\mathcal{V}(t,\tilde X(t))\cdot\sigma(t)\,dB(t)$ on
$(t_n,t_{n+1}]$.  Integrating over $(t,t_{n+1}]$ gives, almost surely,
\begin{equation}\label{eq:app-ds-mart}
  \mathcal{V}\big(t_{n+1},\tilde X(t_{n+1})\big)
  - \mathcal{V}\big(t,\tilde X(t)\big)
  = \int_{t}^{t_{n+1}}\nabla_x \mathcal{V}\big(s,\tilde X(s)\big)\cdot\sigma(s)\,dB(s),
  \qquad t\in(t_n,t_{n+1}].
\end{equation}
Let $(\Omega,\mathcal{F},(\mathcal{F}_t)_{t\in[0,T]},\mathbb{P})$ denote the
filtered probability space supporting the reference process and its
driving Brownian motion $B$, both of which are adapted to
$(\mathcal{F}_t)$.  Under suitable integrability assumptions, the
stochastic integral in~\eqref{eq:app-ds-mart} has zero conditional
mean given $\mathcal{F}_t$, so that
\begin{equation}\label{eq:app-ds-cond-t}
  \mathcal{V}\big(t,\tilde X(t)\big)
  = \E\Big[\mathcal{V}\big(t_{n+1},\tilde X(t_{n+1})\big)\,\Big|\,\mathcal{F}_{t}\Big]
  \quad\text{almost surely},\qquad t\in(t_n,t_{n+1}].
\end{equation}
By the tower property of conditional expectations, for
$t\in(t_n,t_{n+1}]$,
\begin{equation}\label{eq:app-ds-tower}
  \E\Big[\mathcal{V}\big(t,\tilde X(t)\big)\,\Big|\,\mathcal{F}_{t_n}\Big]
  = \E\Big[\mathcal{V}\big(t_{n+1},\tilde X(t_{n+1})\big)\,\Big|\,\mathcal{F}_{t_n}\Big]
  \quad\text{almost surely}.
\end{equation}
Next, evaluating~\eqref{eq:app-ds-V} along the reference process and
letting $t\downarrow t_n$, the two integrals vanish while the last
term is unchanged, so that
\begin{equation}\label{eq:app-ds-limit}
  \lim_{t\downarrow t_n}\mathcal{V}\big(t,\tilde X(t)\big)
  = \mathcal{V}\big(t_n,\tilde X(t_n)\big)
  + F\big(t_n,\tilde X(t_n),\nabla_x \mathcal{V}(t_n,\tilde X(t_n))\big)\,\Delta t_n .
\end{equation}
Then, under suitable integrability
assumptions, passing to the limit $t\downarrow t_n$
in~\eqref{eq:app-ds-tower} and substituting~\eqref{eq:app-ds-limit}
yields
\begin{equation}\label{eq:app-ds-cond}
  \E\Big[\mathcal{V}\big(t_{n+1},\tilde X(t_{n+1})\big)\,\Big|\,\mathcal{F}_{t_n}\Big]
  = \mathcal{V}\big(t_n,\tilde X(t_n)\big)
  + F\big(t_n,\tilde X(t_n),\nabla_x \mathcal{V}(t_n,\tilde X(t_n))\big)\,\Delta t_n .
\end{equation}
The right side of~\eqref{eq:app-ds-cond} is a function of
$\tilde X(t_n)$, so taking conditional expectations of both sides with
respect to the $\sigma$-algebra generated by $\tilde X(t_n)$ and using
the tower property once more gives
\begin{equation*}
  \E\Big[\mathcal{V}\big(t_{n+1},\tilde X(t_{n+1})\big)\,\Big|\,\tilde X(t_n)\Big]
  = \mathcal{V}\big(t_n,\tilde X(t_n)\big)
  + F\big(t_n,\tilde X(t_n),\nabla_x \mathcal{V}(t_n,\tilde X(t_n))\big)\,\Delta t_n .
\end{equation*}
Finally, replacing $\mathcal{V}(t_n,\cdot)$ by $V(t_n,\cdot)$ in view
of~\eqref{eq:app-ds-approx}, we arrive at
\begin{equation*}
  \E\Big[V\big(t_{n+1},\tilde X(t_{n+1})\big)\,\Big|\,\tilde X(t_n)\Big]
  = V\big(t_n,\tilde X(t_n)\big)
  + F\big(t_n,\tilde X(t_n),\nabla_x V(t_n,\tilde X(t_n))\big)\,\Delta t_n,
\end{equation*}
which is the relationship~\eqref{eq:fk-identity} used in
Section~\ref{sec:computational}.  The conditional expectation on the
left is the $L^{2}$ projection of $V(t_{n+1},\tilde X(t_{n+1}))$ onto
the functions of $\tilde X(t_n)$, which is why the pair
$(V(t_n,\cdot),\nabla_xV(t_n,\cdot))$ minimizes the least squares
loss~\eqref{eq:loss-recursive} over the reference process, and why the
recursion proceeds backward in time from the terminal condition
$V(t_N,\cdot) = \hat g(\cdot)$.

\section{Implementation Details of the Computational Method}
\label{app:implementation}

We implement Algorithm~\ref{alg:ds} in Python with the PyTorch package,
following the structure of the code of \citet{beck2021deep}.  The value
function $V(t_n,\cdot)$ and its gradient $\nabla_x V(t_n,\cdot)$ are
represented by separate fully connected networks $H^n(\cdot;\omega_n)$
and $G^n(\cdot;\nu_n)$ with four hidden layers of $100$ neurons and
the ELU activation function, so there are $2N$ networks in total.  The
weights are initialized with the Kaiming uniform scheme, and the
output layers have no bias.  The optimizer is Adam.  This appendix
describes a modification to the loss function of
Section~\ref{sec:computational} and lists the hyperparameters used for
each test problem.

\subsection{A technical modification}

\paragraph{A gradient consistency term in the loss.}
In the loss functions~\eqref{eq:loss-terminal}
and~\eqref{eq:loss-recursive} the gradient network $G^n$ enters only
through the auxiliary function $\hat F$, so nothing in the loss function ties $G^n$ to the gradient of $H^n$.  The gradient is what the policy uses, so we add a term that measures the gap between the two at the sampled
states.  For a batch $\{\tilde X^{(s)}(t_n)\}_{s=1}^{S}$ let
\begin{equation}\label{eq:loss-gap}
  \hat\gamma_n(\omega_n,\nu_n)
  = \frac{1}{S}\sum_{s=1}^{S}
    \Big\|G^n\big(\tilde X^{(s)}(t_n);\nu_n\big)
    - \nabla_x H^n\big(\tilde X^{(s)}(t_n);\omega_n\big)\Big\|^{2},
\end{equation}
where $\nabla_x H^n$ is computed by automatic differentiation. Therefore, the loss minimized at step $n$ is $\hat\ell_n(\omega_n,\nu_n) + \Lambda\,\hat\gamma_n(\omega_n,\nu_n),$
with $\hat\ell_n$ the empirical loss, as in Equation \eqref{eq:empirical-loss-n-1} at
$n = N-1$ and Equation \eqref{eq:empirical-loss-n-2} for $n = 0,1,\ldots,N-2$,
and we set $\Lambda = 1$.
\vspace{-4mm}
\subsection{Hyperparameters used for the test problems}

Table~\ref{tab:hyper-ds} lists the hyperparameters of the deep
splitting solver for the proposed policy of each test problem.  The
three problems share every setting except the reference policy.  Each
run takes about $25$ minutes on an H100 GPU.
\vspace{-2mm}
\begin{table}[H]
\centering
{\footnotesize
\renewcommand{\arraystretch}{0.9}
\begin{tabular}{lccc}
\toprule
Hyperparameters & Main test problem & First variant & Second variant\\
\midrule
Number of hidden layers & 4 & 4 & 4\\
Number of neurons per layer & 100 & 100 & 100\\
Activation function & ELU & ELU & ELU\\
Initialization & Kaiming uniform & Kaiming uniform & Kaiming uniform\\
Number of networks $N$ & 480 & 480 & 480\\
$\Delta t_n = T/N$ & 1 min & 1 min & 1 min\\
Batch size $S$ & 256 & 256 & 256\\
Validation sample & 4{,}096 & 4{,}096 & 4{,}096\\
\midrule
Iterations, final interval & 3{,}000 & 3{,}000 & 3{,}000\\
Iterations, rest of the intervals & up to 2{,}000 & up to 2{,}000 & up to 2{,}000\\
Early stopping & Yes (500 patience iter) & Yes (500 patience iter) & Yes (500 patience iter)\\
Learning rate, final interval & 1e-2 & 1e-2 & 1e-2\\
Learning rate, rest of the intervals & 1e-3 & 1e-3 & 1e-3\\
Learning rate milestones & 500, 1{,}500 & 500, 1{,}500 & 500, 1{,}500\\
Learning rate decay factor $\gamma$ & 0.1 & 0.1 & 0.1\\
Gradient clipping & norm 1 & norm 1 & norm 1\\
\midrule
Reference policy & myopic rule & value-rate rule & myopic rule\\
\bottomrule
\end{tabular}
}
\caption{Summary of the hyperparameters of the deep splitting solver
for the main test problem and its two variants.  The learning rate of
an interval starts at the listed value and is multiplied by $\gamma$ at
each milestone iteration.  The reference policy is used to simulate the training states and to
define the auxiliary function $\hat F$ through the drift networks of
Appendix~\ref{app:offline}.}
\label{tab:hyper-ds}
\end{table}


\section{Offline Approximation of the Auxiliary Function}
\label{app:offline}

This appendix describes the offline approximations of the mappings
$\mathcal{M}$ and $\mathcal{D}$ introduced in
Section~\ref{sec:computational}.  For each grid point $t_n$ of the five-minute
model grid, $n = 0,1,\ldots,M-1$ with $M = 96$, we write
$\mathcal{M}_n(x,p)=\mathcal{M}(t_n,x,p)$ and
$\mathcal{D}_n(x)=\mathcal{D}(t_n,x)$.

The minimization term $\mathcal{M}_n(x,p)$ of~\eqref{eq:M-def} is the
optimal value of the following linear program over the feasible set
$\bar\Psi(t_n,x)$ of Equation~\eqref{eq:Psi-bar}.  Choose
$\hat\psi\in\R^{J}$ so as to
\begin{align}
  \text{Minimize}\quad & \sum_{j\in\calJ} w_j(p)\,\hat\psi_j
    \label{eq:M-term}\\
  \text{subject to}\quad
  & \sum_{j\in\mathcal{B}(i)}\hat\psi_j \ge -x_i,
    && \qquad i\in\calI, \label{eq:M-lp-class}\\
  & \sum_{j\in\mathcal{S}(k)}\hat\psi_j \ge 0,
    && \qquad k\in\calK, \label{eq:M-lp-pool}\\
  & \hat\psi_j \le 0,
    && \qquad j\in\calN(t_n), \label{eq:M-lp-nonbasic}\\
  & \hat\psi_j \le \sqrt{n}\,\psi_j^*(t_n),
    && \qquad j\in\calB(t_n), \label{eq:M-lp-basic}
\end{align}
with the activity weights $w_j(p)$ of Equation~\eqref{eq:weight}.
Evaluating
$\mathcal{D}_n(x)$ requires the reference allocation
$\tilde\psi(t_n,x)$, which for the index rules of
Section~\ref{sec:reference-policies} is the solution of the linear
program~\eqref{eq:ref-lp} over the same feasible set, with the weights
$\tilde w_j$ of the rule in place of $w_j(p)$.

\paragraph{Sampling domains.}
The training data are generated separately for each grid point.  The states are sampled from a box that reflects the range of the prelimit system.  By Equation~\eqref{eq:scaled-state}, the number of class $i$ callers in
the $n$th system is $X^n_i(t) = n\,x^*_i(t) + \sqrt n\,\hat X^n_i(t)$.
Therefore, the scaled state $x_i$ equals $-\sqrt n\,x^*_i(t)$ when
class $i$ is empty in the prelimit system, and it equals
$(\bar H_i - n\,x^*_i(t))/\sqrt n$ when the class has $\bar H_i$
callers.  We take these two values as the endpoints of the sampling
interval, with $\bar H_i$ the largest occupancy that class $i$ is
expected to reach, and sample for $i \in \calI$ as follows:
\begin{equation}\label{eq:state-box}
  x_i^{(m)}\sim\mathrm{Uniform}\big([\underline x_i(t_n),\bar x_i(t_n)]\big),
  \qquad
  \underline x_i(t_n) = -\sqrt n\,x^*_i(t_n),\qquad
  \bar x_i(t_n) = \frac{\bar H_i - n\,x^*_i(t_n)}{\sqrt n},
\end{equation}
independently across classes.  The upper bound $\bar{H}_{i}$ is set using simulation. We simulate $500$ days of the prelimit system under the fastest-server-first rule and under the value-rate rule, record the largest number of class $i$ callers in the system over all days and both rules, and set
\begin{equation}\label{eq:headroom}
  \bar H_i = \max\big\{\lceil 1.5\times(\text{largest occupancy of class } i)\rceil,\ 10\big\},
  \qquad i\in\calI,
\end{equation}
where the lower bound of 10 callers guarantees that the sampling interval in~\eqref{eq:state-box} has width at least  $10/\sqrt{n} = 1$ for the classes with few callers.

The gradients are sampled from a box that is the same at every grid
point $\tau_{m}$,
\begin{equation}\label{eq:grad-box}
  p_i^{(m)}\sim\mathrm{Uniform}\big([-c_i/\theta_i,\ c_i/\theta_i]\big),
  \qquad i\in\calI .
\end{equation}
The half width $c_i/\theta_i$ is an upper bound on the cost that one additional class $i$ caller can add to the value function, since the caller waits for at most $1/\theta_i$ time units in expectation while accruing $c_i$ per unit of time.  We therefore take it as the upper endpoint of the box.  The gradient can also be negative, because a class $i$ caller generates value when served, and so we take the box symmetric.

\paragraph{Approximating the function $\mathcal{M}$.}
At every grid point $\tau_{m}$, we draw $M_{\mathcal{M}} = 10{,}000$ pairs
$(x^{(m)},p^{(m)})$ from~\eqref{eq:state-box}--\eqref{eq:grad-box} and
solve the linear program~\eqref{eq:M-term}--\eqref{eq:M-lp-basic} for
each interval, warm-starting the simplex method from the previous instance, which gives $960{,}000$ observations in total.  Because the
function $\mathcal{M}$ varies slowly in time, we train a single
network $\hat{\mathcal{M}}(\cdot,\cdot,\cdot)$ on all of them, with the
elapsed fraction of the time horizon $t_n/T$ as a third input next to $x$
and $p$, and set
$\hat{\mathcal{M}}_n(x,p) = \hat{\mathcal{M}}(x,p,t_n/T)$.
Subroutine~\ref{sub:offline-M} summarizes the procedure and
Table~\ref{tab:hyper-offline} lists the hyperparameters.

\paragraph{Approximating the mapping $\mathcal{D}$.}
At every grid point $\tau_{m}$, we draw $M_{\mathcal{D}} = 10{,}000$ states from
the box~\eqref{eq:state-box}, compute the reference allocation
$\tilde\psi(t_n,x^{(m)})$ by solving~\eqref{eq:ref-lp}, and obtain the
label $\mathcal{D}_n(x^{(m)})$ from~\eqref{eq:D-def}. We train one network
$\hat{\mathcal{D}}_n$ per grid point, as in
Subroutine~\ref{sub:offline-D}.  We repeat this for each reference
policy of Section~\ref{sec:reference-policies} that we use, which gives
one set of $96$ drift neural networks per reference policy and test problem. The linear programs are solved with the HiGHS solver through the JuMP
package in Julia, and the networks are trained in PyTorch.

\begin{table}[ht]
\centering
{\small
\begin{tabular}{p{5.4cm}p{4.4cm}p{4.4cm}}
\toprule
Hyperparameters & $\hat{\mathcal{M}}(\cdot,\cdot,\cdot)$ network & $\hat{\mathcal{D}}_n(\cdot)$ networks\\
\midrule
Number of networks & 1 & 96\\
Input dimension & $2I + 1 = 53$ & $I = 26$\\
Output dimension & 1 & $I = 26$\\
Number of hidden layers & 4 & 4\\
Number of neurons per layer & 150 & 150\\
Input batch normalization & Yes & Yes\\
Hidden-layer batch normalization & Yes & Yes\\
Activation function and weight initialization & ELU and Kaiming uniform & Leaky ReLU and Kaiming uniform\\
Optimizer & Adam & Adam\\
Batch size & 4{,}096 & 4{,}096\\
Number of epochs & 5{,}000 & 1{,}000\\
Initial learning rate & 1e-2 & 1e-2\\
Learning rate schedule & halved at epochs 150 and 350 & halved after 100 epochs without improvement\\
Train/validation split & 80/20 & 80/20\\
Training observations & 960{,}000 & 10{,}000 per network\\
\bottomrule
\end{tabular}
}
\caption{Summary of the hyperparameters used for the
$\hat{\mathcal{M}}$ and the $\hat{\mathcal{D}}$ networks.  The learning
rate of the $\hat{\mathcal{M}}$ network follows a step schedule and
that of the $\hat{\mathcal{D}}_n$ networks is reduced by half whenever
the validation loss has not improved for $100$ epochs.  The settings
are the same for the three test problems and for every reference
policy.}
\label{tab:hyper-offline}
\end{table}

\begin{subroutine}[H]
\caption{Offline approximation of the minimization terms
  $\mathcal{M}_n(x,p)$}
\label{sub:offline-M}
\begin{minipage}{\textwidth}
\begin{algorithmic}[1]
  \Statex \textbf{Input:} The number of training samples per grid point
    $M_{\mathcal{M}}$, the sampling boxes~\eqref{eq:state-box}
    and~\eqref{eq:grad-box}, the number of epochs $E_{\mathcal{M}}$, a
    batch size $S_{\mathcal{M}}$, a learning rate schedule
    $(\mathrm{lr}_0,\gamma,\text{milestones})$ (we use PyTorch's
    MultiStepLR scheduler, so the learning rate starts at
    $\mathrm{lr}_0$ and is multiplied by $\gamma\in(0,1)$ at each epoch
    listed in milestones), and the network architecture hyperparameters
    (number of layers, neurons per layer, activation function).
  \Statex \textbf{Output:} A trained neural network
    $\hat{\mathcal{M}}(\cdot,\cdot,\cdot)$ with
    $\hat{\mathcal{M}}_n(x,p) = \hat{\mathcal{M}}(x,p,t_n/T)$
    approximating $\mathcal{M}_n(x,p)$ for $n=0,1,\ldots,M-1$.
  \Statex \textit{Data generation:}
  \For{$n\gets 0$ to $M-1$}
    \For{$m=1,\ldots,M_{\mathcal{M}}$}
      \State Sample $x^{(n,m)}$ from~\eqref{eq:state-box} and $p^{(n,m)}$
        from~\eqref{eq:grad-box}.
      \State Solve the linear program
        \eqref{eq:M-term}--\eqref{eq:M-lp-basic} at
        $(x,p)=(x^{(n,m)},p^{(n,m)})$ to obtain
        $\mathcal{M}_n^{(m)}=\mathcal{M}_n(x^{(n,m)},p^{(n,m)})$.
    \EndFor
  \EndFor
  \Statex \textit{Network training:}
  \State Initialize a feedforward neural network
    $\hat{\mathcal{M}}_{\omega}:\R^{2I+1}\to\R$ with parameter vector
    $\omega$.
  \State Pool the $M\cdot M_{\mathcal{M}}$ observations
    $\{(x^{(n,m)},p^{(n,m)},t_n/T),\mathcal{M}_n^{(m)}\}$ over all grid
    points $n$ and samples $m$, and split them into training ($80\%$)
    and validation ($20\%$) sets.
  \For{epoch $=1,\ldots,E_{\mathcal{M}}$}
    \For{each mini-batch of size $S_{\mathcal{M}}$}
      \State Compute the loss
        $\ell(\omega)=\frac{1}{S_{\mathcal{M}}}\sum_{(n,m)}
        \big(\hat{\mathcal{M}}_{\omega}(x^{(n,m)},p^{(n,m)},t_n/T)
        -\mathcal{M}_n^{(m)}\big)^2$ over the mini-batch.
      \State Update $\omega$ using the Adam optimizer.
    \EndFor
    \State Update the learning rate according to the schedule.
  \EndFor
  \State \textbf{return} $\hat{\mathcal{M}}(\cdot,\cdot,\cdot)
    \equiv \hat{\mathcal{M}}_{\omega}(\cdot,\cdot,\cdot)$.
\end{algorithmic}
\end{minipage}
\end{subroutine}

\begin{subroutine}[H]
\caption{Offline approximation of the mappings $\mathcal{D}_n(x)$}
\label{sub:offline-D}
\begin{minipage}{\textwidth}
\begin{algorithmic}[1]
  \Statex \textbf{Input:} The number of training samples per grid point
    $M_{\mathcal{D}}$, the sampling box~\eqref{eq:state-box}, the
    reference policy $\tilde\psi$, the number of epochs
    $E_{\mathcal{D}}$, a batch size $S_{\mathcal{D}}$, an initial
    learning rate $\mathrm{lr}_0$ and a decay factor $\gamma\in(0,1)$
    (we use PyTorch's ReduceLROnPlateau scheduler, so the learning rate
    is multiplied by $\gamma$ whenever the validation loss has not
    improved for a given number of epochs), and the network architecture
    hyperparameters (number of layers, neurons per layer, activation
    function).
  \Statex \textbf{Output:} Trained neural networks
    $\hat{\mathcal{D}}_n(\cdot)$ approximating the mappings
    $\mathcal{D}_n(\cdot)$ of~\eqref{eq:D-def} for $n=0,1,\ldots,M-1$.
  \For{$n\gets 0$ to $M-1$}
    \Statex \quad \textit{Data generation:}
    \For{$m=1,\ldots,M_{\mathcal{D}}$}
      \State Sample $x^{(m)}$ from~\eqref{eq:state-box}.
      \State Compute the reference allocation $\tilde\psi(t_n,x^{(m)})$
        through~\eqref{eq:ref-lp} and set
        $\mathcal{D}_n^{(m)}=\mathcal{D}_n(x^{(m)})$
        via~\eqref{eq:D-def}.
    \EndFor
    \Statex \quad \textit{Network training:}
    \State Initialize a feedforward neural network
      $\hat{\mathcal{D}}_{n,\phi}:\R^{I}\to\R^{I}$ with parameter
      vector $\phi$.
    \State Split the dataset
      $\{x^{(m)},\mathcal{D}_n^{(m)}\}_{m=1}^{M_{\mathcal{D}}}$ into
      training ($80\%$) and validation ($20\%$) sets.
    \For{epoch $=1,\ldots,E_{\mathcal{D}}$}
      \For{each mini-batch $\{x^{(m)},\mathcal{D}_n^{(m)}\}$ of size
        $S_{\mathcal{D}}$}
        \State Compute the loss
          $\ell(\phi)=\frac{1}{S_{\mathcal{D}}}\sum_m
          \big\|\hat{\mathcal{D}}_{n,\phi}(x^{(m)})
          -\mathcal{D}_n^{(m)}\big\|^2$.
        \State Update $\phi$ using the Adam optimizer.
      \EndFor
      \State Update the learning rate according to the validation loss.
    \EndFor
  \EndFor
  \State \textbf{return} $\hat{\mathcal{D}}_n(\cdot)$ for
    $n=0,1,\ldots,M-1$.
\end{algorithmic}
\end{minipage}
\end{subroutine}

\section{Derivation of the Proposed Prelimit Policy}
\label{app:prelimit-policy}
We derive the prelimit linear
program~\eqref{eq:prelimit-policy}--\eqref{eq:pp-pos} by substituting
the scaling relations~\eqref{eq:scaled-state}
and~\eqref{eq:control-prelimit} into the linear
program~\eqref{eq:limit-policy} over $\bar\Psi(t_n,x)$ and simplifying
each component one at a time.
Throughout we evaluate this program at $x=\hat X^n(t)$ and
read~\eqref{eq:control-prelimit} as
\begin{equation}\label{eq:app-pp-invert}
  \hat\psi_j = \frac{n\psi_j^*(t)-\psi_j^n(t)}{\sqrt n},
  \qquad j\in\calJ,
\end{equation}
where the term of order $o(1)$ is suppressed.

\paragraph{Objective.}
Substituting~\eqref{eq:app-pp-invert} into the objective
of~\eqref{eq:limit-policy} and writing $w_j(p)$ for the activity
weight of~\eqref{eq:weight},
\begin{equation*}
  \sum_{j\in\calJ}w_j(p)\,\hat\psi_j
  = \sqrt n\sum_{j\in\calJ}w_j(p)\,\psi_j^*(t)
  - \frac{1}{\sqrt n}\sum_{j\in\calJ}w_j(p)\,\psi_j^n(t).
\end{equation*}
The first sum is constant with respect to the decision variables
$\psi_j^n(t)$ and the second carries the coefficient $-1/\sqrt n$.  Minimizing this expression is therefore equivalent
to maximizing
\begin{equation*}
  \sum_{j\in\calJ}w_j(p)\,\psi_j^n(t),
\end{equation*}
which yields~\eqref{eq:prelimit-policy}.

\paragraph{Constraint~\eqref{eq:M-lp-class} $\to$~\eqref{eq:pp-class}.}
Substituting~\eqref{eq:app-pp-invert} and~\eqref{eq:scaled-state} into
the class constraint $\sum_{j\in\mathcal{B}(i)}\hat\psi_j\ge-x_i$ gives
\begin{equation*}
  \sum_{j\in\mathcal{B}(i)}
    \frac{n\psi_j^*(t)-\psi_j^n(t)}{\sqrt n}
  \ \ge\ -\,\frac{X_i^n(t)-n\,x_i^*(t)}{\sqrt n}.
\end{equation*}
Multiplying through by $\sqrt n$ gives
\begin{equation*}
  n\sum_{j\in\mathcal{B}(i)}\psi_j^*(t)
  - \sum_{j\in\mathcal{B}(i)}\psi_j^n(t)
  \ \ge\ n\,x_i^*(t) - X_i^n(t).
\end{equation*}
The nominal allocations satisfy
$x_i^*(t)=\sum_{j\in\mathcal{B}(i)}\psi_j^*(t)$, so the terms of order
$n$ cancel on the two sides:
\begin{equation*}
  \sum_{j\in\mathcal{B}(i)}\psi_j^n(t) \le X_i^n(t),
\end{equation*}
which yields~\eqref{eq:pp-class}.

\paragraph{Constraint~\eqref{eq:M-lp-pool} $\to$~\eqref{eq:pp-pool}.}
Substituting~\eqref{eq:app-pp-invert} into the pool constraint
$\sum_{j\in\mathcal{S}(k)}\hat\psi_j\ge0$ and multiplying by $\sqrt n$
gives
\begin{equation*}
  \sum_{j\in\mathcal{S}(k)}\psi_j^n(t)
  \ \le\ n\sum_{j\in\mathcal{S}(k)}\psi_j^*(t).
\end{equation*}
Assumption~\ref{ass:ht} gives $A\xi^*(t)=e$, so that the nominal
allocations satisfy $\sum_{j\in\mathcal{S}(k)}\psi_j^*(t)=\nu_k(t)$,
and $N_k^n(t)=n\nu_k(t)$ by~\eqref{eq:scaling-staff}.  The right side
is therefore $N_k^n(t)$, which yields~\eqref{eq:pp-pool}.

\paragraph{Constraints~\eqref{eq:M-lp-nonbasic}
  and~\eqref{eq:M-lp-basic} $\to$~\eqref{eq:pp-pos}.}
A nonbasic activity has $\psi_j^*(t)=0$, so the constraint
$\hat\psi_j\le0$ reads $-\psi_j^n(t)/\sqrt n\le0$, that is,
$\psi_j^n(t)\ge0$.  For a basic activity, the bound
$\hat\psi_j\le\sqrt n\,\psi_j^*(t)$ reads
$n\psi_j^*(t)-\psi_j^n(t)\le n\psi_j^*(t)$ after multiplying by
$\sqrt n$, and the nominal terms cancel, again leaving
$\psi_j^n(t)\ge0$.  The two sign constraints of $\bar\Psi(t_n,x)$
thus collapse to~\eqref{eq:pp-pos}.

\newpage

\section{Uniqueness of the Nominal Plan}\label{app:uniqueness}

Assumption~\ref{ass:ht} requires that, at each grid point $\tau_{m}$, the planning problem~\eqref{eq:spp-lp}--\eqref{eq:spp-lp-pos} have a unique optimal
solution that fully utilizes every server pool. This appendix describes
how we verify these properties for the calibrated test problems.

Fix a grid point $\tau_m$ and omit it from the notation. Let
$\sigma_k \ge 0$ denote the unused fraction of pool $k$. The capacity
constraints become
\[
  \sum_{j\in\mathcal{S}(k)}\xi_j+\sigma_k=1,
  \qquad k\in\calK.
\]
Together with the $I$ flow balance constraints, these give $I+K$
equations in $J+K$ nonnegative variables.

Let $\xi^*$ be the optimal solution obtained by the simplex method.
Let $y\in\R^I$ and $z\in\R^K$ be the corresponding optimal dual
variables for the flow balance and capacity constraints. The reduced
cost of activity $j$ is
\begin{equation}\label{eq:reduced-cost}
  d_j=\tilde v_j-\nu_{s(j)}R_j^\top y-z_{s(j)},
  \qquad j\in\calJ,
\end{equation}
where $R_j$ is the $j$th column of $R$. The reduced cost of
$\sigma_k$ is $-z_k$. Optimality gives $d_j\le 0$, $z_k\ge 0$,
and $d_j=0$ whenever $\xi_j^*>0$.

At each grid point, we check two conditions:
\begin{enumerate}
\item $d_j<0$ for every activity not used by the optimal plan
  ($\xi_j^*=0$).
\item $z_k>0$ for every pool $k$.
\end{enumerate}
The second condition implies, by complementary slackness, that
$\sigma_k^*=0$ for every pool. Therefore, the optimal plan fully
utilizes every pool.

To establish uniqueness, consider any other feasible plan $\xi$
with unused capacity $\sigma$. Its objective value satisfies
\[
  \tilde v\cdot\xi-\tilde v\cdot\xi^*
  =
  \sum_{j:\xi_j^*=0}d_j\xi_j-\sum_k z_k\sigma_k.
\]
Under the two conditions above, this difference is negative if
$\xi$ uses an activity outside the optimal plan or leaves any
capacity unused. Therefore, any equally good plan must use only
the activities used by $\xi^*$ and fully utilize every pool.

In our test problems, the optimal plan has exactly $I+K=35$ basic
activities at every grid point. Their columns form the simplex basis
and are linearly independent. Therefore, the $I+K$ flow balance and
capacity equations determine their allocations uniquely, and any
equally good plan must equal $\xi^*$. This verifies uniqueness and
determines the basic set $\calB(\tau_m)$ of Section~\ref{sec:bcp}.

\section{Data Used for the Test Problems}\label{app:data}

This appendix presents the class-level and pool-level estimates
underlying Section~\ref{sec:calibration}.  In every graph, the
resolution of the horizontal axis is five minutes and the solid line is
the average
across the 84 business days of July--October 2002, the dashed
lines show $99\%$ confidence intervals.

\subsection{Arrival Rates by Class}\label{app:data-arrivals}

\begin{figure}[H]
     \centering
     \begin{minipage}{0.48\textwidth}
         \centering
         \includegraphics[width=\linewidth]{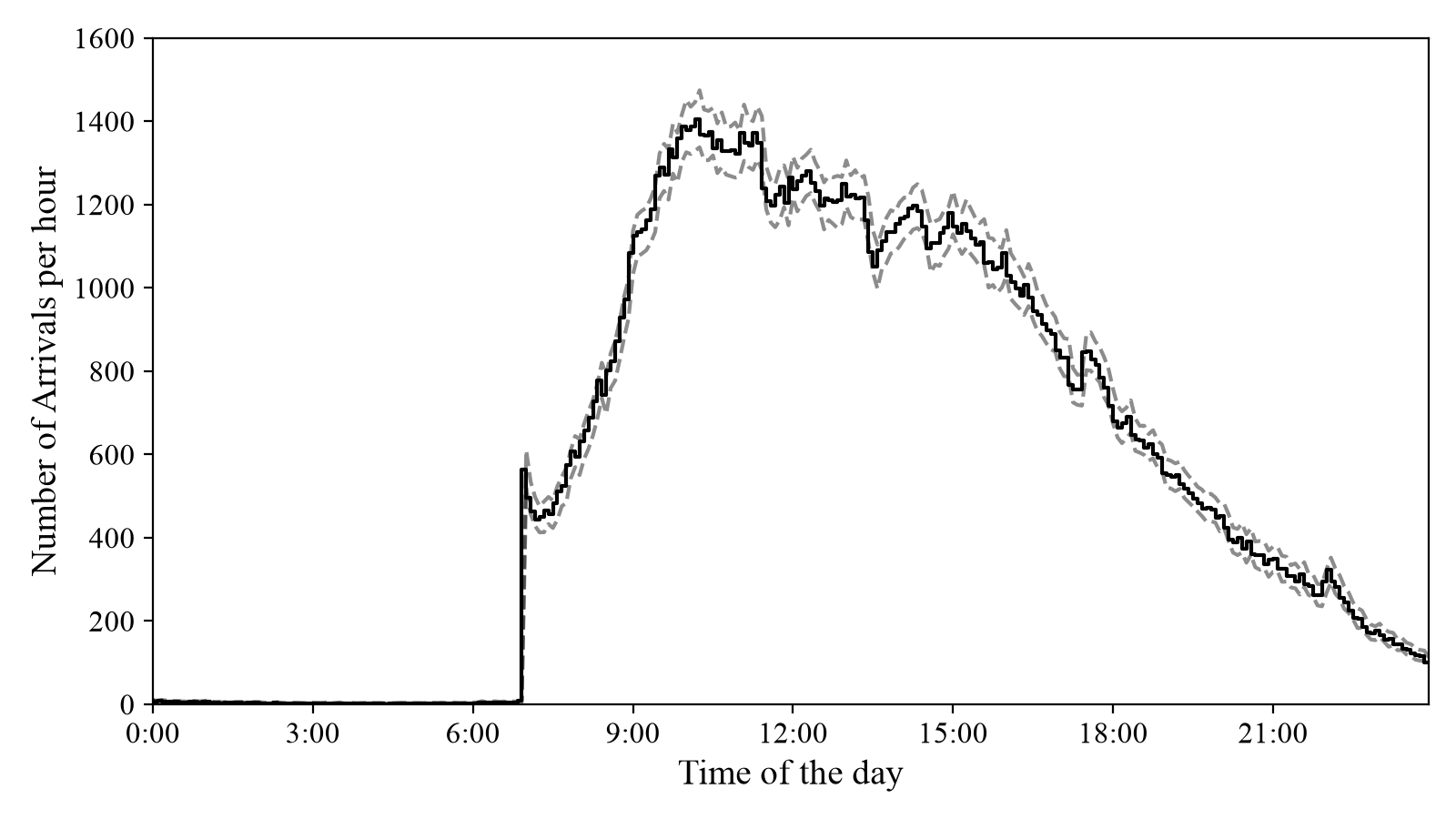}
         \captionsetup{font={footnotesize}}
         \caption{Hourly arrival rates for Retail (Node: 1)}
         \label{graph_arrival_first}
     \end{minipage}
     \hfill
     \begin{minipage}{0.48\textwidth}
         \centering
         \includegraphics[width=\linewidth]{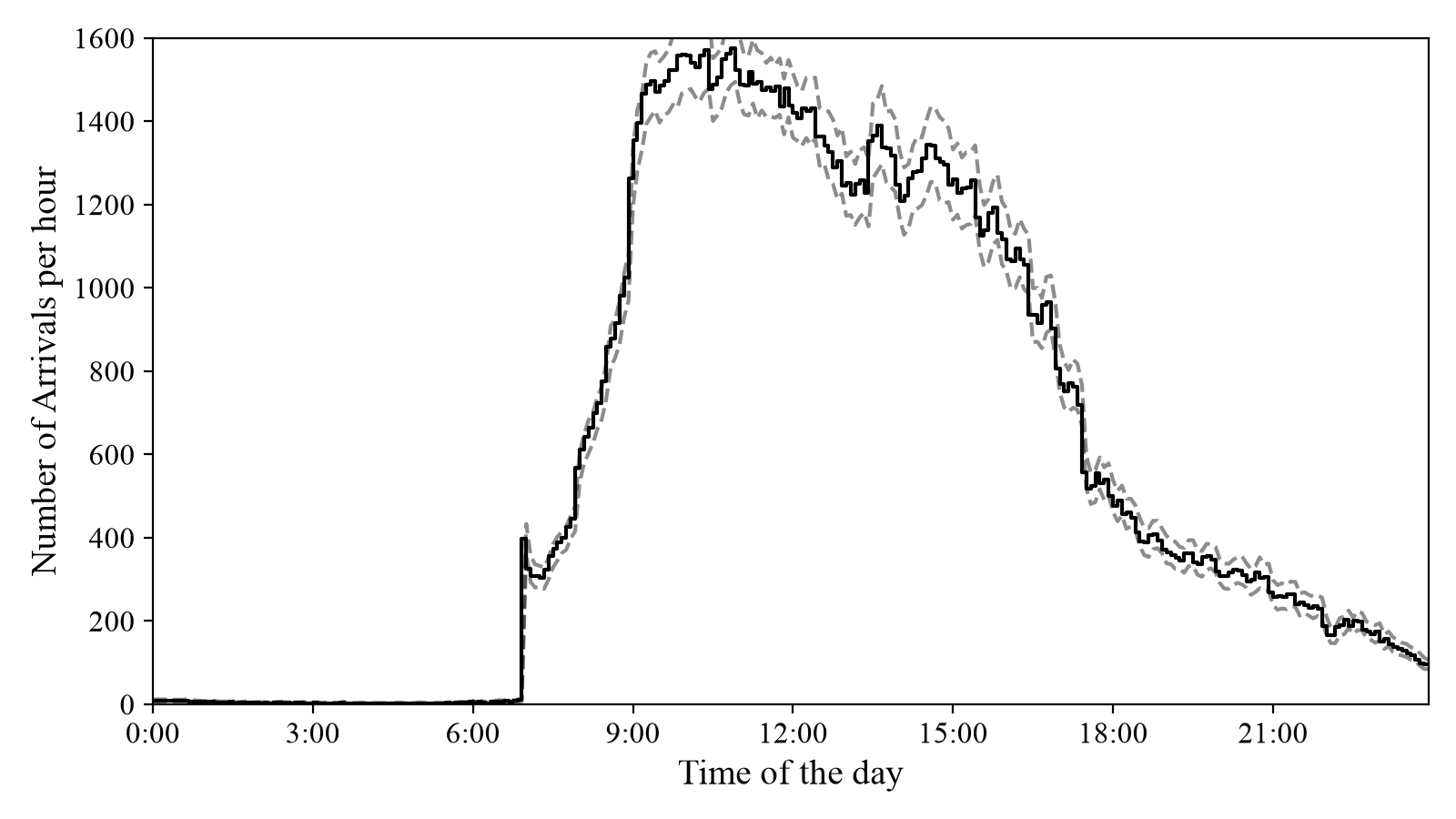}
         \captionsetup{font={footnotesize}}
         \caption{Hourly arrival rates for Retail (Node: 2)}
     \end{minipage}
\end{figure}

\begin{figure}[H]
     \centering
     \begin{minipage}{0.48\textwidth}
         \centering
         \includegraphics[width=\linewidth]{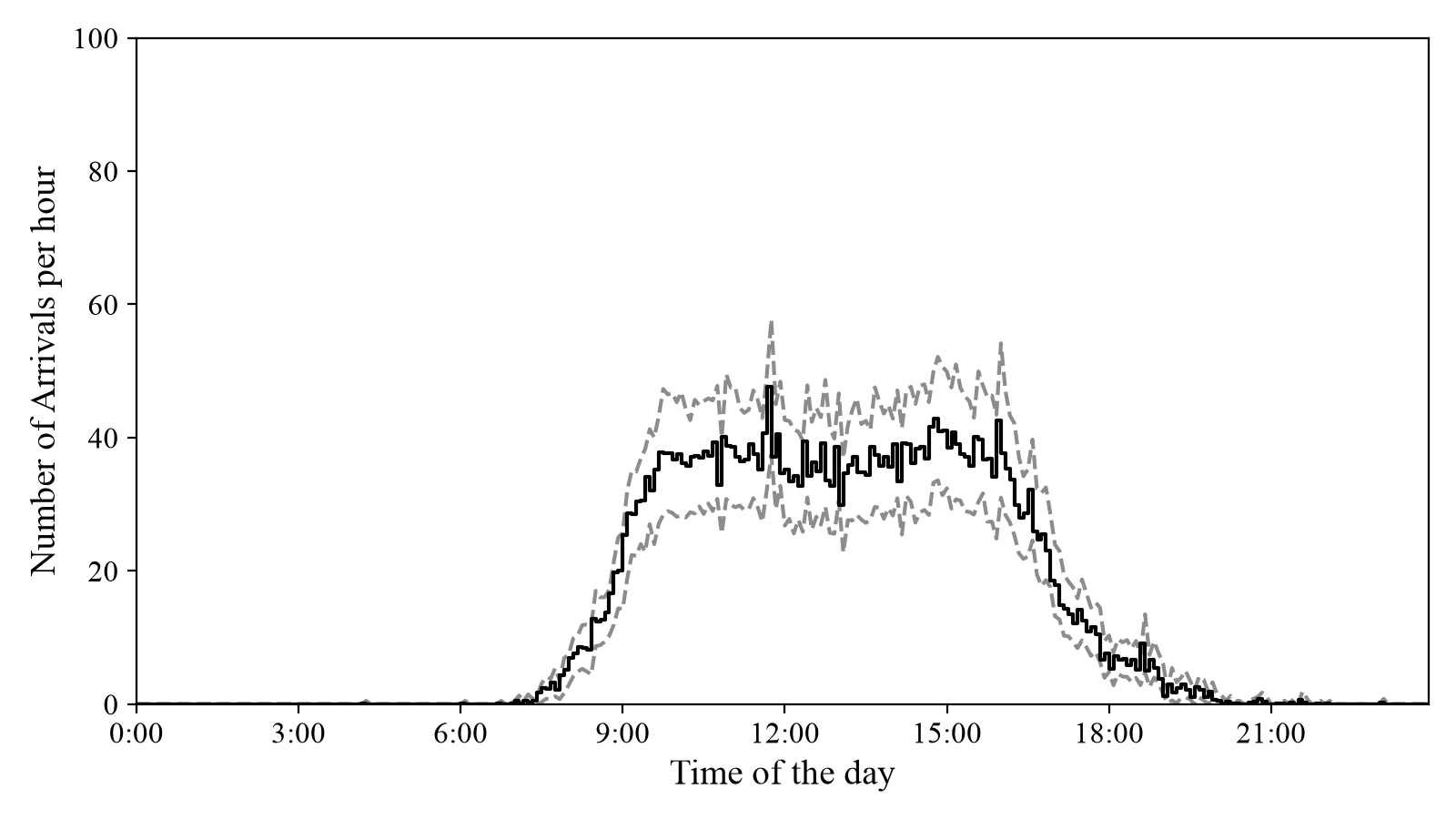}
         \captionsetup{font={footnotesize}}
         \caption{Hourly arrival rates for Priority Service}
         \label{graph_arrival_last}
     \end{minipage}
     \hfill
     \begin{minipage}{0.48\textwidth}
     \end{minipage}
\end{figure}

Figures~\ref{graph_arrival_first}--\ref{graph_arrival_last} display the
average hourly arrival pattern of each original class.  To improve
visibility while keeping classes comparable, the vertical axes share
common limits within three call-volume tiers: the three Retail node
classes (0--1600), the medium-volume classes Premier, Business, Consumer
Loans, Online Banking, and Telesales (0--500), and the small-volume
classes with less than $1\%$ of aggregate arrivals (0--100).

\begin{figure}[H]
     \centering
     \begin{minipage}{0.48\textwidth}
         \centering
         \includegraphics[width=\linewidth]{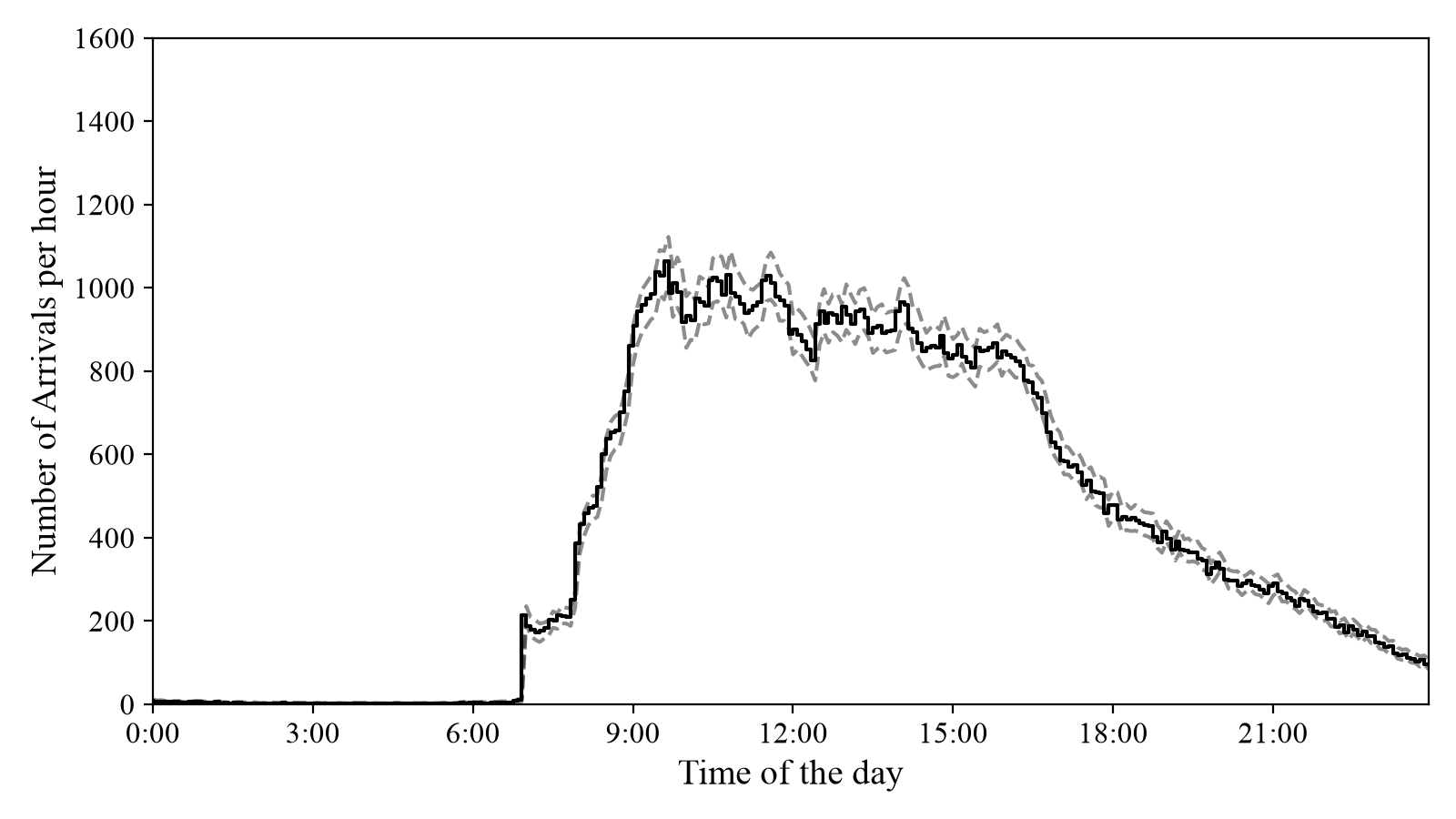}
         \captionsetup{font={footnotesize}}
         \caption{Hourly arrival rates for Retail (Node: 3)}
     \end{minipage}
     \hfill
     \begin{minipage}{0.48\textwidth}
         \centering
         \includegraphics[width=\linewidth]{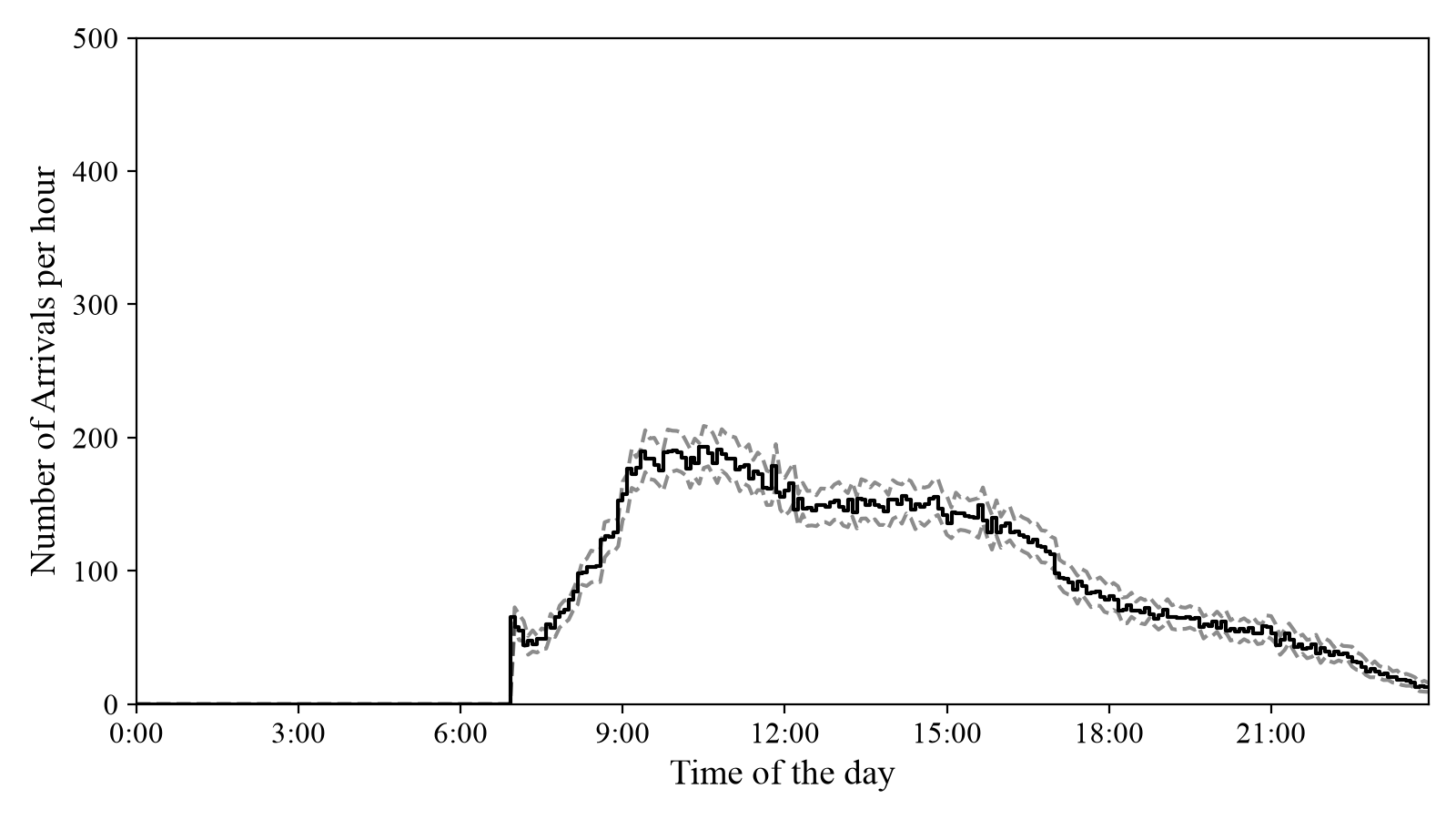}
         \captionsetup{font={footnotesize}}
         \caption{Hourly arrival rates for Premier}
     \end{minipage}
\end{figure}
\begin{figure}[H]
     \centering
     \begin{minipage}{0.48\textwidth}
         \centering
         \includegraphics[width=\linewidth]{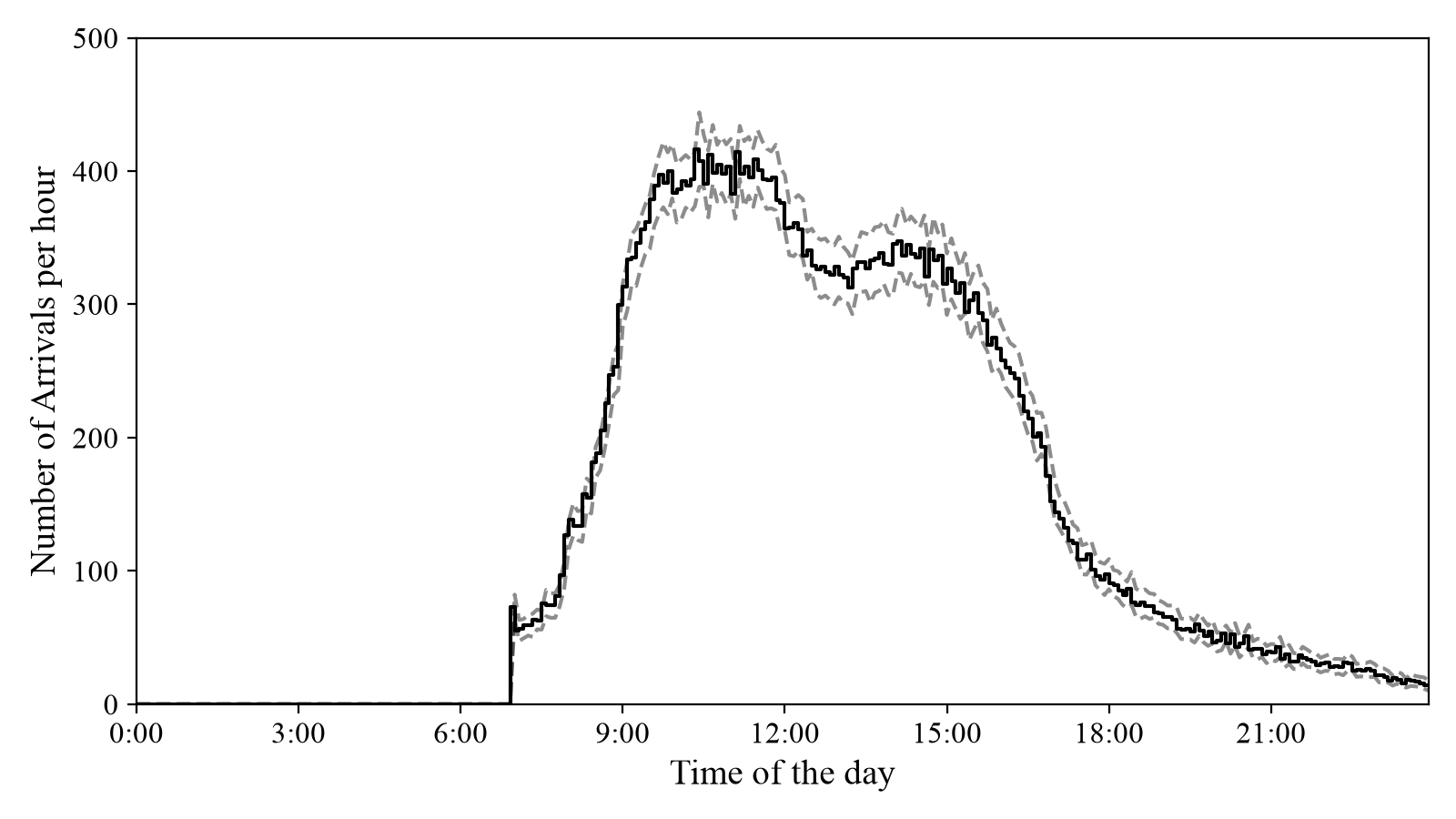}
         \captionsetup{font={footnotesize}}
         \caption{Hourly arrival rates for Business}
     \end{minipage}
     \hfill
     \begin{minipage}{0.48\textwidth}
         \centering
         \includegraphics[width=\linewidth]{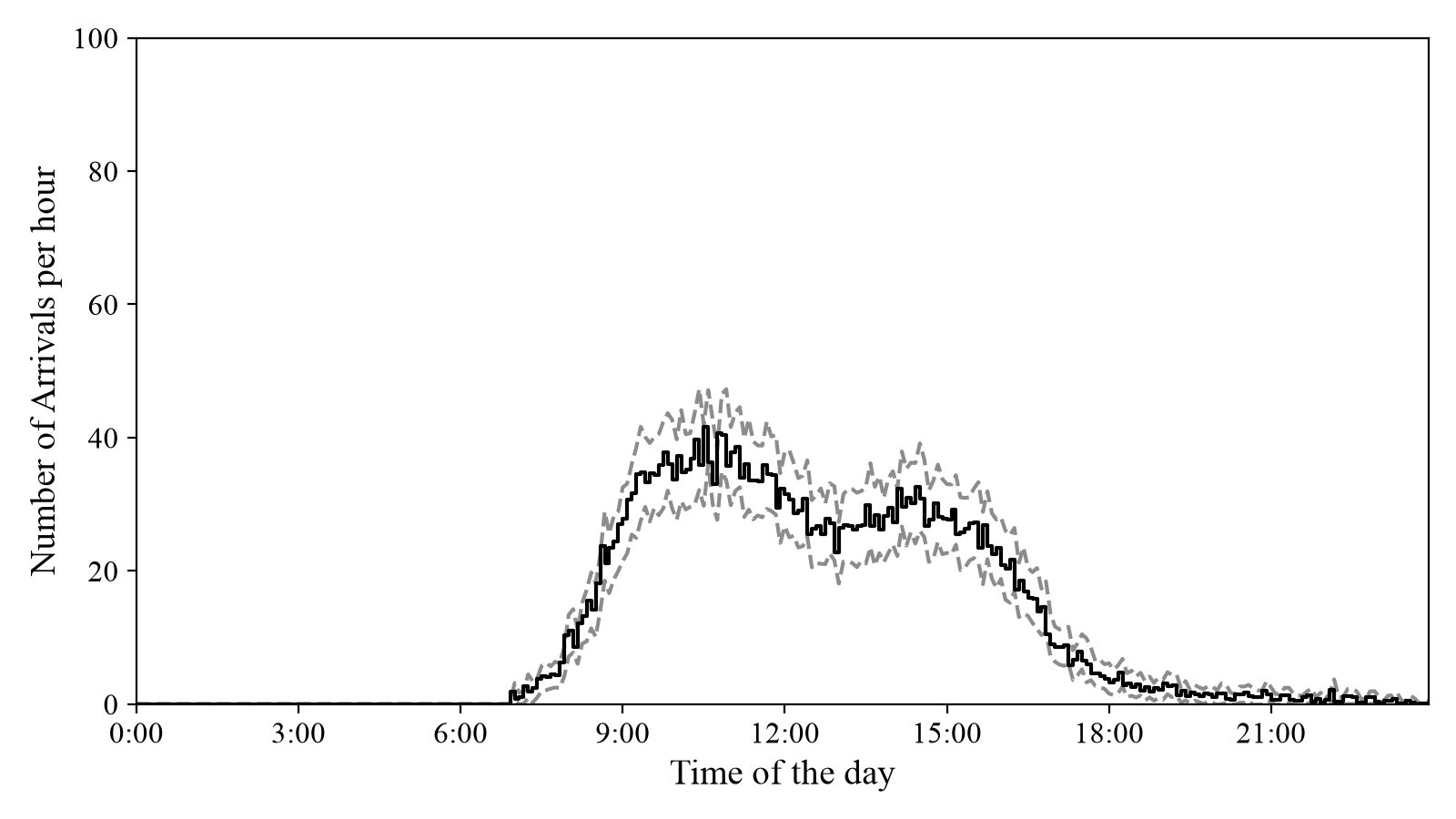}
         \captionsetup{font={footnotesize}}
         \caption{Hourly arrival rates for Platinum}
     \end{minipage}
\end{figure}
\begin{figure}[H]
     \centering
     \begin{minipage}{0.48\textwidth}
         \centering
         \includegraphics[width=\linewidth]{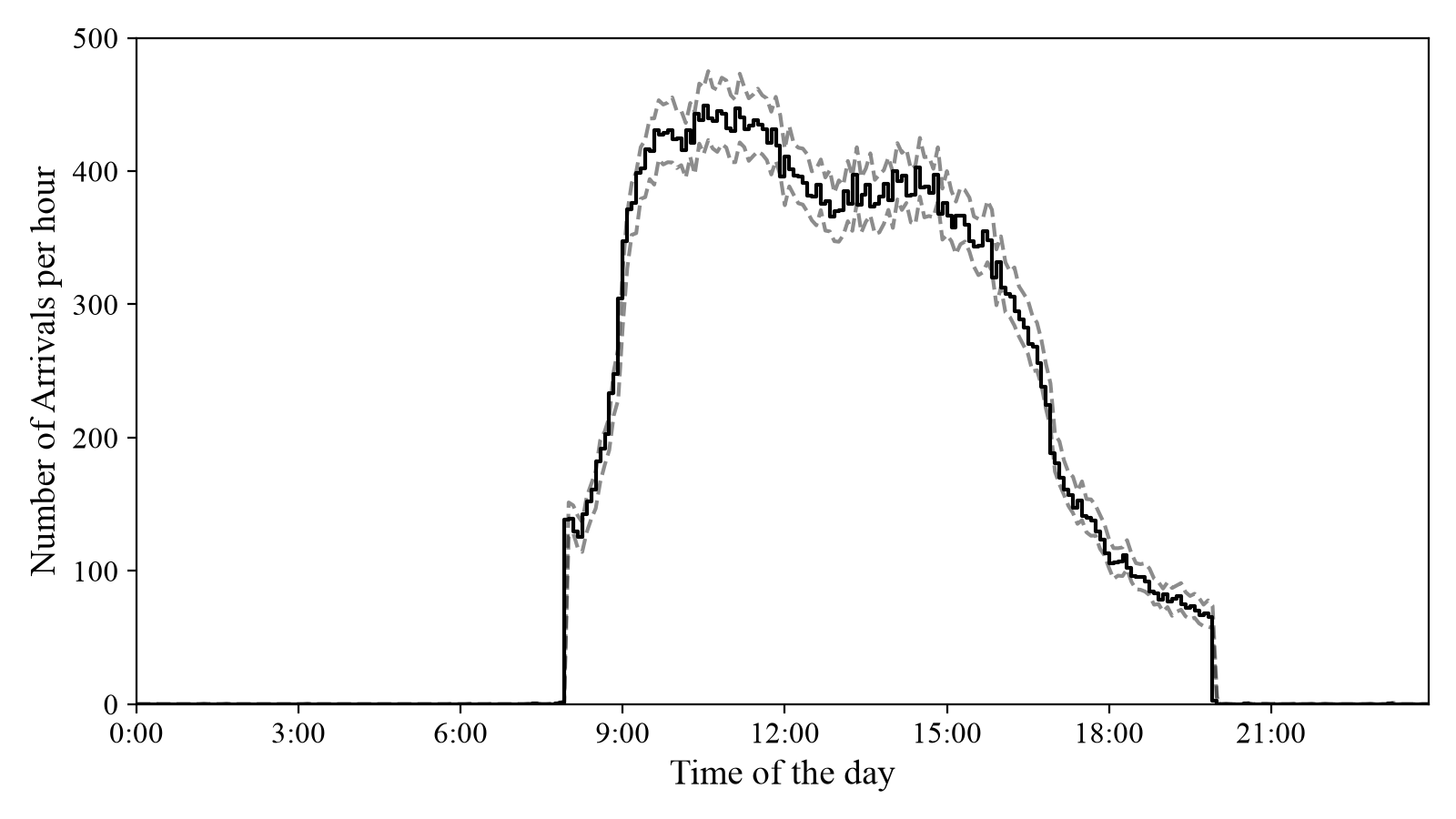}
         \captionsetup{font={footnotesize}}
         \caption{Hourly arrival rates for Consumer Loans}
     \end{minipage}
     \hfill
     \begin{minipage}{0.48\textwidth}
         \centering
         \includegraphics[width=\linewidth]{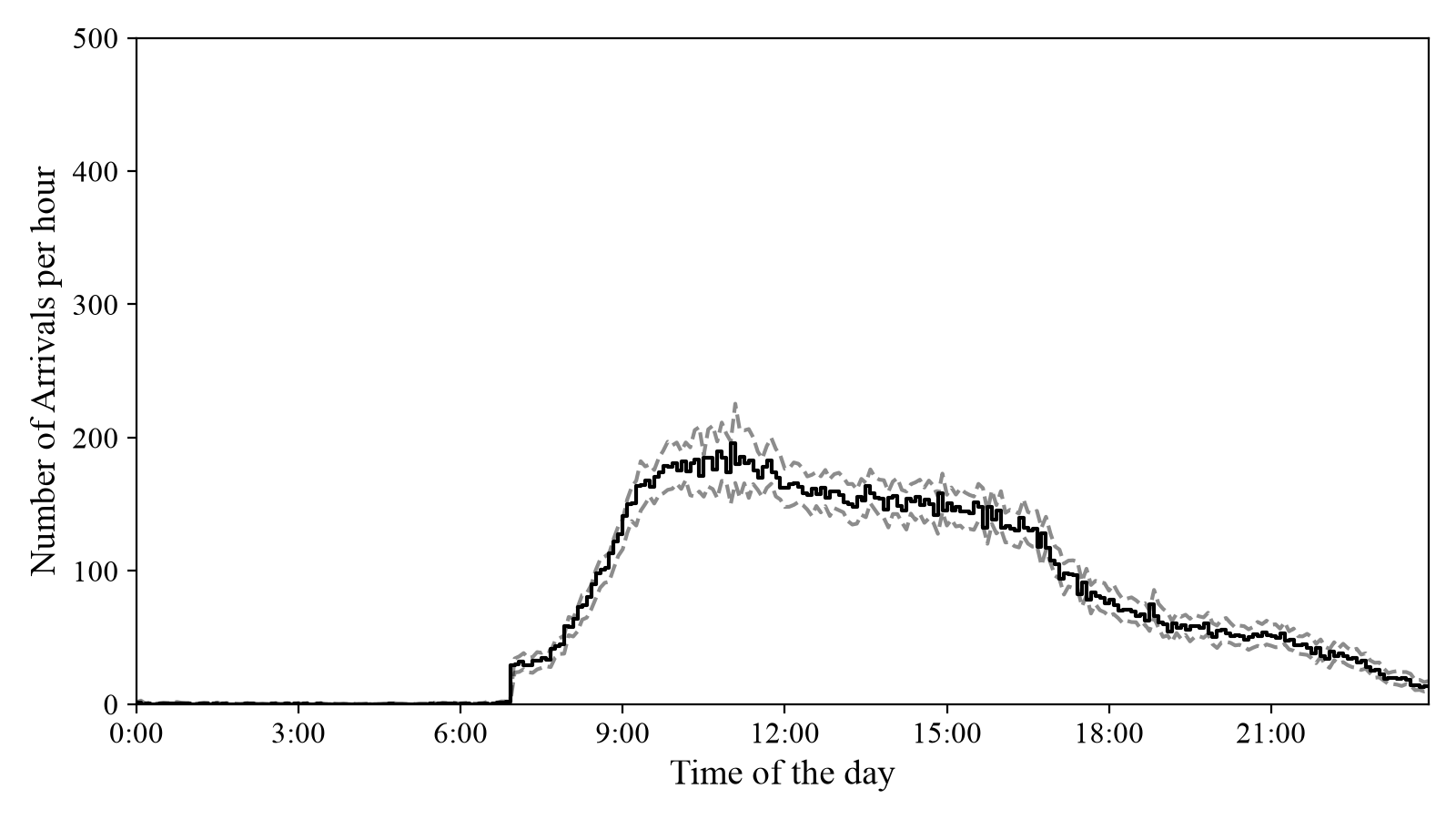}
         \captionsetup{font={footnotesize}}
         \caption{Hourly arrival rates for Online Banking}
     \end{minipage}
\end{figure}
\begin{figure}[H]
     \centering
     \begin{minipage}{0.48\textwidth}
         \centering
         \includegraphics[width=\linewidth]{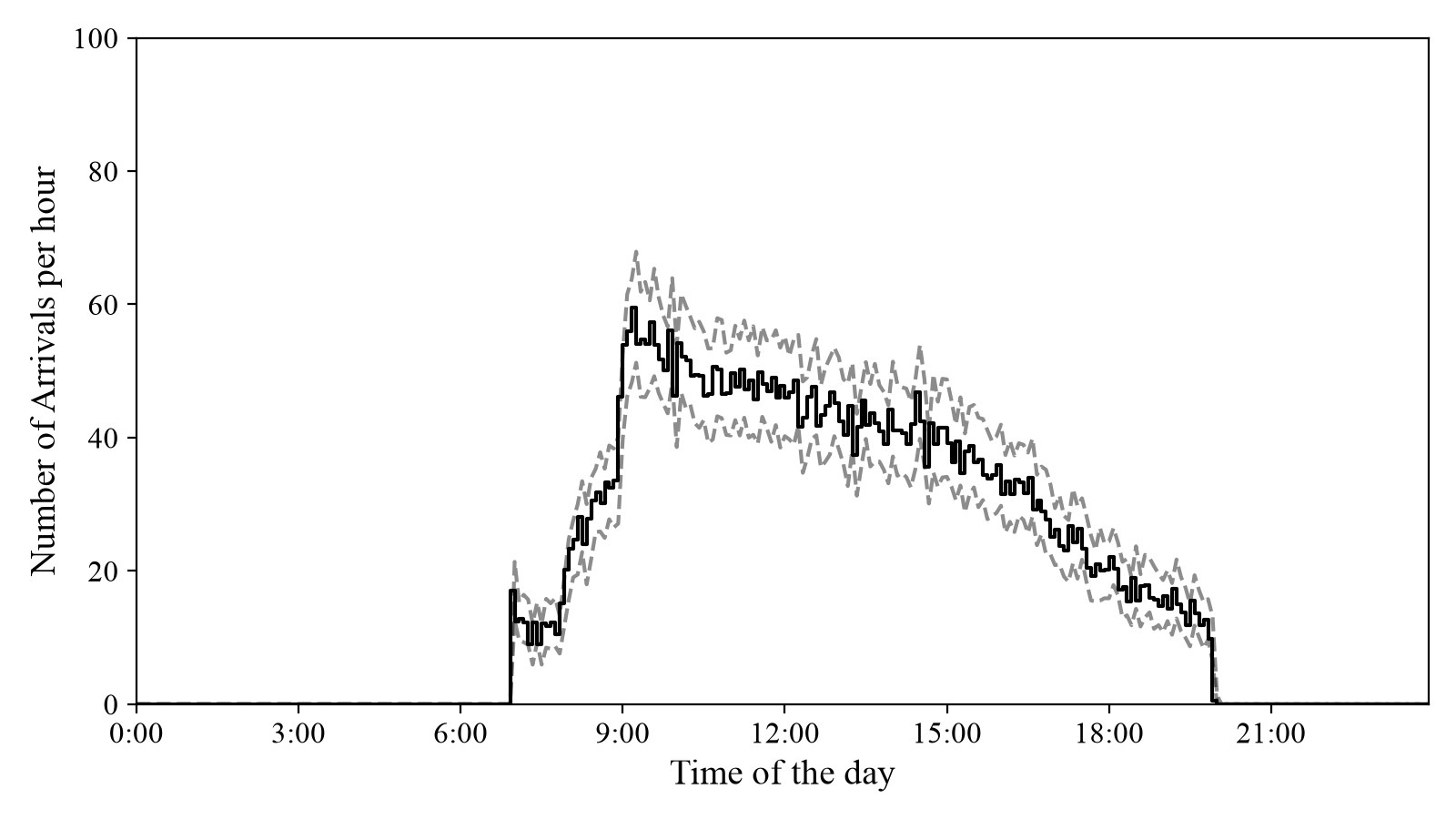}
         \captionsetup{font={footnotesize}}
         \caption{Hourly arrival rates for EBO}
     \end{minipage}
     \hfill
     \begin{minipage}{0.48\textwidth}
         \centering
         \includegraphics[width=\linewidth]{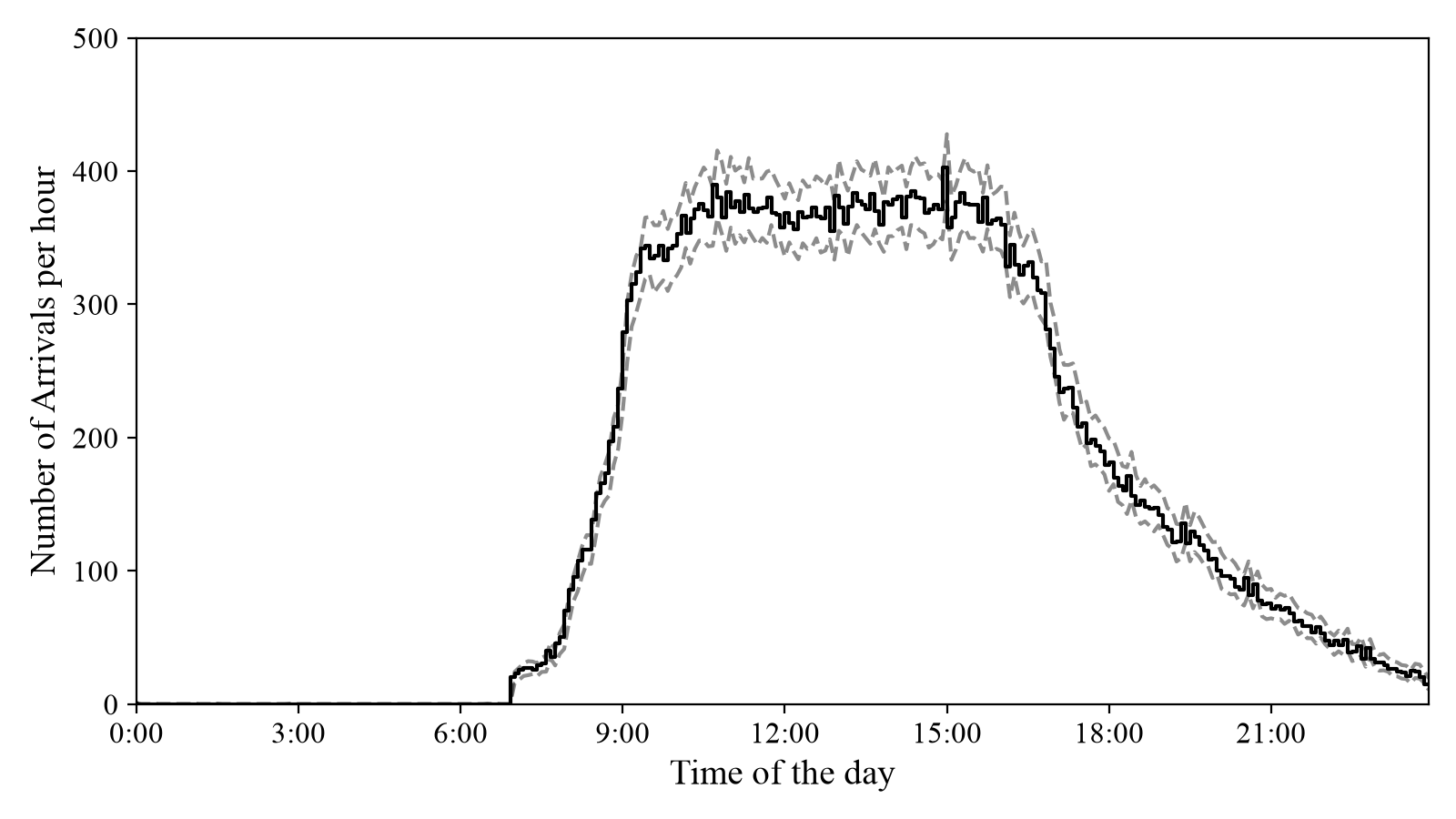}
         \captionsetup{font={footnotesize}}
         \caption{Hourly arrival rates for Telesales}
     \end{minipage}
\end{figure}
\begin{figure}[H]
     \centering
     \begin{minipage}{0.48\textwidth}
         \centering
         \includegraphics[width=\linewidth]{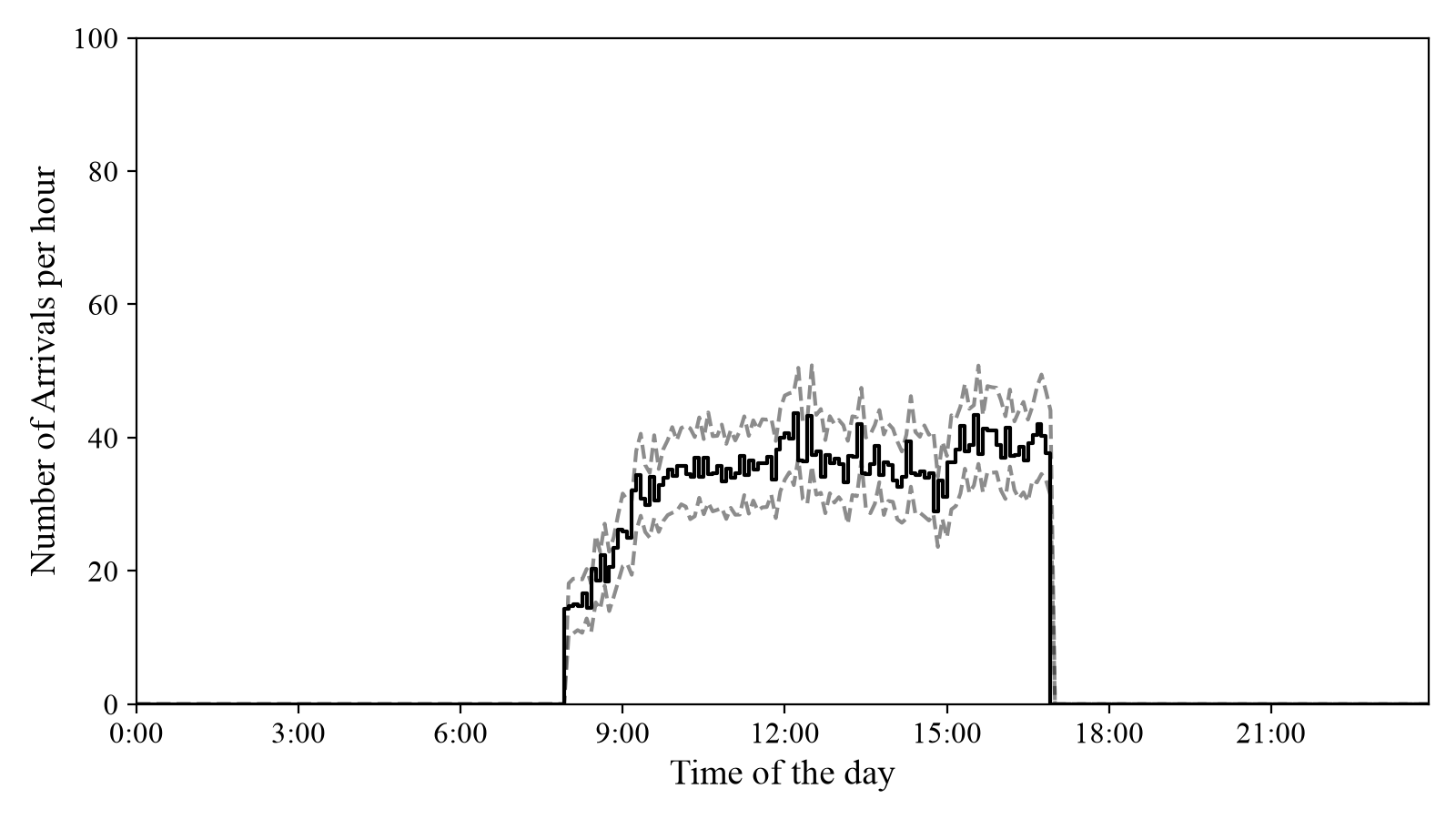}
         \captionsetup{font={footnotesize}}
         \caption{Hourly arrival rates for Subanco}
     \end{minipage}
     \hfill
     \begin{minipage}{0.48\textwidth}
         \centering
         \includegraphics[width=\linewidth]{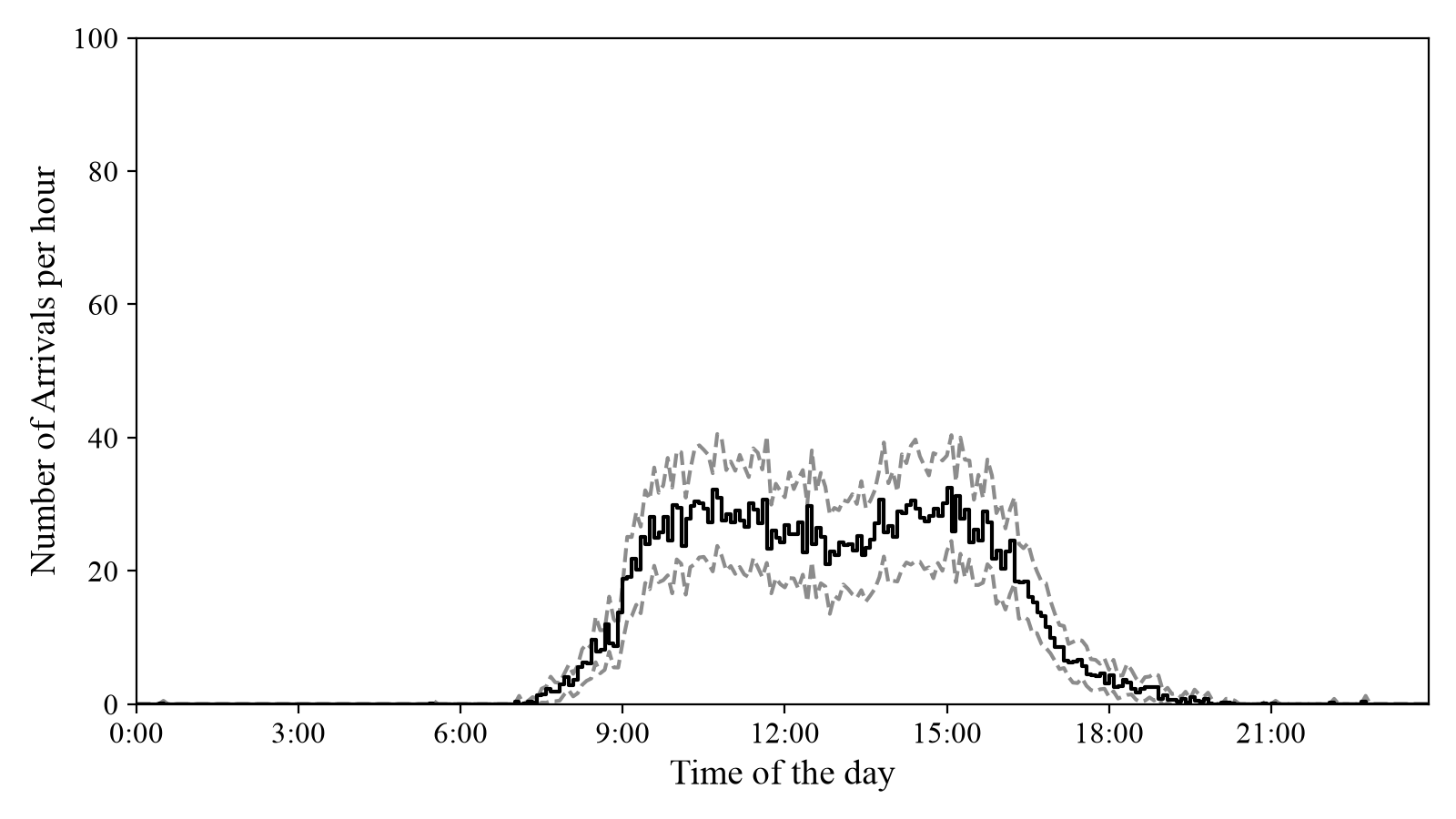}
         \captionsetup{font={footnotesize}}
         \caption{Hourly arrival rates for Case Quality}
     \end{minipage}
\end{figure}

\begin{figure}[H]
\centering
\includegraphics[width=0.8\linewidth]{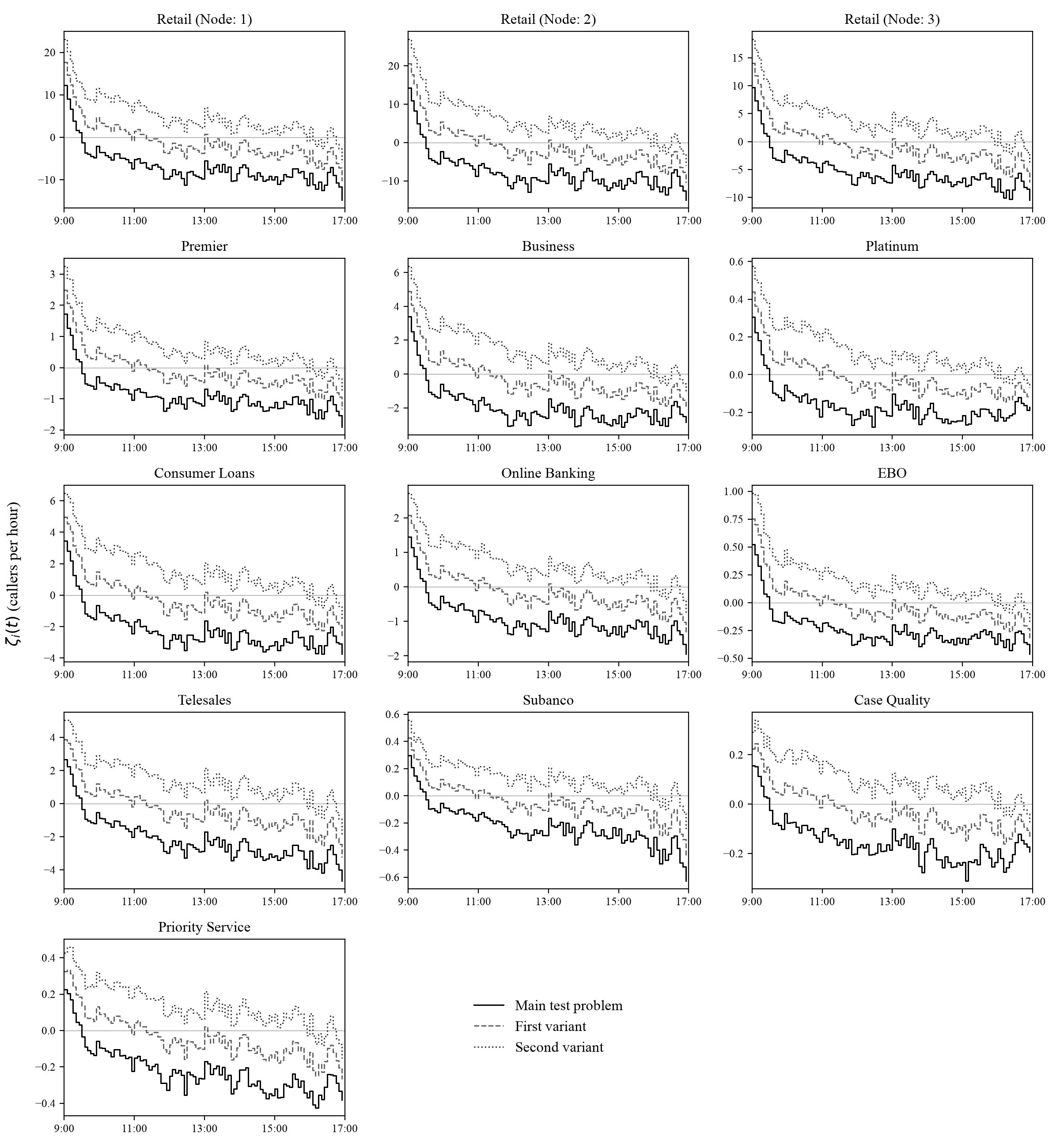}
\caption{Second-order terms $\zeta_i(\cdot)$ of the thirteen original
classes for the main test problem and its two variants.  The resolution
of the horizontal axis is five minutes.  The vertical axes are scaled to
the magnitude of the term of each class.}
\label{fig:zeta-classes}
\end{figure}

\vspace{-10mm}
\paragraph{The second-order terms $\zeta_i(\cdot)$.}
Figure~\ref{fig:zeta-classes} shows the second-order term of every
original class in the main test problem and its two variants, from
Equation~\eqref{eq:zeta-calib}.  Each variant curve of a class exceeds the curve of the main test
problem by $(\beta-1)\lambda^n_i(\cdot)/\sqrt n$, with $\beta = 1.05$
and $\beta = 1.10$, as in Equation~\eqref{eq:zeta-variant}, so the
variants shift the term of every class upward in proportion to its
arrival rate.  The rework classes have no external arrivals and their
second-order terms are zero.

\subsection{Staffing Levels by Server Pool}\label{app:data-staffing}

\begin{figure}[H]
     \centering
     \begin{minipage}{0.48\textwidth}
         \centering
         \includegraphics[width=\linewidth]{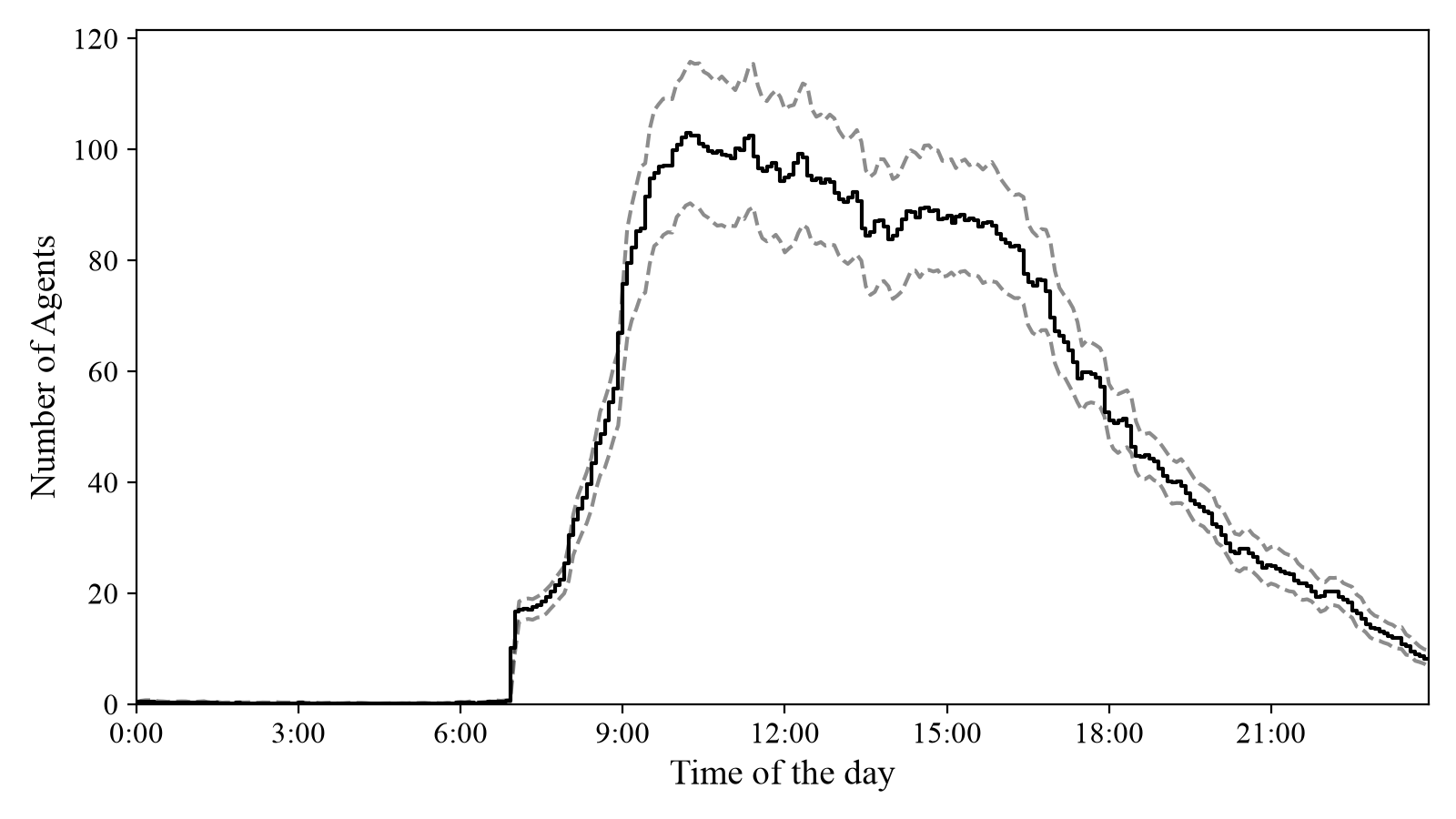}
         \captionsetup{font={footnotesize}}
         \caption{Staffing level of pool 1}
         \label{graph_staffing_first}
     \end{minipage}
     \hfill
     \begin{minipage}{0.48\textwidth}
         \centering
         \includegraphics[width=\linewidth]{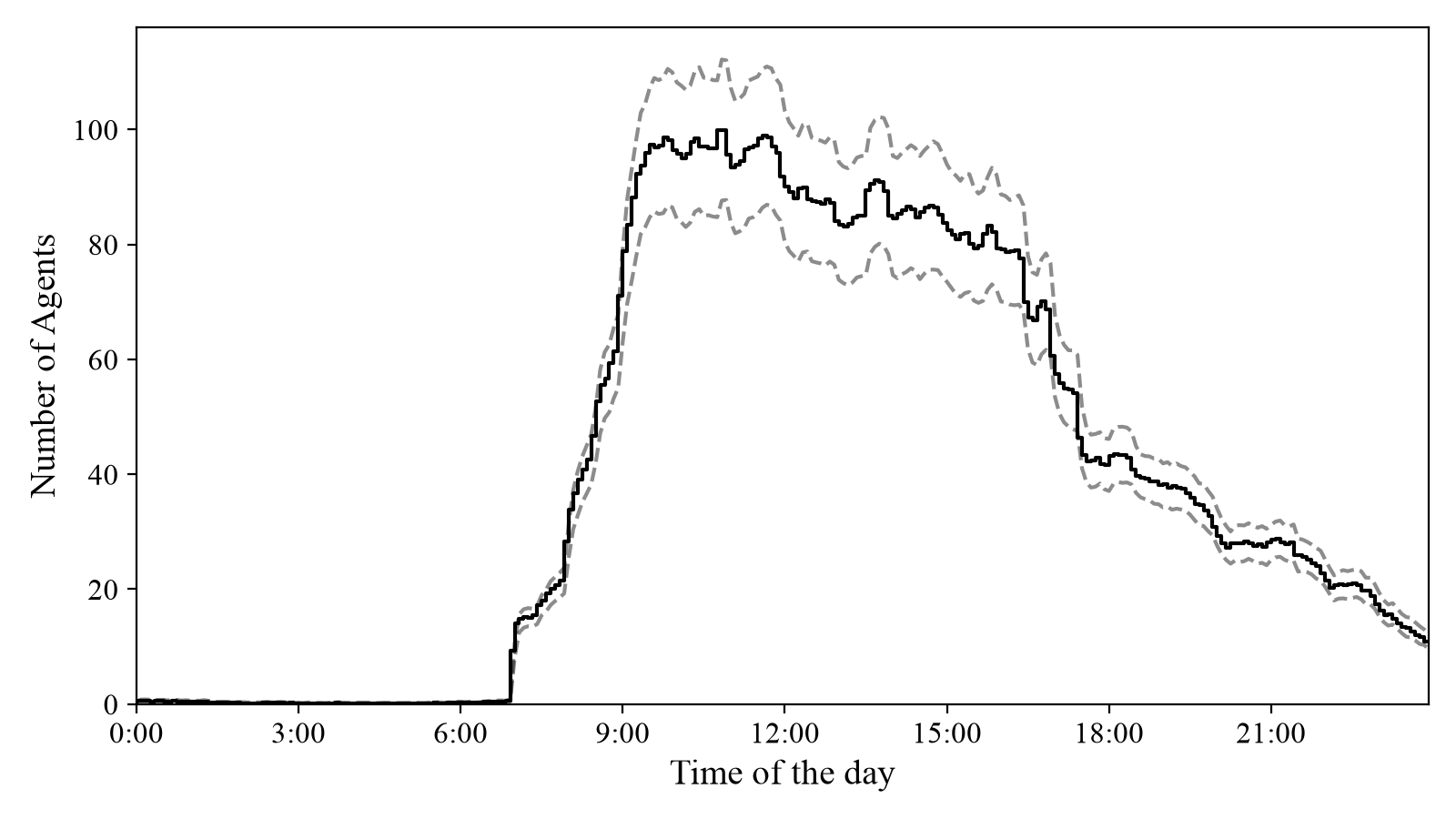}
         \captionsetup{font={footnotesize}}
         \caption{Staffing level of pool 2}
     \end{minipage}
\end{figure}

\begin{figure}[H]
     \centering
     \begin{minipage}{0.48\textwidth}
         \centering
         \includegraphics[width=\linewidth]{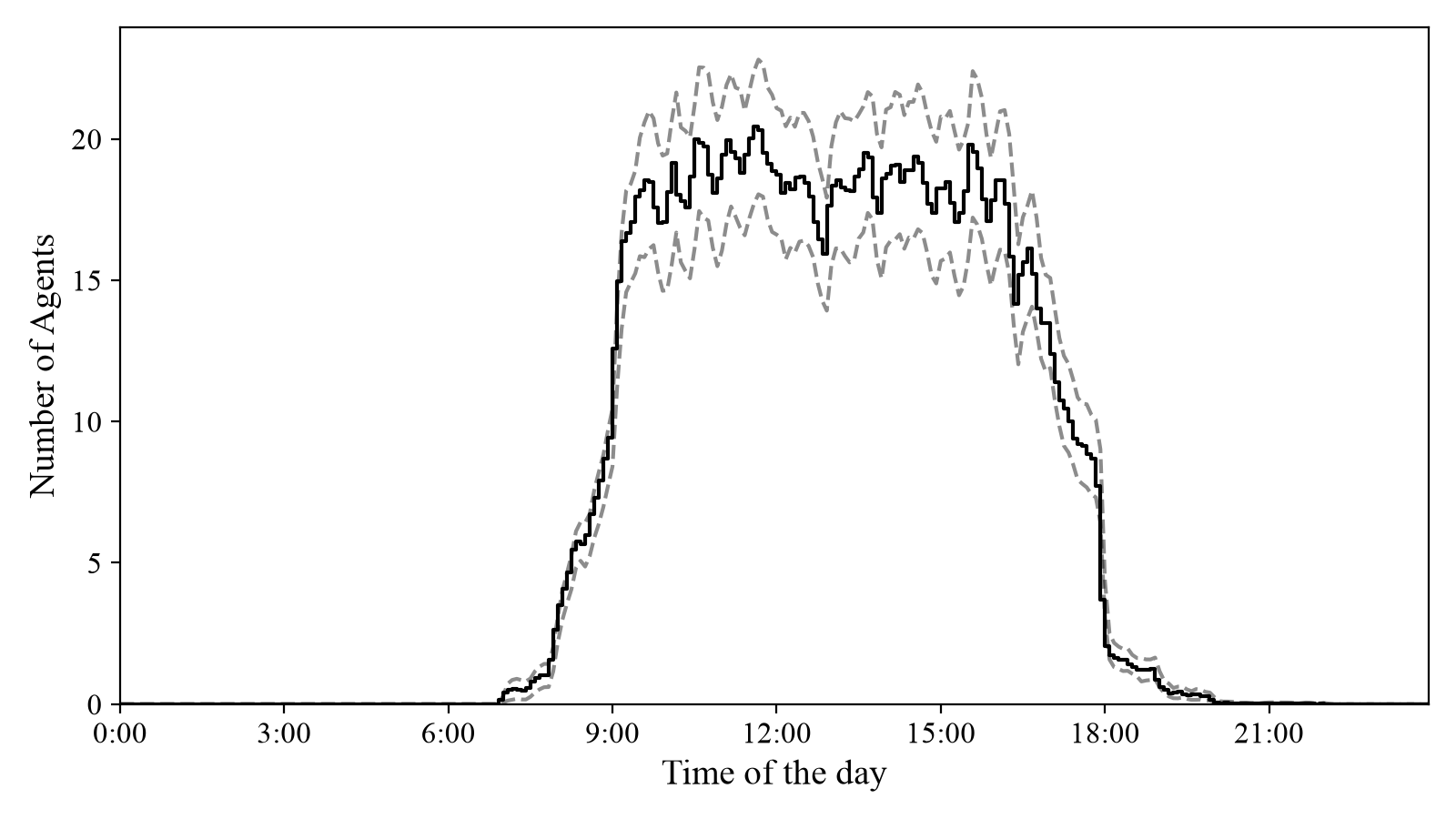}
         \captionsetup{font={footnotesize}}
         \caption{Staffing level of pool 9}
         \label{graph_staffing_last}
     \end{minipage}
     \hfill
     \begin{minipage}{0.48\textwidth}
     \end{minipage}
\end{figure}

Figures~\ref{graph_staffing_first}--\ref{graph_staffing_last} display the
average number of agents on duty in each of the nine server pools over
the day.

\begin{figure}[H]
     \centering
     \begin{minipage}{0.48\textwidth}
         \centering
         \includegraphics[width=\linewidth]{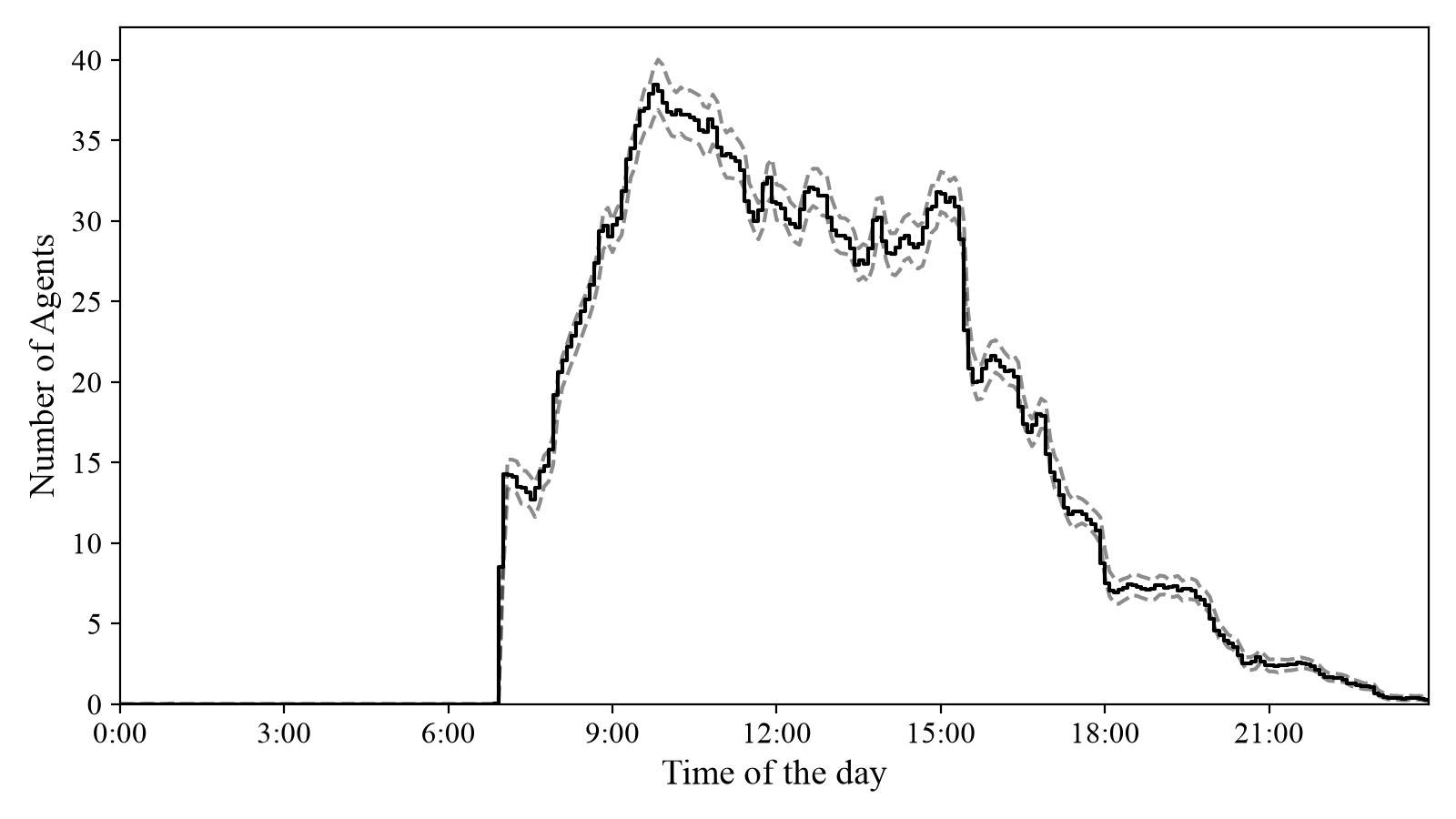}
         \captionsetup{font={footnotesize}}
         \caption{Staffing level of pool 3}
     \end{minipage}
     \hfill
     \begin{minipage}{0.48\textwidth}
         \centering
         \includegraphics[width=\linewidth]{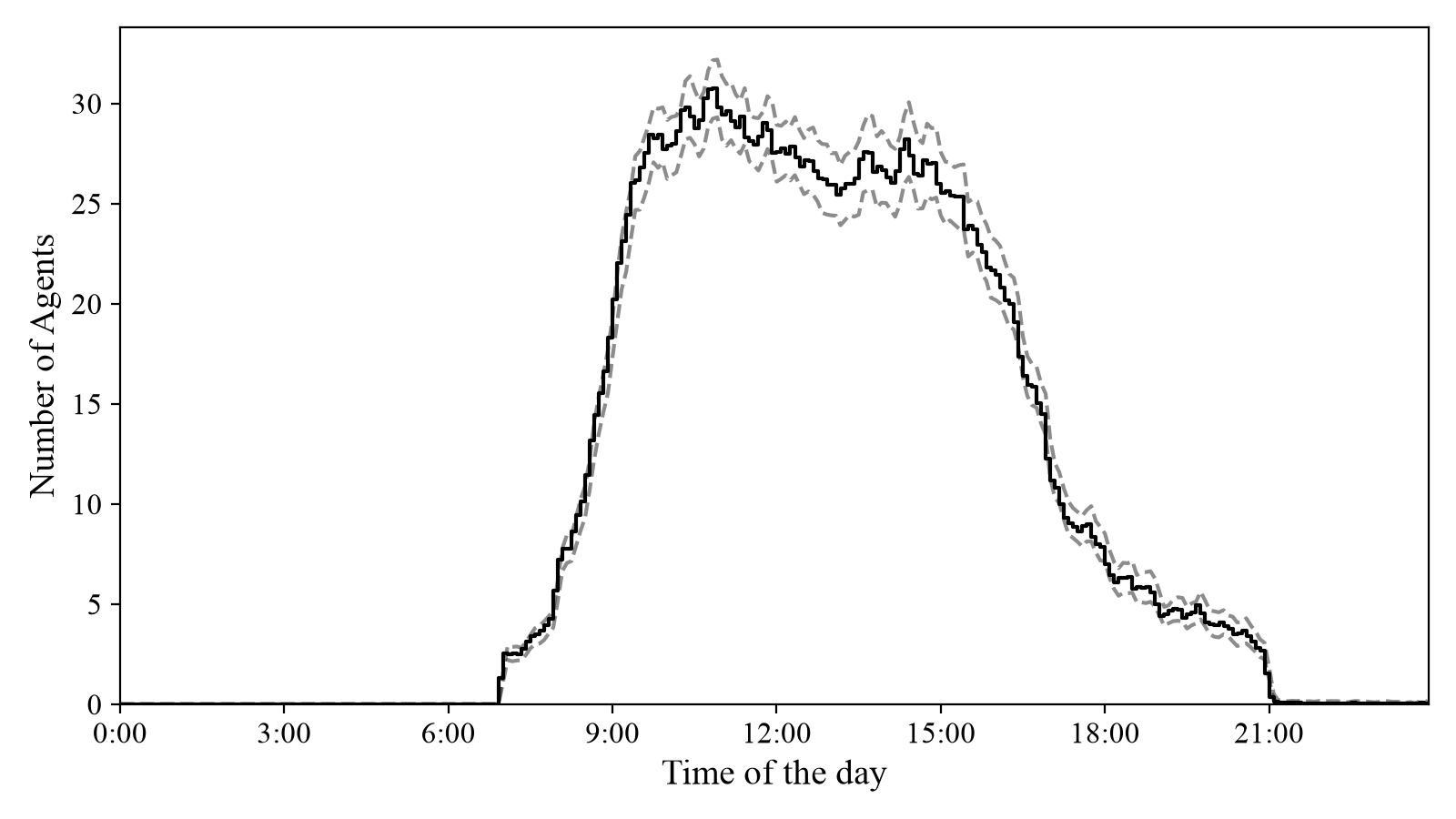}
         \captionsetup{font={footnotesize}}
         \caption{Staffing level of pool 4}
     \end{minipage}
\end{figure}
\begin{figure}[H]
     \centering
     \begin{minipage}{0.48\textwidth}
         \centering
         \includegraphics[width=\linewidth]{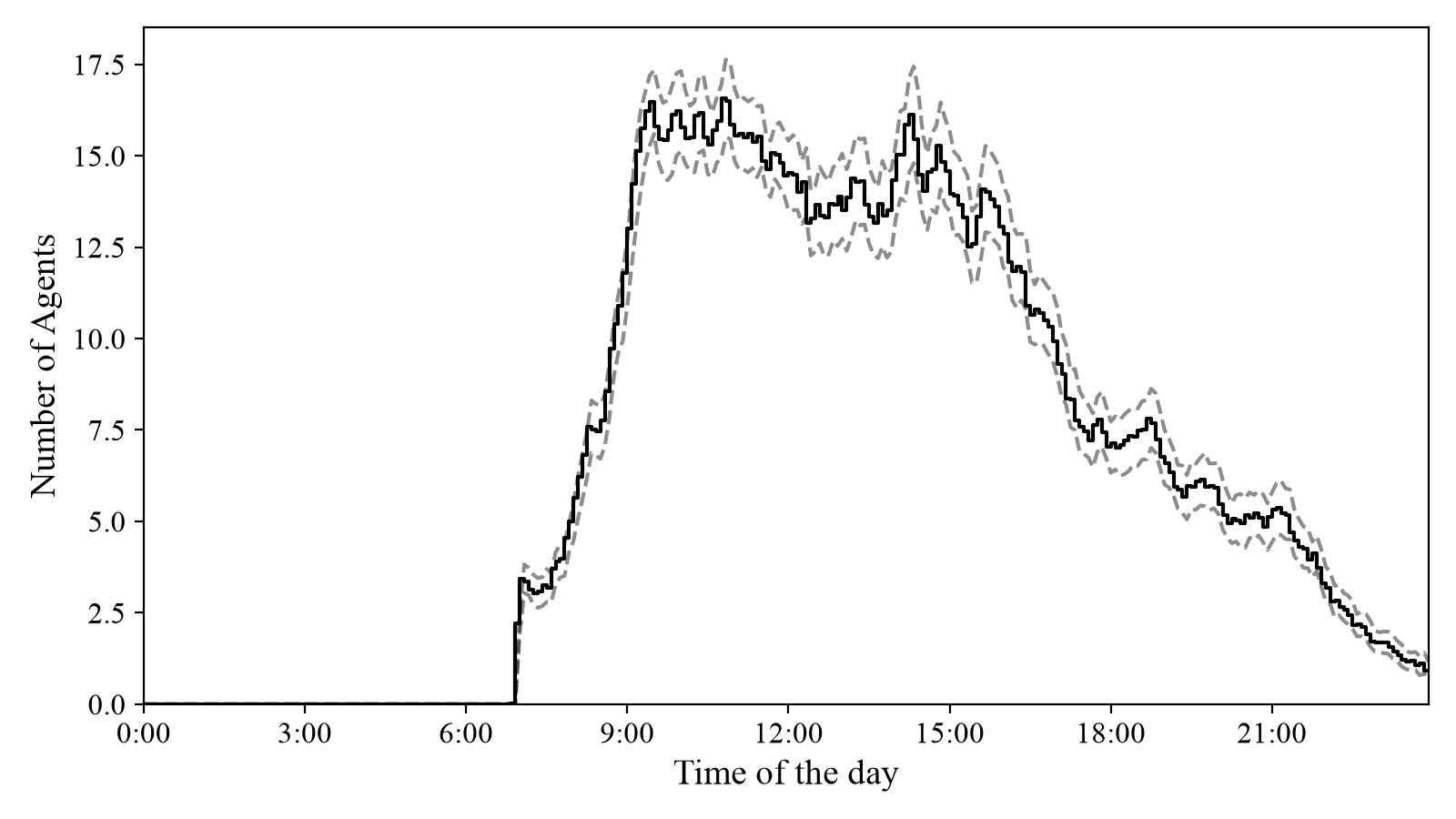}
         \captionsetup{font={footnotesize}}
         \caption{Staffing level of pool 5}
     \end{minipage}
     \hfill
     \begin{minipage}{0.48\textwidth}
         \centering
         \includegraphics[width=\linewidth]{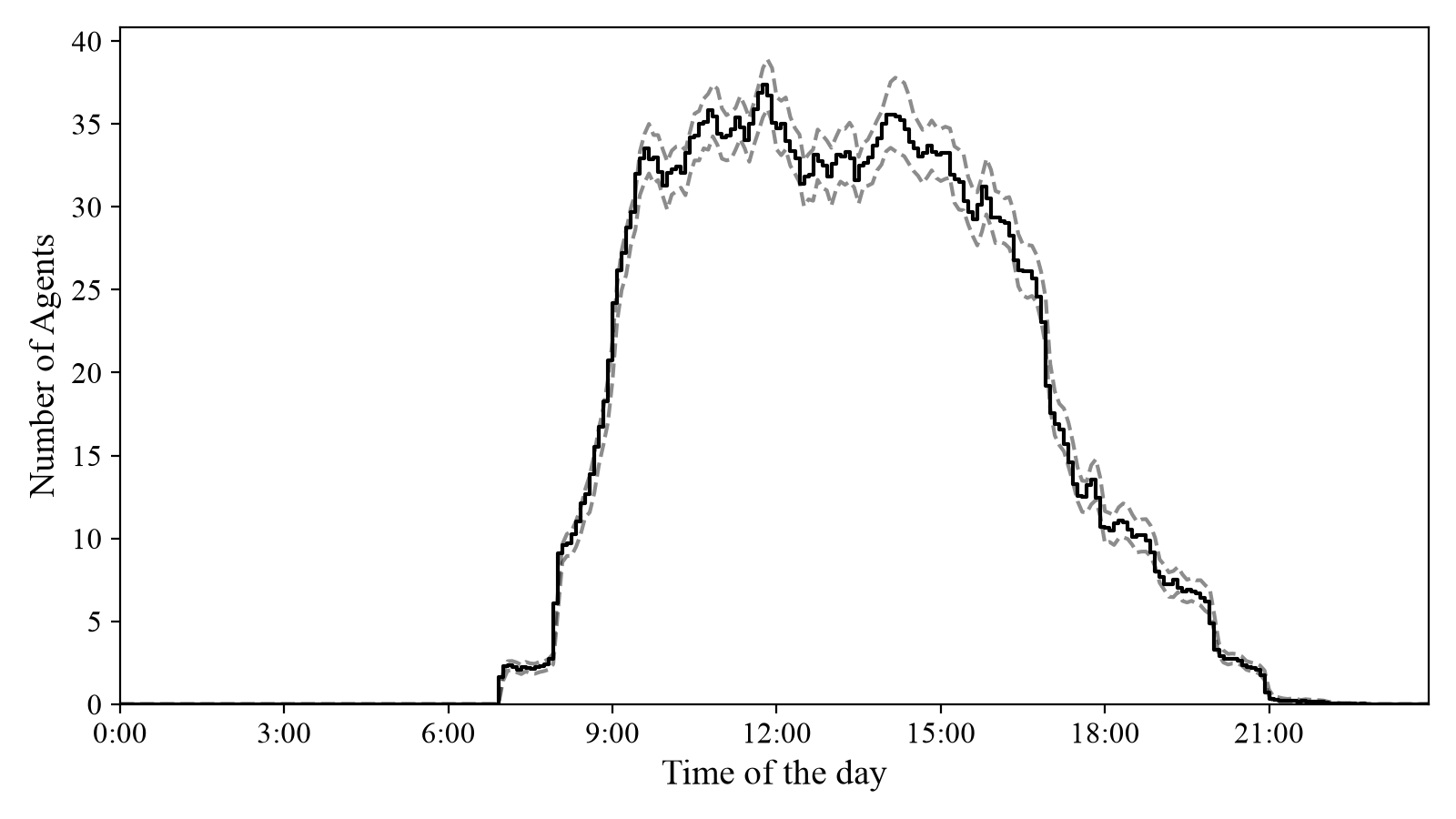}
         \captionsetup{font={footnotesize}}
         \caption{Staffing level of pool 6}
     \end{minipage}
\end{figure}
\begin{figure}[H]
     \centering
     \begin{minipage}{0.48\textwidth}
         \centering
         \includegraphics[width=\linewidth]{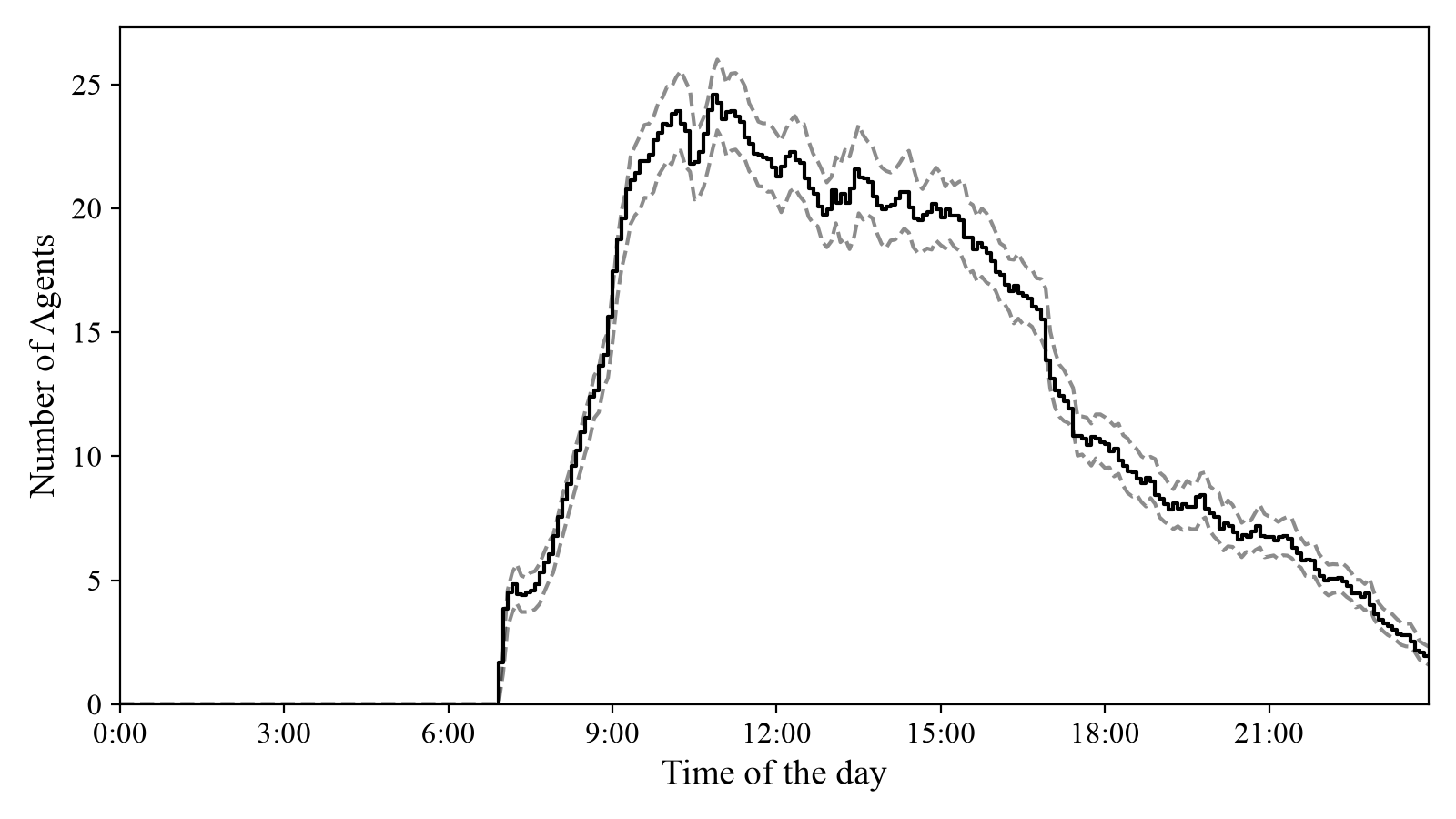}
         \captionsetup{font={footnotesize}}
         \caption{Staffing level of pool 7}
     \end{minipage}
     \hfill
     \begin{minipage}{0.48\textwidth}
         \centering
         \includegraphics[width=\linewidth]{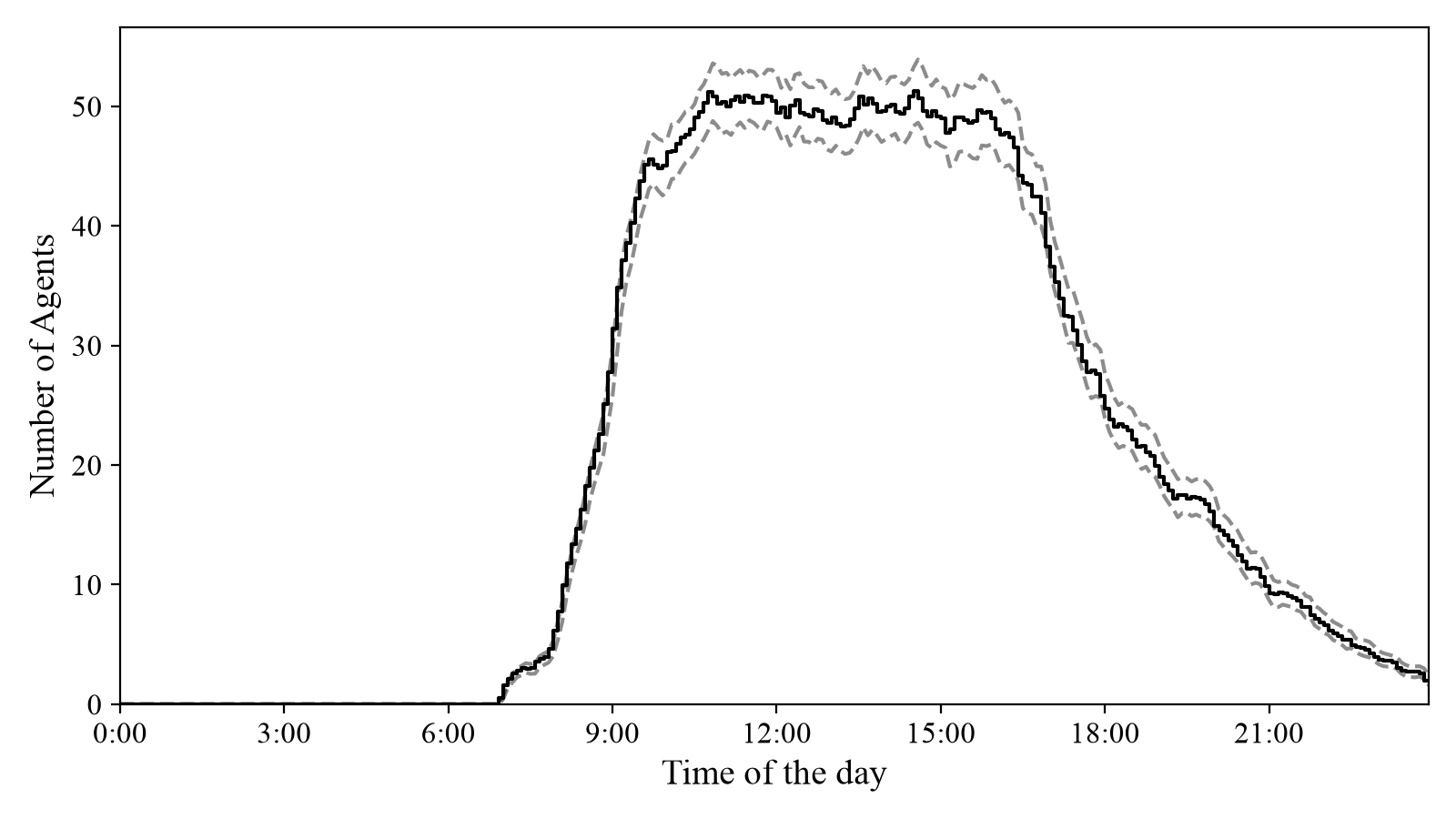}
         \captionsetup{font={footnotesize}}
         \caption{Staffing level of pool 8}
     \end{minipage}
\end{figure}
\vspace{-10mm}

\subsection{Activities of the Network}\label{app:data-activities}
Table~\ref{tab:activities} lists the $J=62$ activities of the network
and fixes their indexing.  Following the ordering notation of
Section~\ref{sec:model}, activities $1,\ldots,41$ serve the original
classes and activities $42,\ldots,62$ serve the rework classes, ordered
by server pool and, within a pool, by class.  For each activity~$j$, the
table reports the server pool $s(j)$ and the class $b(j)$.

\begin{table}[H]
\centering
\scriptsize
\setlength{\tabcolsep}{4.5pt}
\renewcommand{\arraystretch}{0.85}
\begin{tabular}{rcc@{\hskip 18pt}rcc}
\toprule
\multicolumn{3}{c}{Original-class activities ($\calJ_o$)} &
\multicolumn{3}{c}{Rework-class activities ($\calJ_r$)}\\
\cmidrule(r{18pt}){1-3}\cmidrule{4-6}
$j$ & $s(j)$ & $b(j)$ & $j$ & $s(j)$ & $b(j)$\\
\midrule
1 & 1 & R1 & 42 & 1 & R1\\
2 & 1 & R2 & 43 & 1 & R2\\
3 & 1 & R3 & 44 & 1 & R3\\
4 & 1 & Bus & 45 & 2 & R2\\
5 & 1 & Tel & 46 & 2 & R3\\
6 & 2 & R2 & 47 & 2 & Sub\\
7 & 2 & R3 & 48 & 3 & R1\\
8 & 2 & Pre & 49 & 3 & R2\\
9 & 2 & Bus & 50 & 3 & EBO\\
10 & 3 & R1 & 51 & 4 & Bus\\
11 & 3 & R2 & 52 & 4 & Plt\\
12 & 3 & Pre & 53 & 5 & Pre\\
13 & 3 & Bus & 54 & 6 & R3\\
14 & 3 & EBO & 55 & 6 & CL\\
15 & 4 & R2 & 56 & 7 & OB\\
16 & 4 & R3 & 57 & 8 & Tel\\
17 & 4 & Bus & 58 & 9 & R2\\
18 & 4 & Plt & 59 & 9 & R3\\
19 & 4 & Tel & 60 & 9 & Sub\\
20 & 5 & R2 & 61 & 9 & CQ\\
21 & 5 & R3 & 62 & 9 & PS\\
22 & 5 & Pre &  &  & \\
23 & 5 & Bus &  &  & \\
24 & 5 & CL &  &  & \\
25 & 6 & R1 &  &  & \\
26 & 6 & R3 &  &  & \\
27 & 6 & CL &  &  & \\
28 & 7 & R1 &  &  & \\
29 & 7 & R3 &  &  & \\
30 & 7 & Pre &  &  & \\
31 & 7 & Bus &  &  & \\
32 & 7 & CL &  &  & \\
33 & 7 & OB &  &  & \\
34 & 8 & R3 &  &  & \\
35 & 8 & Tel &  &  & \\
36 & 9 & R1 &  &  & \\
37 & 9 & R3 &  &  & \\
38 & 9 & Pre &  &  & \\
39 & 9 & Sub &  &  & \\
40 & 9 & CQ &  &  & \\
41 & 9 & PS &  &  & \\
\bottomrule
\end{tabular}
\caption{The $62$ activities of the network.}
\label{tab:activities}
\end{table}
\newpage
\subsection{Class--Pool Service Shares}\label{app:data-shares}

Tables~\ref{tab:shares-orig} and~\ref{tab:shares-rework} report, for
each class, the percentage of its services performed by each server pool (according to raw data), in the original and rework layers, respectively.  The percentages
are computed over all service observations of the class in the
corresponding layer.
\begin{table}[H]
\centering
\footnotesize
\setlength{\tabcolsep}{4.5pt}
\begin{tabular}{lrrrrrrrrr}
\toprule
Class & \multicolumn{9}{c}{Pool}\\
\cmidrule{2-10}
 & 1 & 2 & 3 & 4 & 5 & 6 & 7 & 8 & 9\\
\midrule
Retail (Node: 1) & 78.5 & -- & 15.8 & -- & -- & 1.8 & 0.5 & -- & 3.3\\
Retail (Node: 2) & 24.4 & 66.3 & 7.1 & 0.8 & 1.4 & -- & -- & -- & --\\
Retail (Node: 3) & 20.3 & 62.8 & -- & 0.8 & 2.1 & 4.4 & 1.8 & 0.1 & 7.6\\
Premier & 0.8 & 2.6 & 0.6 & 0.1 & 94.1 & -- & 0.1 & -- & 1.6\\
Business & 0.5 & 3.1 & 0.1 & 95.0 & 1.1 & -- & 0.2 & -- & --\\
Platinum & 0.2 & 0.5 & -- & 99.3 & -- & -- & -- & -- & --\\
Consumer Loans & 0.2 & -- & -- & -- & 0.8 & 98.9 & 0.1 & 0.1 & --\\
Online Banking & -- & 0.1 & 0.2 & -- & -- & -- & 99.7 & -- & --\\
EBO & 0.1 & 0.1 & 99.6 & -- & -- & -- & 0.1 & -- & --\\
Telesales & 0.5 & -- & -- & 0.8 & -- & -- & -- & 98.6 & --\\
Subanco & 0.1 & -- & -- & -- & -- & -- & -- & -- & 99.9\\
Case Quality & -- & -- & -- & -- & -- & -- & -- & -- & 100.0\\
Priority Service & -- & -- & -- & -- & -- & -- & -- & -- & 100.0\\
\bottomrule
\end{tabular}
\caption{Percentage of original class services performed by
each server pool in the data.}
\label{tab:shares-orig}
\end{table}

\begin{table}[H]
\centering
\footnotesize
\setlength{\tabcolsep}{4.5pt}
\begin{tabular}{lrrrrrrrrr}
\toprule
Class & \multicolumn{9}{c}{Pool}\\
\cmidrule{2-10}
 & 1 & 2 & 3 & 4 & 5 & 6 & 7 & 8 & 9\\
\midrule
Retail (Node: 1) & 79.1 & 0.1 & 16.8 & -- & -- & 1.8 & 0.2 & -- & 1.8\\
Retail (Node: 2) & 25.2 & 64.1 & 6.5 & 1.1 & 0.9 & -- & -- & -- & 2.2\\
Retail (Node: 3) & 21.2 & 61.7 & -- & 0.8 & 2.0 & 4.5 & 1.3 & 1.3 & 7.3\\
Premier & 0.3 & 1.7 & 0.1 & 0.2 & 96.1 & -- & -- & -- & 1.5\\
Business & 0.2 & 0.6 & -- & 99.1 & -- & -- & -- & 0.1 & --\\
Platinum & -- & 0.6 & -- & 98.4 & -- & -- & -- & -- & 0.9\\
Consumer Loans & -- & -- & -- & -- & -- & 99.7 & 0.1 & 0.1 & --\\
Online Banking & -- & -- & 0.3 & -- & -- & -- & 99.6 & -- & --\\
EBO & -- & -- & 99.9 & -- & -- & -- & 0.1 & -- & --\\
Telesales & -- & -- & -- & -- & -- & -- & -- & 99.9 & --\\
Subanco & 0.1 & 12.0 & -- & -- & -- & 5.2 & 0.1 & 1.0 & 81.6\\
Case Quality & -- & -- & -- & 0.1 & -- & -- & -- & -- & 99.9\\
Priority Service & -- & 0.1 & 0.1 & -- & -- & -- & -- & -- & 99.9\\
\bottomrule
\end{tabular}
\caption{Percentage of rework class services performed by
each server pool in the data.}
\label{tab:shares-rework}
\end{table}
\newpage
\subsection{Service Rates by Activity}\label{app:data-service}

Tables~\ref{tab:mu-orig} and~\ref{tab:mu-rework} report the estimated
hourly service rates $\mu_j$ for the activities serving the original and
rework classes, respectively.  Dashes denote (class, pool) pairs that are
not activities of the network, that is, pairs that account for less than
$0.1\%$ of the pool's services in the original layer or fewer than $100$
services in the rework layer.

\begin{table}[H]
\centering
\footnotesize
\setlength{\tabcolsep}{3.5pt}
\begin{tabular}{crrrrrrrrrrrrr}
\toprule
Pool & R1 & R2 & R3 & Pre & Bus & Plt & CL & OB & EBO & Tel & Sub & CQ & PS\\
\midrule
1 & 16.57 & 16.53 & 15.84 & -- & 17.26 & -- & -- & -- & -- & 17.74 & -- & -- & --\\
2 & -- & 16.33 & 16.80 & 13.22 & 16.52 & -- & -- & -- & -- & -- & -- & -- & --\\
3 & 16.17 & 15.21 & -- & 12.58 & 17.38 & -- & -- & -- & 9.01 & -- & -- & -- & --\\
4 & -- & 16.89 & 16.78 & -- & 15.68 & 15.44 & -- & -- & -- & 15.43 & -- & -- & --\\
5 & -- & 16.33 & 16.32 & 13.55 & 16.91 & -- & 13.90 & -- & -- & -- & -- & -- & --\\
6 & 17.46 & -- & 18.30 & -- & -- & -- & 15.15 & -- & -- & -- & -- & -- & --\\
7 & 16.61 & -- & 15.24 & 12.13 & 14.35 & -- & 13.18 & 10.68 & -- & -- & -- & -- & --\\
8 & -- & -- & 27.78 & -- & -- & -- & -- & -- & -- & 9.39 & -- & -- & --\\
9 & 16.82 & -- & 16.75 & 15.21 & -- & -- & -- & -- & -- & -- & 10.73 & 11.34 & 11.00\\
\bottomrule
\end{tabular}
\caption{Hourly service rates for the $41$ activities serving the
original classes.  Rows are server pools, columns
original classes.}
\label{tab:mu-orig}
\end{table}

\begin{table}[H]
\centering
\footnotesize
\setlength{\tabcolsep}{3.5pt}
\begin{tabular}{crrrrrrrrrrrrr}
\toprule
Pool & R1 & R2 & R3 & Pre & Bus & Plt & CL & OB & EBO & Tel & Sub & CQ & PS\\
\midrule
1 & 14.27 & 13.14 & 13.94 & -- & -- & -- & -- & -- & -- & -- & -- & -- & --\\
2 & -- & 13.21 & 14.31 & -- & -- & -- & -- & -- & -- & -- & 9.33 & -- & --\\
3 & 14.03 & 11.84 & -- & -- & -- & -- & -- & -- & 6.53 & -- & -- & -- & --\\
4 & -- & -- & -- & -- & 10.31 & 12.95 & -- & -- & -- & -- & -- & -- & --\\
5 & -- & -- & -- & 9.75 & -- & -- & -- & -- & -- & -- & -- & -- & --\\
6 & -- & -- & 15.61 & -- & -- & -- & 11.88 & -- & -- & -- & -- & -- & --\\
7 & -- & -- & -- & -- & -- & -- & -- & 8.05 & -- & -- & -- & -- & --\\
8 & -- & -- & -- & -- & -- & -- & -- & -- & -- & 6.12 & -- & -- & --\\
9 & -- & 9.31 & 11.40 & -- & -- & -- & -- & -- & -- & -- & 8.82 & 12.88 & 11.46\\
\bottomrule
\end{tabular}
\caption{Hourly service rates for the $21$ activities serving the rework
classes.  Rows are server pools, columns rework
classes.}
\label{tab:mu-rework}
\end{table}
\newpage
Table~\ref{tab:mu-activities} lists the same service rate estimates by
activity, following the activity indexing of Table~\ref{tab:activities}.

\begin{table}[H]
\centering
\footnotesize
\setlength{\tabcolsep}{4.5pt}
\begin{tabular}{rlr@{\hskip 18pt}rlr}
\toprule
\multicolumn{3}{c}{Original-class activities ($\calJ_o$)} &
\multicolumn{3}{c}{Rework-class activities ($\calJ_r$)}\\
\cmidrule(r{18pt}){1-3}\cmidrule{4-6}
$j$ & $(s(j),\,b(j))$ & $\mu_j$ & $j$ & $(s(j),\,b(j))$ & $\mu_j$\\
\midrule
1 & (1, R1) & 16.57 & 42 & (1, R1) & 14.27\\
2 & (1, R2) & 16.53 & 43 & (1, R2) & 13.14\\
3 & (1, R3) & 15.84 & 44 & (1, R3) & 13.94\\
4 & (1, Bus) & 17.26 & 45 & (2, R2) & 13.21\\
5 & (1, Tel) & 17.74 & 46 & (2, R3) & 14.31\\
6 & (2, R2) & 16.33 & 47 & (2, Sub) & 9.33\\
7 & (2, R3) & 16.80 & 48 & (3, R1) & 14.03\\
8 & (2, Pre) & 13.22 & 49 & (3, R2) & 11.84\\
9 & (2, Bus) & 16.52 & 50 & (3, EBO) & 6.53\\
10 & (3, R1) & 16.17 & 51 & (4, Bus) & 10.31\\
11 & (3, R2) & 15.21 & 52 & (4, Plt) & 12.95\\
12 & (3, Pre) & 12.58 & 53 & (5, Pre) & 9.75\\
13 & (3, Bus) & 17.38 & 54 & (6, R3) & 15.61\\
14 & (3, EBO) & 9.01 & 55 & (6, CL) & 11.88\\
15 & (4, R2) & 16.89 & 56 & (7, OB) & 8.05\\
16 & (4, R3) & 16.78 & 57 & (8, Tel) & 6.12\\
17 & (4, Bus) & 15.68 & 58 & (9, R2) & 9.31\\
18 & (4, Plt) & 15.44 & 59 & (9, R3) & 11.40\\
19 & (4, Tel) & 15.43 & 60 & (9, Sub) & 8.82\\
20 & (5, R2) & 16.33 & 61 & (9, CQ) & 12.88\\
21 & (5, R3) & 16.32 & 62 & (9, PS) & 11.46\\
22 & (5, Pre) & 13.55 &  & &\\
23 & (5, Bus) & 16.91 &  & &\\
24 & (5, CL) & 13.90 &  & &\\
25 & (6, R1) & 17.46 &  & &\\
26 & (6, R3) & 18.30 &  & &\\
27 & (6, CL) & 15.15 &  & &\\
28 & (7, R1) & 16.61 &  & &\\
29 & (7, R3) & 15.24 &  & &\\
30 & (7, Pre) & 12.13 &  & &\\
31 & (7, Bus) & 14.35 &  & &\\
32 & (7, CL) & 13.18 &  & &\\
33 & (7, OB) & 10.68 &  & &\\
34 & (8, R3) & 27.78 &  & &\\
35 & (8, Tel) & 9.39 &  & &\\
36 & (9, R1) & 16.82 &  & &\\
37 & (9, R3) & 16.75 &  & &\\
38 & (9, Pre) & 15.21 &  & &\\
39 & (9, Sub) & 10.73 &  & &\\
40 & (9, CQ) & 11.34 &  & &\\
41 & (9, PS) & 11.00 &  & &\\
\bottomrule
\end{tabular}
\caption{Hourly service rates by activity.
Activities are indexed as in Table~\ref{tab:activities}.}
\label{tab:mu-activities}
\end{table}

\subsection{Routing Probabilities}\label{app:data-routing}

Table~\ref{tab:P-or} reports the estimated routing
probabilities~\eqref{eq:phat} in the form of the routing matrix $P$
of~\eqref{eq:P-block}.  By its block structure, $P$ has a single
nonzero block $P^{o}$, whose rows are the activities serving
the original classes and whose columns are the rework classes.  The
remaining blocks are zero because a rework completion exits the system
and no completion is routed into an original class.  Row~$j$ of Table~\ref{tab:P-or} gives the
probability that a first agent service delivered through activity~$j$
is followed by a second agent service of each rework class, and the
exit probability is $q_j = 1-\sum_{l\in\calI_r}P_{jl}$.

\begin{table}[H]
\centering
\footnotesize
\setlength{\tabcolsep}{2.5pt}
\begin{tabular}{rlrrrrrrrrrrrrr}
\toprule
$j$ & $(s(j),\,b(j))$ & R1 & R2 & R3 & Pre & Bus & Plt & CL & OB & EBO & Tel & Sub & CQ & PS\\
\midrule
1 & (1, R1) & 0.45 & 0.01 & 0.01 & -- & -- & -- & 1.16 & 0.68 & 0.86 & -- & 0.03 & -- & --\\
2 & (1, R2) & 0.02 & 0.51 & 0.01 & 0.04 & 0.33 & -- & -- & -- & 0.73 & -- & -- & -- & --\\
3 & (1, R3) & 0.05 & 0.02 & 0.51 & 0.07 & 0.46 & -- & 1.29 & 0.76 & -- & 1.09 & -- & -- & --\\
4 & (1, Bus) & 0.07 & 0.36 & 0.07 & -- & 1.07 & -- & -- & 0.21 & -- & -- & -- & -- & --\\
5 & (1, Tel) & -- & -- & -- & -- & -- & -- & 1.75 & 0.34 & -- & -- & -- & -- & --\\
6 & (2, R2) & 0.01 & 0.41 & 0.01 & 0.05 & 0.34 & -- & -- & -- & 0.80 & -- & -- & -- & --\\
7 & (2, R3) & 0.04 & 0.02 & 0.49 & 0.07 & 0.38 & -- & 1.25 & 0.92 & -- & 1.05 & -- & -- & --\\
8 & (2, Pre) & -- & 0.11 & 0.03 & 2.88 & 0.08 & -- & 0.91 & 1.06 & 0.24 & 0.29 & -- & -- & --\\
9 & (2, Bus) & 0.02 & 0.24 & 0.04 & -- & 6.12 & -- & 0.05 & 0.28 & 0.07 & 0.12 & -- & -- & --\\
10 & (3, R1) & 0.35 & 0.01 & 0.01 & -- & -- & -- & 1.12 & 0.57 & 0.03 & -- & 0.03 & -- & --\\
11 & (3, R2) & 0.03 & 0.55 & 0.01 & 0.06 & 0.23 & -- & -- & -- & 0.01 & -- & -- & -- & --\\
12 & (3, Pre) & -- & -- & -- & 3.41 & -- & -- & -- & -- & -- & -- & -- & -- & --\\
13 & (3, Bus) & -- & -- & -- & -- & 1.55 & -- & -- & -- & -- & -- & -- & -- & --\\
14 & (3, EBO) & 0.03 & 0.03 & -- & 0.01 & 0.01 & -- & 0.02 & 0.14 & 0.44 & -- & 0.03 & -- & --\\
15 & (4, R2) & 0.01 & 0.51 & 0.03 & 0.04 & 0.03 & -- & -- & -- & 0.52 & -- & -- & -- & --\\
16 & (4, R3) & 0.03 & 0.01 & 0.26 & 0.07 & 0.01 & -- & 1.51 & 0.66 & -- & 1.03 & -- & -- & --\\
17 & (4, Bus) & -- & 0.02 & 0.01 & -- & 0.41 & -- & 0.08 & 0.15 & 0.16 & 0.21 & -- & -- & --\\
18 & (4, Plt) & -- & -- & -- & -- & 0.07 & 1.57 & -- & -- & 0.10 & -- & -- & -- & --\\
19 & (4, Tel) & -- & -- & -- & -- & 0.04 & -- & -- & -- & 0.08 & 0.21 & -- & -- & --\\
20 & (5, R2) & 0.02 & 0.35 & 0.02 & 0.04 & 0.16 & -- & -- & -- & 0.05 & -- & -- & -- & --\\
21 & (5, R3) & 0.02 & 0.01 & 0.50 & 0.08 & 0.29 & -- & 0.21 & 0.39 & -- & 2.57 & -- & -- & --\\
22 & (5, Pre) & -- & -- & -- & 0.89 & 0.10 & -- & 0.07 & 0.42 & 0.02 & 0.36 & -- & -- & --\\
23 & (5, Bus) & 0.03 & 0.07 & 0.03 & 0.03 & 14.17 & -- & 0.07 & 0.21 & 0.03 & 0.83 & -- & -- & --\\
24 & (5, CL) & -- & -- & -- & 0.13 & 0.18 & -- & 13.24 & 0.36 & -- & 4.31 & -- & -- & --\\
25 & (6, R1) & 0.56 & 0.01 & 0.02 & -- & -- & -- & 0.03 & 0.75 & 0.80 & -- & 0.02 & -- & --\\
26 & (6, R3) & 0.09 & 0.02 & 0.79 & 0.05 & 0.65 & -- & 0.04 & 0.95 & -- & 1.13 & -- & -- & --\\
27 & (6, CL) & 0.03 & -- & 0.04 & 0.01 & 0.17 & -- & 0.31 & 0.11 & 0.02 & 2.46 & 0.02 & -- & --\\
28 & (7, R1) & 0.42 & -- & 0.03 & -- & -- & -- & 1.24 & 0.10 & 1.12 & -- & 0.02 & -- & --\\
29 & (7, R3) & 0.05 & 0.03 & 0.26 & 0.03 & 0.34 & -- & 1.16 & 0.01 & -- & 1.17 & -- & -- & --\\
30 & (7, Pre) & -- & -- & -- & 0.95 & 0.95 & -- & 4.27 & -- & -- & 0.47 & -- & -- & --\\
31 & (7, Bus) & -- & -- & 0.22 & -- & 16.59 & -- & 0.90 & -- & -- & 0.45 & -- & -- & --\\
32 & (7, CL) & -- & -- & -- & -- & -- & -- & 0.26 & -- & -- & -- & -- & -- & --\\
33 & (7, OB) & 0.06 & -- & 0.05 & 0.01 & 0.35 & -- & 1.51 & 0.39 & 0.12 & 0.46 & 0.02 & -- & --\\
34 & (8, R3) & -- & -- & -- & -- & -- & -- & -- & -- & -- & -- & -- & -- & --\\
35 & (8, Tel) & 0.01 & -- & 0.08 & 0.02 & 0.14 & -- & 0.09 & 0.03 & -- & 0.34 & -- & -- & --\\
36 & (9, R1) & 0.52 & 0.03 & 0.01 & -- & -- & -- & 1.40 & 0.64 & 0.69 & -- & -- & -- & --\\
37 & (9, R3) & 0.02 & 0.02 & 0.41 & 0.12 & 0.71 & -- & 1.33 & 0.72 & -- & 1.00 & -- & -- & --\\
38 & (9, Pre) & -- & -- & -- & 3.68 & 0.08 & -- & 2.72 & 1.80 & -- & 1.00 & -- & -- & --\\
39 & (9, Sub) & 0.02 & -- & -- & -- & 0.04 & -- & 0.44 & 0.05 & 0.13 & 5.29 & 1.97 & -- & --\\
40 & (9, CQ) & 0.01 & -- & -- & -- & 0.05 & -- & 0.01 & 0.05 & 0.07 & -- & -- & 7.70 & 0.07\\
41 & (9, PS) & -- & -- & -- & -- & -- & -- & -- & 0.01 & 0.02 & -- & -- & -- & 8.01\\
\bottomrule
\end{tabular}
\caption{The nonzero block $P^{o}$ of the estimated routing
matrix~\eqref{eq:P-block}, in percentages.  Rows are
the original-class activities $j\in\calJ_o$ as indexed in
Table~\ref{tab:activities} and columns are the rework classes
$l\in\calI_r$, labeled by the class abbreviations of
Table~\ref{tab:P-class}.}
\label{tab:P-or}
\end{table}

Table~\ref{tab:P-class} aggregates the activity-level estimates to the
class level,
\begin{equation}\label{eq:P-class}
  P_{il} = \sum_{j\in\mathcal{B}(i)} \alpha_{ij}\,P_{jl},
  \qquad i\in\calI_o,\ l\in\calI_r,
\end{equation}
where $\alpha_{ij}$ is the share of the services of class $i$ performed
by activity $j$, reported in Table~\ref{tab:shares-orig}.  Rows are original
classes and columns rework classes.  Each row sums to the rework
percentage reported in Table~\ref{tab:stats}, and the diagonal gives
same-type rework.

\begin{table}[H]
\centering
\footnotesize
\setlength{\tabcolsep}{3pt}
\begin{tabular}{lrrrrrrrrrrrrr}
\toprule
$i$\,$\backslash$\,$l$ & R1 & R2 & R3 & Pre & Bus & Plt & CL & OB & EBO & Tel & Sub & CQ & PS\\
\midrule
Retail (Node: 1) (R1) & 0.44 & 0.01 & 0.01 & -- & -- & -- & 1.14 & 0.66 & 0.73 & -- & 0.03 & -- & --\\
Retail (Node: 2) (R2) & 0.02 & 0.45 & 0.01 & 0.05 & 0.33 & -- & -- & -- & 0.72 & -- & -- & -- & --\\
Retail (Node: 3) (R3) & 0.04 & 0.02 & 0.49 & 0.07 & 0.43 & -- & 1.19 & 0.84 & -- & 1.09 & -- & -- & --\\
Premier (Pre) & -- & -- & -- & 1.02 & 0.10 & -- & 0.15 & 0.46 & 0.03 & 0.37 & -- & -- & --\\
Business (Bus) & -- & 0.03 & 0.01 & -- & 0.77 & -- & 0.08 & 0.15 & 0.15 & 0.21 & -- & -- & --\\
Platinum (Plt) & -- & -- & -- & -- & 0.07 & 1.58 & -- & -- & 0.10 & -- & -- & -- & --\\
Consumer Loans (CL) & 0.03 & -- & 0.04 & 0.01 & 0.17 & -- & 0.43 & 0.11 & 0.02 & 2.47 & 0.02 & -- & --\\
Online Banking (OB) & 0.06 & -- & 0.05 & 0.01 & 0.35 & -- & 1.51 & 0.39 & 0.12 & 0.46 & 0.02 & -- & --\\
EBO & 0.03 & 0.03 & -- & 0.01 & 0.01 & -- & 0.02 & 0.14 & 0.44 & -- & 0.03 & -- & --\\
Telesales (Tel) & 0.01 & -- & 0.08 & 0.02 & 0.13 & -- & 0.10 & 0.03 & -- & 0.34 & -- & -- & --\\
Subanco (Sub) & 0.02 & -- & -- & -- & 0.04 & -- & 0.44 & 0.05 & 0.13 & 5.29 & 1.97 & -- & --\\
Case Quality (CQ) & 0.01 & -- & -- & -- & 0.05 & -- & 0.01 & 0.05 & 0.07 & -- & -- & 7.70 & 0.07\\
Priority Service (PS) & -- & -- & -- & -- & -- & -- & -- & 0.01 & 0.02 & -- & -- & -- & 8.01\\
\bottomrule
\end{tabular}
\caption{Estimated class-level routing probabilities $P_{il}$ (in percentages)
of Equation~\eqref{eq:P-class}, aggregated over the activities serving
each original class.  Rows are the original classes $i\in\calI_o$, with
their abbreviations in parentheses, and columns are the rework classes
$l\in\calI_r$, labeled by the same abbreviations.}
\label{tab:P-class}
\end{table}
\section{Scaled and Centered Objective of the Test Problems}
\label{app:scaled-objective}

The expected net values in
Tables~\ref{tab:results-main}--\ref{tab:results-summary} are dominated
by the value of the nominal plan, which is of order $n$.  The policies differ at the order $\sqrt n$, which is
the scale of the Brownian control problem of Section~\ref{sec:bcp} and
of the HJB equation that our computational method solves.  To compare
the policies on the diffusion scale, we report the centered and scaled process
$\hat\Pi^n(T)$ of Equation~\eqref{eq:scaled-reward-process} at the end
of the horizon.  The nominal plan $\xi^*(\cdot)$, the limiting staffing
levels $\nu(\cdot)$, and the mean activity values $\bv_j$ are the same
in the three test problems, so the cumulative value earned by the
nominal plan is the same in all three,
\begin{equation}\label{eq:nominal-value-T}
  n\int_0^T \tilde v(s)\cdot\xi^*(s)\,ds
  = n\sum_{m=0}^{M-1}\sum_{j\in\calJ}\bv_j\,\mu_j\,\nu_{s(j)}(\tau_m)\,
    \xi^*_j(\tau_m)\,(\tau_{m+1}-\tau_m)
  = \$143{,}894.2.
\end{equation}
For each policy, the scaled cost is
\begin{equation}\label{eq:scaled-cost-T}
  \E\big[\hat\Pi^n(T)\big]
  = \frac{n\int_0^T \tilde v(s)\cdot\xi^*(s)\,ds - \E[\Pi^n(T)]}{\sqrt n},
\end{equation}
with $n = 100$, where $\E[\Pi^n(T)]$ is the expected net value
reported in Tables~\ref{tab:results-main}--\ref{tab:results-second}.
Table~\ref{tab:results-scaled} reports the scaled cost of our policy
and of every benchmark policy in the three test problems.

\begin{table}[ht]
\centering
{\small
\begin{tabular}{lccc}
\toprule
Policy & Main test problem & First variant & Second variant\\
\midrule
Our policy & $861.45 \pm 1.53$ & $421.46 \pm 1.54$ & $-28.28 \pm 1.51$\\
Nominal-plan rule & $875.26 \pm 1.59$ & $488.10 \pm 1.83$ & $501.57 \pm 3.21$\\
Ab.-adj.\ myopic & $1{,}267.85 \pm 1.71$ & $675.22 \pm 1.68$ & $128.86 \pm 1.65$\\
Myopic & $1{,}269.00 \pm 1.69$ & $674.16 \pm 1.68$ & $123.03 \pm 1.64$\\
Value-rate & $1{,}272.19 \pm 1.68$ & $677.37 \pm 1.67$ & $126.07 \pm 1.63$\\
FSF & $2{,}424.16 \pm 1.91$ & $1{,}751.17 \pm 1.91$ & $1{,}122.46 \pm 1.81$\\
$c\mu/\theta$ & $2{,}425.90 \pm 1.91$ & $1{,}759.21 \pm 1.90$ & $1{,}142.77 \pm 1.78$\\
$c\mu$ & $2{,}472.53 \pm 1.93$ & $1{,}797.76 \pm 1.90$ & $1{,}169.10 \pm 1.79$\\
\midrule
Performance gap & $13.81 \pm 2.04$ & $66.63 \pm 2.18$ & $151.30 \pm 2.05$\\
\bottomrule
\end{tabular}
}
\caption{The scaled cost of Equation~\eqref{eq:scaled-cost-T} of our
policy and of the benchmark policies in the three test problems, with
the half-width of the $99\%$ confidence interval.  The last row reports
the gap between the best benchmark and our policy, the scaled cost of
the best benchmark minus that of our policy, with the half-width of its
$99\%$ confidence intervals.  The
best benchmark is the nominal-plan rule in the main test problem and
the first variant, and the myopic rule in the second variant.}
\label{tab:results-scaled}
\end{table}

Table~\ref{tab:results-scaled} shows that the expected net value of the
best benchmark is close to that of our policy because every policy
earns the value of the nominal plan up to a correction of order
$\sqrt n$.  The second-order performance separates the policies.  Our
policy reduces the scaled cost of the best benchmark by
$13.81$ in the main test problem and by $66.63$ in
the first variant, which are $1.6\%$ and
$13.7\%$ of the scaled cost of the best benchmark.
In the second variant, the scaled cost of our policy is negative, so
our policy earns more than the value of the nominal plan, while the
scaled cost of the best benchmark is $123.03$.  A negative scaled cost is
possible because the prelimit arrival rates of the second variant
exceed $n\lambda(\cdot)$ on most intervals.

\end{document}